\documentclass[11pt]{article}

\usepackage[final]{acl}
\usepackage{geometry}
\usepackage{authblk}
\usepackage{pdflscape}
\usepackage{subcaption}
\usepackage{makecell}
\usepackage{multirow}
\usepackage{dblfloatfix}
\usepackage[most]{tcolorbox}
\definecolor{high}{RGB}{0, 102, 202} 
\newtcolorbox{prompt}[2][]{
    colback=gray!20,
    colframe=white,
    fonttitle=\bfseries\small,
    boxrule=0.4mm,
    fontupper=\small, 
    fontlower=\small,
    coltitle=white,
    title=#2,
    #1,breakable
}
\usepackage[T1,T2A]{fontenc}
\usepackage[utf8]{inputenc}
\usepackage[ukrainian,english]{babel}
\usepackage{CJKutf8}
\newcolumntype{C}[1]{>{\centering\arraybackslash}p{#1}}
\usepackage{times}
\usepackage{latexsym}
\usepackage{booktabs}
\usepackage[T1]{fontenc}
\usepackage[utf8]{inputenc}
\usepackage{microtype}
\usepackage{inconsolata}
\usepackage{graphicx}
\usepackage{float}
\usepackage{makecell}
\usepackage[table]{xcolor}
\usepackage{xspace}

\newcommand*{\gemini}{\texttt{Gemini3.1-Pro}\xspace}
\newcommand*{\qwen}{\texttt{Qwen3-Omni}\xspace}
\newcommand*{\gemma}{\texttt{Gemma4-12B}\xspace}

\newcommand{\blfootnote}[1]{%
  \begingroup
  \renewcommand{\thefootnote}{}\footnote{#1}%
  \addtocounter{footnote}{-1}%
  \endgroup
}

\title{VoxSumm: A Multilingual Corpus of Long-Form Spoken News for Joint Summarization and Translation}

\author{
Yejin Jeon$^{1,2\dagger}$,
Marie Maltais$^{1,2\dagger}$,
Virginia Ceccatelli$^{1,2}$,
Min Ma$^{3}$,
David Ifeoluwa Adelani$^{1,2,4}$ \\
$^{1}$Mila - Quebec AI Institute 
$^{2}$McGill University, Canada \\
$^{3}$Google DeepMind 
$^{4}$Canada CIFAR AI Chair \\
}

\begin{document}
\maketitle
\blfootnote{$\dagger$ denotes equal contribution.}
\begin{abstract}
As information increasingly traverses linguistic boundaries, users require concise cross-lingual representations of long-form content. Nevertheless, long-document summarization research remains text-centric, whereas multilingual speech research has largely prioritized translation, preserving source content rather than compressing it. We address this methodological gap by formalizing \textbf{joint speech summarization and translation (JSumT)}: the generation of a succinct, faithful target-language summary directly from a long spoken document in a source language. We additionally introduce \textbf{\textsc{VoxSumm}}, the first multilingual and cross-lingual benchmark for this task, comprising 10,045 BBC article-summary pairs across 24 languages and encompassing approximately 703 hours of speech data. Our evaluation of representative speech-language models reveals pronounced variation across models and generation settings: Gemini3.1-Pro demonstrates the greatest consistency, summarization into English generally surpasses generation into non-English target languages, and translating an entire document before summarization compounds instruction-following failures. Through the release of \textsc{VoxSumm}, we establish a foundation for developing and evaluating multilingual systems capable of jointly interpreting, compressing, and translating long-form speech.
\end{abstract}

\section{Introduction}

Spoken language has emerged as a primary medium for communication and information consumption. 
With the rapid expansion of digital media, the availability of such spoken content has grown substantially, while individual recordings have become increasingly longer and more information-rich. This abundance creates significant challenges for listeners who must efficiently comprehend, identify, and retain the most relevant information from long-form spoken contents \citep{clifton-etal-2020-100000,retkowski-etal-2025-summarizing}. Speech summarization aims to address this challenge by transforming extended spoken recordings into concise overview of an article 
that preserve their most essential content. 

Despite its practical significance, long-form speech summarization has received substantially less attention than its text-based counterpart \citep{retkowski-etal-2025-summarizing}. It is important to note that long-form speech summarization is not merely text summarization applied to a different input modality. Rather, spoken language conveys information through both lexical content and paralinguistic signals, including prosody, emphasis, hesitation patterns, pronunciation variation, and disfluencies. These acoustic cues can meaningfully influence how information is interpreted and summarized. In fact, access to the original speech signal, rather than a transcript alone, has been shown to affect both content selection and factual consistency in summary generation \citep{sharma-etal-2024-speech}. Beyond these representational differences, long-form speech introduces significant computational challenges. Acoustic sequences are substantially longer than their textual counterparts, resulting in increased memory requirements and greater training and inference complexity, thereby complicating both model development and evaluation \citep{kano-etal-2023-long,sharma-etal-2024-r}.

Beyond the challenges associated with the speech modality, existing research is also constrained by the limited availability of resources that support broad linguistic diversity. Although real-world data has become increasingly multilingual, summarization research remains largely concentrated on a small subset of high-resource languages, particularly English \citep{ICLR2024_f7b77476}. While there are multilingual summarization benchmarks such as XL-Sum \citep{hasan-etal-2021-xl} and CrossSum \citep{bhattacharjee-etal-2023-crosssum}, which provide extensive language coverage, these are restricted to textual domains. Meanwhile, speech-based multilingual datasets such as MuST-C \citep{di-gangi-etal-2019-mustc},  CoVoST~2 \citep{wang-etal-2021-covost}, and Fleurs~\cite{conneau2023fleurs} primarily target speech translation, where the objective is to preserve the complete semantic content of the source utterance rather than identify and compress its most salient information. As a result, existing resources do not enable the evaluation of whether a model can comprehend long spoken documents, extract their key information, and generate concise summaries across languages.

To address these gaps, we formalize the \emph{joint speech summarization and translation (JSumT)} task, which requires a model to generate a faithful and concise summary in a target language when given a long spoken document in a source language. This setting captures a realistic scenario in which users consume spoken content produced in one language but require condensed information in another. To support systematic evaluation for the JSumT task, we present \textsc{VoxSumm}, a multilingual long-form speech summarization benchmark that is constructed from long-form BBC text-based news articles and their cross-lingual summaries. Specifically, \textsc{VoxSumm} contains approximately 703 hours of 10,045 article-summary pairs spanning 24 languages. Using this benchmark, we investigate representative speech language models, and the impact of different methodologies, including zero-shot, few-shot, and Chain-of-Thought (CoT)-based prompting. Finally, to facilitate future multilingual speech summarization research, we release the \textsc{VoxSumm} dataset and accompanying code.

\vspace{-5pt}
\section{Related Work}
\vspace{-5pt}
Summarization research has primarily developed in the text domain, progressing from sentence-level benchmarks such as CNN/DailyMail \citep{hermann-etal-2015-teaching} and XSum \citep{narayan-etal-2018-dont} to long-context and multilingual settings. Recent studies have explored long-document summarization through long-context language models and hierarchical architectures \citep{ICLR2024_f7b77476}, while multilingual benchmarks such as MLSUM \citep{scialom-etal-2020-mlsum}, XL-Sum \citep{hasan-etal-2021-xl}, WikiLingua \citep{ladhak-etal-2020-wikilingua}, and CrossSum \citep{bhattacharjee-etal-2023-crosssum} have enabled cross-lingual summarization through large-scale article-summary collections. However, these resources are limited to textual inputs and do not address the challenges of long-form speech summarization.

Speech introduces unique challenges, including the absence of explicit structural cues such as punctuation and paragraph boundaries and the presence of paralinguistic information that is often lost in ASR-based processing \citep{rehbein-etal-2020-improving,zechner-waibel-2000-diasumm,sharma-etal-2024-speech}. Multilingual spoken language research has therefore focused primarily on speech translation, with approaches ranging from cascaded ASR--translation pipelines to speech-to-unit and end-to-end systems \citep{qian2023polyvoice,pu2025empowering,barrault2023seamless,alastruey2026omnilingual}.

In contrast, speech summarization aims to extract and compress salient information from spoken documents. Prior studies have explored multimodal summarization \citep{palaskar-etal-2019-multimodal}, meeting summarization \citep{zhong-etal-2021-qmsum}, efficient modeling of long speech sequences \citep{kano-etal-2023-long}, and relevance-based content selection \citep{sharma-etal-2024-r}. Recent work further shows that speech-based summarization differs from transcript-based summarization in information selection and factual consistency, highlighting the importance of modeling speech beyond transcription \citep{sharma-etal-2024-speech}. Nevertheless, existing speech summarization resources remain largely monolingual, while multilingual speech benchmarks focus predominantly on translation rather than compression. We address this gap by introducing the \emph{joint speech summarization and translation (JSumT)} task and \textsc{VoxSumm}, a benchmark for multilingual and cross-lingual long-form speech summarization.

\vspace{-5pt}
\section{VoxSumm Dataset}
\label{sec:dataset}
\vspace{-5pt}

\begin{table}[t]
\centering
\resizebox{0.74\linewidth}{!}{%
\begin{tabular}{lrrr}
\toprule
\textbf{Language} & \textbf{Hours} & \textbf{CER ($\downarrow$)} & \textbf{NISQA ($\uparrow$)} \\
\midrule
Amharic           &26.69 & 21.55 & 3.87 \\
Arabic            &26.68 & 4.26  & 4.34 \\
Bengali           &31.30 & 7.50  & 4.38 \\
Chinese           &26.79 & 9.46 & 4.58 \\
French            &	32.75 & 3.61  & 4.40 \\
Gujarati          &	34.32 & 6.80  & 4.42 \\
Hindi             &	26.34 & 4.86  & 4.35 \\
Indonesian        &24.30 & 5.19  & 4.41 \\
Japanese          &32.28 & 9.84  & 4.72 \\
Korean            &	26.67 & 5.98  & 4.46\\
Kyrgyz            &	28.83 & 4.98  & 4.43 \\
Persian           &	31.56 & 4.55  & 4.41 \\
Portuguese        &	33.23 & 3.52  & 4.41 \\
Punjabi           &31.50 & 10.44 & 4.37 \\
Russian           &26.36 & 3.98  & 4.37 \\
Sinhala           &	24.93 & 7.17  & 4.48 \\
Spanish           &33.91 & 3.32  & 4.42 \\
Swahili           &	23.93 & 4.40  & 4.31 \\
Tamil             &29.88 & 7.13  & 4.38\\
Telugu            &	31.02 & 8.38  & 4.42 \\
Thai              &	34.38 & 9.28  & 4.40 \\
Turkish           &	24.00 & 4.71  & 4.42 \\
Ukrainian         &	23.47 & 4.05  & 4.29 \\
Vietnamese        &	27.35 & 4.70  & 4.41 \\
\midrule
\textbf{Average} &\textbf{28.90} & \textbf{6.65} & \textbf{4.39} \\
\bottomrule
\end{tabular}%
}
\vspace{-2mm}
\caption{\textbf{Dataset statistics and accepted quality criteria}: CER and NISQA are calculated over every language. CER is conducted with \texttt{omniASR\_CTC\_1B\_v2}.}
\label{tab:cer_nisqa}
\end{table}

We construct \textsc{VoxSumm} as an English-centric multilingual benchmark for long-form speech summarization across languages. The construction pipeline consists of three stages: (1) collecting cross-lingual article-summary pairs, (2) synthesizing both source articles and target summaries into speech, and (3) validating the generated speech through automatic and human evaluation.

\vspace{-5pt}
\subsection{Multilingual Article--Summary Collection}
\label{sec:splits}

The textual foundation of \textsc{VoxSumm} is derived from CC BY-NCSA4.0 CrossSum \citep{bhattacharjee-etal-2023-crosssum}, a large-scale cross-lingual summarization dataset built by aligning monolingual article-summary pairs from XL-Sum \citep{hasan-etal-2021-xl} across languages. Specifically, CrossSum identifies articles reporting on the same underlying event across different languages by computing multilingual sentence-embedding similarity (via LaBSE) between article summaries, and pairs each source-language article with the summary of its most semantically similar counterpart in a different target language. This yields cross-lingual article-summary pairs spanning a wide range of languages, from which we construct \textsc{VoxSumm}.

From this resource, we retain only instances for which the source URL, target URL, article body, and summary are all present and non-empty. Additionally, since CrossSum's language coverage varies substantially in size, we exclude any language with fewer than 205 paired articles to ensure sufficient data per language for meaningful evaluation.
Note that each instance pairs a source-language article with a URL to its aligned counterpart in a target language, which serves as an identifier for retrieving that article's professionally written summary. As a result, each raw instance inherently encodes only a single direction: a source article paired with a target-language summary. However, our objective is to support \emph{bidirectional} multilingual summarization evaluation, e.g., producing both an \textit{English-to-French} and a \textit{French-to-English} summarization instance from the same underlying article pair, rather than treating the two directions as unrelated content. This requires a way to recognize when two directional instances originate from the same source--target article pair, so that their roles can be swapped consistently.

\begin{figure}[t]
    \centering
    \includegraphics[width=1\linewidth]{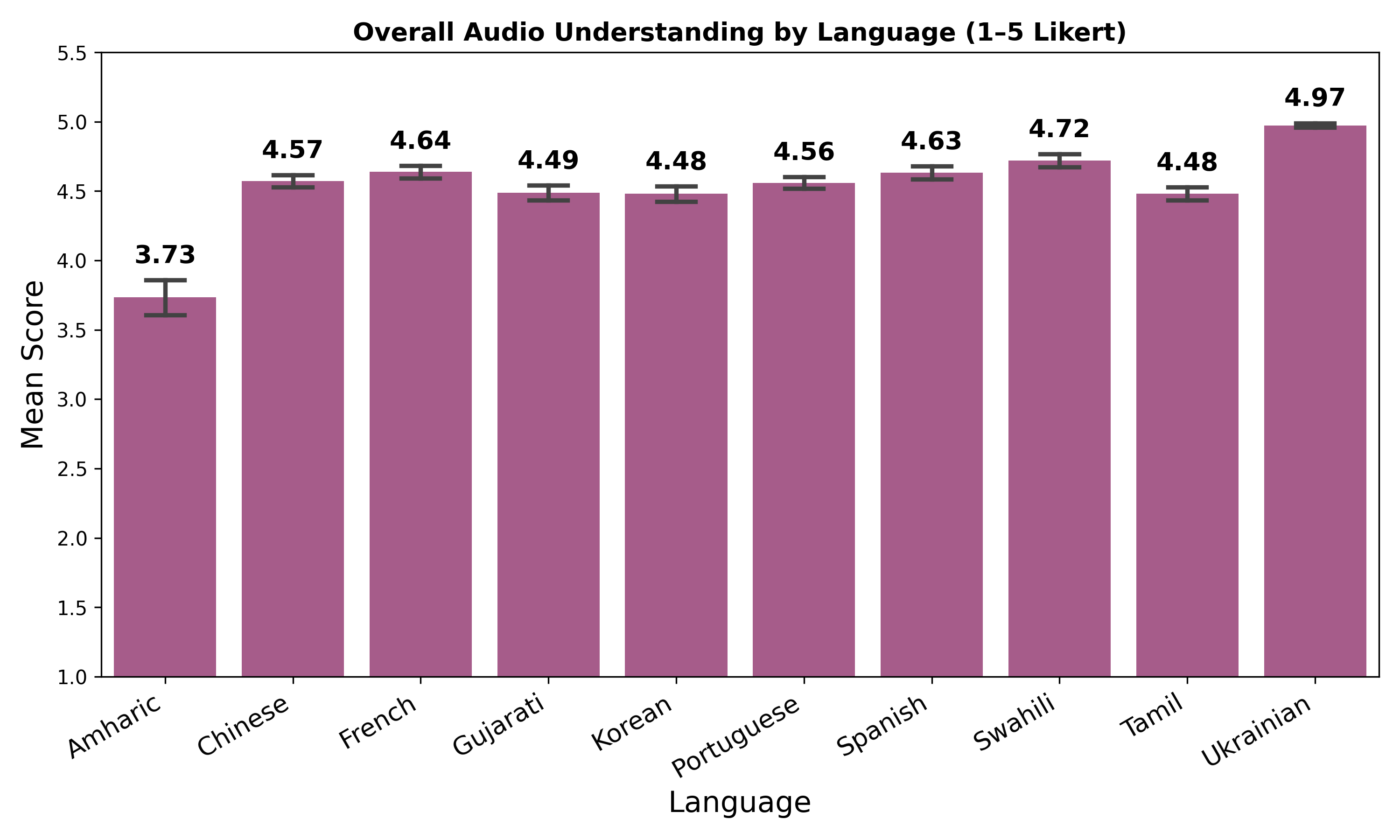}
    \vspace{-5mm}
\caption{\textbf{Human subjective listening evaluation of audio quality}. This is measured using a 1-5 Likert scale averaged over a sample of 50 utterances per language.}
    \label{fig:external_human_quality}
\end{figure}

\begin{table*}[t]
\centering
\resizebox{0.75\textwidth}{!}{%
\begin{tabular}{ l *{12}{C{1.4cm}} }
\toprule
\multirow{2}{*}{\textbf{Pair}} & \multicolumn{4}{c}{\textbf{\gemini}} & \multicolumn{4}{c}{\textbf{Gemma}} & \multicolumn{4}{c}{\textbf{\qwen}} \\
\cmidrule(lr){2-5} \cmidrule(lr){6-9} \cmidrule(lr){10-13}
 & FS & ZS & CoT & Avg & FS & ZS & CoT & Avg & FS & ZS & CoT & Avg \\
\midrule
am & \cellcolor{purple!50} \makecell{0.780 \\[-2pt] {\footnotesize (0.374)}}  & \cellcolor{purple!48} \makecell{0.716 \\[-2pt] {\footnotesize (0.274)}}  & \cellcolor{purple!46} \makecell{0.713 \\[-2pt] {\footnotesize (0.276)}}  & \cellcolor{gray!15} \makecell{0.736 \\[-2pt] {\footnotesize (0.308)}}  & \cellcolor{purple!39} \makecell{0.624 \\[-2pt] {\footnotesize (0.220)}}  & \cellcolor{purple!41} \makecell{0.623 \\[-2pt] {\footnotesize (0.214)}}  & \cellcolor{purple!41} \makecell{0.623 \\[-2pt] {\footnotesize (0.206)}}  & \cellcolor{gray!15} \makecell{0.623 \\[-2pt] {\footnotesize (0.213)}}  & \cellcolor{purple!12} \makecell{0.589 \\[-2pt] {\footnotesize (0.194)}}  & \cellcolor{purple!23} \makecell{0.637 \\[-2pt] {\footnotesize (0.184)}}  & \cellcolor{purple!21} \makecell{0.627 \\[-2pt] {\footnotesize (0.172)}}  & \cellcolor{gray!15} \makecell{0.618 \\[-2pt] {\footnotesize (0.184)}} \\[3pt]
ar & \cellcolor{purple!52} \makecell{0.792 \\[-2pt] {\footnotesize (0.428)}}  & \cellcolor{purple!52} \makecell{0.756 \\[-2pt] {\footnotesize (0.309)}}  & \cellcolor{purple!52} \makecell{0.759 \\[-2pt] {\footnotesize (0.282)}}  & \cellcolor{gray!15} \makecell{0.769 \\[-2pt] {\footnotesize (0.340)}}  & \cellcolor{purple!52} \makecell{0.658 \\[-2pt] {\footnotesize (0.251)}}  & \cellcolor{purple!52} \makecell{0.685 \\[-2pt] {\footnotesize (0.233)}}  & \cellcolor{purple!52} \makecell{0.682 \\[-2pt] {\footnotesize (0.216)}}  & \cellcolor{gray!15} \makecell{0.675 \\[-2pt] {\footnotesize (0.233)}}  & \cellcolor{purple!52} \makecell{0.745 \\[-2pt] {\footnotesize (0.350)}}  & \cellcolor{purple!52} \makecell{0.729 \\[-2pt] {\footnotesize (0.245)}}  & \cellcolor{purple!50} \makecell{0.714 \\[-2pt] {\footnotesize (0.194)}}  & \cellcolor{gray!15} \makecell{0.729 \\[-2pt] {\footnotesize (0.263)}} \\[3pt]
bn & \cellcolor{purple!23} \makecell{0.716 \\[-2pt] {\footnotesize (0.336)}}  & \cellcolor{purple!23} \makecell{0.672 \\[-2pt] {\footnotesize (0.232)}}  & \cellcolor{purple!21} \makecell{0.671 \\[-2pt] {\footnotesize (0.246)}}  & \cellcolor{gray!15} \makecell{0.686 \\[-2pt] {\footnotesize (0.271)}}  & \cellcolor{purple!10} \makecell{0.560 \\[-2pt] {\footnotesize (0.203)}}  & \cellcolor{purple!29} \makecell{0.608 \\[-2pt] {\footnotesize (0.206)}}  & \cellcolor{purple!22} \makecell{0.597 \\[-2pt] {\footnotesize (0.199)}}  & \cellcolor{gray!15} \makecell{0.588 \\[-2pt] {\footnotesize (0.203)}}  & \cellcolor{purple!23} \makecell{0.646 \\[-2pt] {\footnotesize (0.242)}}  & \cellcolor{purple!21} \makecell{0.634 \\[-2pt] {\footnotesize (0.203)}}  & \cellcolor{purple!23} \makecell{0.627 \\[-2pt] {\footnotesize (0.183)}}  & \cellcolor{gray!15} \makecell{0.636 \\[-2pt] {\footnotesize (0.210)}} \\[3pt]
zh & \cellcolor{purple!54} \makecell{0.802 \\[-2pt] {\footnotesize (0.496)}}  & \cellcolor{purple!54} \makecell{0.778 \\[-2pt] {\footnotesize (0.388)}}  & \cellcolor{purple!54} \makecell{0.775 \\[-2pt] {\footnotesize (0.401)}}  & \cellcolor{gray!15} \makecell{0.785 \\[-2pt] {\footnotesize (0.428)}}  & \cellcolor{purple!54} \makecell{0.716 \\[-2pt] {\footnotesize (0.279)}}  & \cellcolor{purple!54} \makecell{0.723 \\[-2pt] {\footnotesize (0.286)}}  & \cellcolor{purple!54} \makecell{0.726 \\[-2pt] {\footnotesize (0.243)}}  & \cellcolor{gray!15} \makecell{0.722 \\[-2pt] {\footnotesize (0.269)}}  & \cellcolor{purple!54} \makecell{0.770 \\[-2pt] {\footnotesize (0.410)}}  & \cellcolor{purple!54} \makecell{0.763 \\[-2pt] {\footnotesize (0.330)}}  & \cellcolor{purple!54} \makecell{0.751 \\[-2pt] {\footnotesize (0.261)}}  & \cellcolor{gray!15} \makecell{0.762 \\[-2pt] {\footnotesize (0.334)}} \\[3pt]
fr & \cellcolor{purple!44} \makecell{0.745 \\[-2pt] {\footnotesize (0.380)}}  & \cellcolor{purple!46} \makecell{0.715 \\[-2pt] {\footnotesize (0.281)}}  & \cellcolor{purple!43} \makecell{0.710 \\[-2pt] {\footnotesize (0.267)}}  & \cellcolor{gray!15} \makecell{0.724 \\[-2pt] {\footnotesize (0.309)}}  & \cellcolor{purple!41} \makecell{0.625 \\[-2pt] {\footnotesize (0.219)}}  & \cellcolor{purple!48} \makecell{0.641 \\[-2pt] {\footnotesize (0.217)}}  & \cellcolor{purple!43} \makecell{0.635 \\[-2pt] {\footnotesize (0.203)}}  & \cellcolor{gray!15} \makecell{0.634 \\[-2pt] {\footnotesize (0.213)}}  & \cellcolor{purple!46} \makecell{0.701 \\[-2pt] {\footnotesize (0.326)}}  & \cellcolor{purple!48} \makecell{0.684 \\[-2pt] {\footnotesize (0.227)}}  & \cellcolor{purple!43} \makecell{0.664 \\[-2pt] {\footnotesize (0.199)}}  & \cellcolor{gray!15} \makecell{0.683 \\[-2pt] {\footnotesize (0.251)}} \\[3pt]
gu & \cellcolor{purple!10} \makecell{0.679 \\[-2pt] {\footnotesize (0.321)}}  & \cellcolor{purple!10} \makecell{0.644 \\[-2pt] {\footnotesize (0.260)}}  & \cellcolor{purple!10} \makecell{0.651 \\[-2pt] {\footnotesize (0.263)}}  & \cellcolor{gray!15} \makecell{0.658 \\[-2pt] {\footnotesize (0.281)}}  & \cellcolor{purple!12} \makecell{0.568 \\[-2pt] {\footnotesize (0.214)}}  & \cellcolor{purple!10} \makecell{0.575 \\[-2pt] {\footnotesize (0.207)}}  & \cellcolor{purple!12} \makecell{0.581 \\[-2pt] {\footnotesize (0.204)}}  & \cellcolor{gray!15} \makecell{0.575 \\[-2pt] {\footnotesize (0.208)}}  & \cellcolor{purple!16} \makecell{0.620 \\[-2pt] {\footnotesize (0.220)}}  & \cellcolor{purple!12} \makecell{0.612 \\[-2pt] {\footnotesize (0.204)}}  & \cellcolor{purple!14} \makecell{0.606 \\[-2pt] {\footnotesize (0.185)}}  & \cellcolor{gray!15} \makecell{0.613 \\[-2pt] {\footnotesize (0.203)}} \\[3pt]
hi & \cellcolor{purple!18} \makecell{0.694 \\[-2pt] {\footnotesize (0.343)}}  & \cellcolor{purple!14} \makecell{0.656 \\[-2pt] {\footnotesize (0.250)}}  & \cellcolor{purple!14} \makecell{0.657 \\[-2pt] {\footnotesize (0.259)}}  & \cellcolor{gray!15} \makecell{0.669 \\[-2pt] {\footnotesize (0.284)}}  & \cellcolor{purple!18} \makecell{0.578 \\[-2pt] {\footnotesize (0.251)}}  & \cellcolor{purple!16} \makecell{0.594 \\[-2pt] {\footnotesize (0.217)}}  & \cellcolor{purple!16} \makecell{0.587 \\[-2pt] {\footnotesize (0.201)}}  & \cellcolor{gray!15} \makecell{0.586 \\[-2pt] {\footnotesize (0.223)}}  & \cellcolor{purple!29} \makecell{0.656 \\[-2pt] {\footnotesize (0.296)}}  & \cellcolor{purple!15} \makecell{0.623 \\[-2pt] {\footnotesize (0.216)}}  & \cellcolor{purple!20} \makecell{0.622 \\[-2pt] {\footnotesize (0.192)}}  & \cellcolor{gray!15} \makecell{0.633 \\[-2pt] {\footnotesize (0.235)}} \\[3pt]
id & \cellcolor{purple!33} \makecell{0.721 \\[-2pt] {\footnotesize (0.488)}}  & \cellcolor{purple!29} \makecell{0.686 \\[-2pt] {\footnotesize (0.357)}}  & \cellcolor{purple!27} \makecell{0.676 \\[-2pt] {\footnotesize (0.331)}}  & \cellcolor{gray!15} \makecell{0.695 \\[-2pt] {\footnotesize (0.392)}}  & \cellcolor{purple!29} \makecell{0.607 \\[-2pt] {\footnotesize (0.251)}}  & \cellcolor{purple!37} \makecell{0.616 \\[-2pt] {\footnotesize (0.252)}}  & \cellcolor{purple!35} \makecell{0.620 \\[-2pt] {\footnotesize (0.231)}}  & \cellcolor{gray!15} \makecell{0.614 \\[-2pt] {\footnotesize (0.245)}}  & \cellcolor{purple!41} \makecell{0.686 \\[-2pt] {\footnotesize (0.422)}}  & \cellcolor{purple!37} \makecell{0.662 \\[-2pt] {\footnotesize (0.285)}}  & \cellcolor{purple!29} \makecell{0.641 \\[-2pt] {\footnotesize (0.201)}}  & \cellcolor{gray!15} \makecell{0.663 \\[-2pt] {\footnotesize (0.303)}} \\[3pt]
ja & \cellcolor{purple!48} \makecell{0.768 \\[-2pt] {\footnotesize (0.458)}}  & \cellcolor{purple!50} \makecell{0.721 \\[-2pt] {\footnotesize (0.326)}}  & \cellcolor{purple!50} \makecell{0.734 \\[-2pt] {\footnotesize (0.334)}}  & \cellcolor{gray!15} \makecell{0.741 \\[-2pt] {\footnotesize (0.373)}}  & \cellcolor{purple!50} \makecell{0.644 \\[-2pt] {\footnotesize (0.226)}}  & \cellcolor{purple!50} \makecell{0.651 \\[-2pt] {\footnotesize (0.226)}}  & \cellcolor{purple!50} \makecell{0.656 \\[-2pt] {\footnotesize (0.210)}}  & \cellcolor{gray!15} \makecell{0.650 \\[-2pt] {\footnotesize (0.221)}}  & \cellcolor{purple!50} \makecell{0.718 \\[-2pt] {\footnotesize (0.341)}}  & \cellcolor{purple!50} \makecell{0.708 \\[-2pt] {\footnotesize (0.263)}}  & \cellcolor{purple!52} \makecell{0.714 \\[-2pt] {\footnotesize (0.210)}}  & \cellcolor{gray!15} \makecell{0.713 \\[-2pt] {\footnotesize (0.271)}} \\[3pt]
ko & \cellcolor{purple!29} \makecell{0.717 \\[-2pt] {\footnotesize (0.367)}}  & \cellcolor{purple!31} \makecell{0.687 \\[-2pt] {\footnotesize (0.300)}}  & \cellcolor{purple!33} \makecell{0.690 \\[-2pt] {\footnotesize (0.320)}}  & \cellcolor{gray!15} \makecell{0.698 \\[-2pt] {\footnotesize (0.329)}}  & \cellcolor{purple!43} \makecell{0.630 \\[-2pt] {\footnotesize (0.249)}}  & \cellcolor{purple!33} \makecell{0.614 \\[-2pt] {\footnotesize (0.220)}}  & \cellcolor{purple!29} \makecell{0.608 \\[-2pt] {\footnotesize (0.196)}}  & \cellcolor{gray!15} \makecell{0.617 \\[-2pt] {\footnotesize (0.222)}}  & \cellcolor{purple!43} \makecell{0.690 \\[-2pt] {\footnotesize (0.302)}}  & \cellcolor{purple!41} \makecell{0.666 \\[-2pt] {\footnotesize (0.252)}}  & \cellcolor{purple!31} \makecell{0.641 \\[-2pt] {\footnotesize (0.203)}}  & \cellcolor{gray!15} \makecell{0.666 \\[-2pt] {\footnotesize (0.252)}} \\[3pt]
ky & \cellcolor{purple!41} \makecell{0.737 \\[-2pt] {\footnotesize (0.380)}}  & \cellcolor{purple!43} \makecell{0.710 \\[-2pt] {\footnotesize (0.272)}}  & \cellcolor{purple!44} \makecell{0.712 \\[-2pt] {\footnotesize (0.288)}}  & \cellcolor{gray!15} \makecell{0.720 \\[-2pt] {\footnotesize (0.313)}}  & \cellcolor{purple!46} \makecell{0.634 \\[-2pt] {\footnotesize (0.201)}}  & \cellcolor{purple!43} \makecell{0.635 \\[-2pt] {\footnotesize (0.210)}}  & \cellcolor{purple!39} \makecell{0.622 \\[-2pt] {\footnotesize (0.197)}}  & \cellcolor{gray!15} \makecell{0.630 \\[-2pt] {\footnotesize (0.203)}}  & \cellcolor{purple!37} \makecell{0.678 \\[-2pt] {\footnotesize (0.213)}}  & \cellcolor{purple!33} \makecell{0.660 \\[-2pt] {\footnotesize (0.188)}}  & \cellcolor{purple!41} \makecell{0.660 \\[-2pt] {\footnotesize (0.163)}}  & \cellcolor{gray!15} \makecell{0.666 \\[-2pt] {\footnotesize (0.188)}} \\[3pt]
fa & \cellcolor{purple!35} \makecell{0.724 \\[-2pt] {\footnotesize (0.487)}}  & \cellcolor{purple!35} \makecell{0.691 \\[-2pt] {\footnotesize (0.354)}}  & \cellcolor{purple!35} \makecell{0.694 \\[-2pt] {\footnotesize (0.355)}}  & \cellcolor{gray!15} \makecell{0.703 \\[-2pt] {\footnotesize (0.399)}}  & \cellcolor{purple!33} \makecell{0.609 \\[-2pt] {\footnotesize (0.262)}}  & \cellcolor{purple!39} \makecell{0.618 \\[-2pt] {\footnotesize (0.231)}}  & \cellcolor{purple!33} \makecell{0.612 \\[-2pt] {\footnotesize (0.202)}}  & \cellcolor{gray!15} \makecell{0.613 \\[-2pt] {\footnotesize (0.232)}}  & \cellcolor{purple!27} \makecell{0.655 \\[-2pt] {\footnotesize (0.313)}}  & \cellcolor{purple!35} \makecell{0.662 \\[-2pt] {\footnotesize (0.257)}}  & \cellcolor{purple!33} \makecell{0.642 \\[-2pt] {\footnotesize (0.204)}}  & \cellcolor{gray!15} \makecell{0.653 \\[-2pt] {\footnotesize (0.258)}} \\[3pt]
pt & \cellcolor{purple!12} \makecell{0.679 \\[-2pt] {\footnotesize (0.407)}}  & \cellcolor{purple!18} \makecell{0.658 \\[-2pt] {\footnotesize (0.319)}}  & \cellcolor{purple!18} \makecell{0.658 \\[-2pt] {\footnotesize (0.337)}}  & \cellcolor{gray!15} \makecell{0.665 \\[-2pt] {\footnotesize (0.354)}}  & \cellcolor{purple!35} \makecell{0.613 \\[-2pt] {\footnotesize (0.238)}}  & \cellcolor{purple!26} \makecell{0.605 \\[-2pt] {\footnotesize (0.250)}}  & \cellcolor{purple!27} \makecell{0.607 \\[-2pt] {\footnotesize (0.217)}}  & \cellcolor{gray!15} \makecell{0.608 \\[-2pt] {\footnotesize (0.235)}}  & \cellcolor{purple!25} \makecell{0.649 \\[-2pt] {\footnotesize (0.324)}}  & \cellcolor{purple!29} \makecell{0.647 \\[-2pt] {\footnotesize (0.290)}}  & \cellcolor{purple!25} \makecell{0.636 \\[-2pt] {\footnotesize (0.223)}}  & \cellcolor{gray!15} \makecell{0.644 \\[-2pt] {\footnotesize (0.279)}} \\[3pt]
pa & \cellcolor{purple!16} \makecell{0.686 \\[-2pt] {\footnotesize (0.316)}}  & \cellcolor{purple!12} \makecell{0.652 \\[-2pt] {\footnotesize (0.240)}}  & \cellcolor{purple!20} \makecell{0.665 \\[-2pt] {\footnotesize (0.252)}}  & \cellcolor{gray!15} \makecell{0.668 \\[-2pt] {\footnotesize (0.270)}}  & \cellcolor{purple!14} \makecell{0.573 \\[-2pt] {\footnotesize (0.207)}}  & \cellcolor{purple!12} \makecell{0.585 \\[-2pt] {\footnotesize (0.208)}}  & \cellcolor{purple!10} \makecell{0.577 \\[-2pt] {\footnotesize (0.198)}}  & \cellcolor{gray!15} \makecell{0.578 \\[-2pt] {\footnotesize (0.204)}}  & \cellcolor{purple!18} \makecell{0.626 \\[-2pt] {\footnotesize (0.257)}}  & \cellcolor{purple!19} \makecell{0.629 \\[-2pt] {\footnotesize (0.218)}}  & \cellcolor{purple!16} \makecell{0.613 \\[-2pt] {\footnotesize (0.193)}}  & \cellcolor{gray!15} \makecell{0.623 \\[-2pt] {\footnotesize (0.223)}} \\[3pt]
ru & \cellcolor{purple!43} \makecell{0.743 \\[-2pt] {\footnotesize (0.539)}}  & \cellcolor{purple!41} \makecell{0.705 \\[-2pt] {\footnotesize (0.394)}}  & \cellcolor{purple!41} \makecell{0.702 \\[-2pt] {\footnotesize (0.377)}}  & \cellcolor{gray!15} \makecell{0.717 \\[-2pt] {\footnotesize (0.437)}}  & \cellcolor{purple!37} \makecell{0.624 \\[-2pt] {\footnotesize (0.285)}}  & \cellcolor{purple!46} \makecell{0.638 \\[-2pt] {\footnotesize (0.284)}}  & \cellcolor{purple!44} \makecell{0.636 \\[-2pt] {\footnotesize (0.246)}}  & \cellcolor{gray!15} \makecell{0.633 \\[-2pt] {\footnotesize (0.272)}}  & \cellcolor{purple!48} \makecell{0.707 \\[-2pt] {\footnotesize (0.456)}}  & \cellcolor{purple!44} \makecell{0.677 \\[-2pt] {\footnotesize (0.298)}}  & \cellcolor{purple!44} \makecell{0.665 \\[-2pt] {\footnotesize (0.222)}}  & \cellcolor{gray!15} \makecell{0.683 \\[-2pt] {\footnotesize (0.325)}} \\[3pt]
si & \cellcolor{purple!46} \makecell{0.754 \\[-2pt] {\footnotesize (0.405)}}  & \cellcolor{purple!44} \makecell{0.711 \\[-2pt] {\footnotesize (0.241)}}  & \cellcolor{purple!48} \makecell{0.717 \\[-2pt] {\footnotesize (0.253)}}  & \cellcolor{gray!15} \makecell{0.727 \\[-2pt] {\footnotesize (0.300)}}  & \cellcolor{purple!23} \makecell{0.595 \\[-2pt] {\footnotesize (0.201)}}  & \cellcolor{purple!23} \makecell{0.603 \\[-2pt] {\footnotesize (0.196)}}  & \cellcolor{purple!25} \makecell{0.604 \\[-2pt] {\footnotesize (0.189)}}  & \cellcolor{gray!15} \makecell{0.601 \\[-2pt] {\footnotesize (0.195)}}  & \cellcolor{purple!10} \makecell{0.570 \\[-2pt] {\footnotesize (0.178)}}  & \cellcolor{purple!10} \makecell{0.592 \\[-2pt] {\footnotesize (0.175)}}  & \cellcolor{purple!10} \makecell{0.568 \\[-2pt] {\footnotesize (0.172)}}  & \cellcolor{gray!15} \makecell{0.577 \\[-2pt] {\footnotesize (0.175)}} \\[3pt]
es & \cellcolor{purple!25} \makecell{0.716 \\[-2pt] {\footnotesize (0.421)}}  & \cellcolor{purple!37} \makecell{0.694 \\[-2pt] {\footnotesize (0.333)}}  & \cellcolor{purple!37} \makecell{0.695 \\[-2pt] {\footnotesize (0.340)}}  & \cellcolor{gray!15} \makecell{0.702 \\[-2pt] {\footnotesize (0.364)}}  & \cellcolor{purple!48} \makecell{0.640 \\[-2pt] {\footnotesize (0.227)}}  & \cellcolor{purple!44} \makecell{0.635 \\[-2pt] {\footnotesize (0.245)}}  & \cellcolor{purple!46} \makecell{0.639 \\[-2pt] {\footnotesize (0.210)}}  & \cellcolor{gray!15} \makecell{0.638 \\[-2pt] {\footnotesize (0.227)}}  & \cellcolor{purple!44} \makecell{0.697 \\[-2pt] {\footnotesize (0.387)}}  & \cellcolor{purple!46} \makecell{0.679 \\[-2pt] {\footnotesize (0.284)}}  & \cellcolor{purple!48} \makecell{0.669 \\[-2pt] {\footnotesize (0.216)}}  & \cellcolor{gray!15} \makecell{0.682 \\[-2pt] {\footnotesize (0.296)}} \\[3pt]
sw & \cellcolor{purple!31} \makecell{0.718 \\[-2pt] {\footnotesize (0.400)}}  & \cellcolor{purple!25} \makecell{0.678 \\[-2pt] {\footnotesize (0.278)}}  & \cellcolor{purple!25} \makecell{0.675 \\[-2pt] {\footnotesize (0.283)}}  & \cellcolor{gray!15} \makecell{0.690 \\[-2pt] {\footnotesize (0.320)}}  & \cellcolor{purple!20} \makecell{0.588 \\[-2pt] {\footnotesize (0.213)}}  & \cellcolor{purple!18} \makecell{0.595 \\[-2pt] {\footnotesize (0.212)}}  & \cellcolor{purple!20} \makecell{0.597 \\[-2pt] {\footnotesize (0.202)}}  & \cellcolor{gray!15} \makecell{0.593 \\[-2pt] {\footnotesize (0.209)}}  & \cellcolor{purple!14} \makecell{0.596 \\[-2pt] {\footnotesize (0.214)}}  & \cellcolor{purple!19} \makecell{0.629 \\[-2pt] {\footnotesize (0.202)}}  & \cellcolor{purple!12} \makecell{0.596 \\[-2pt] {\footnotesize (0.186)}}  & \cellcolor{gray!15} \makecell{0.607 \\[-2pt] {\footnotesize (0.201)}} \\[3pt]
ta & \cellcolor{purple!27} \makecell{0.717 \\[-2pt] {\footnotesize (0.285)}}  & \cellcolor{purple!27} \makecell{0.683 \\[-2pt] {\footnotesize (0.226)}}  & \cellcolor{purple!29} \makecell{0.685 \\[-2pt] {\footnotesize (0.234)}}  & \cellcolor{gray!15} \makecell{0.695 \\[-2pt] {\footnotesize (0.248)}}  & \cellcolor{purple!21} \makecell{0.590 \\[-2pt] {\footnotesize (0.203)}}  & \cellcolor{purple!20} \makecell{0.598 \\[-2pt] {\footnotesize (0.202)}}  & \cellcolor{purple!18} \makecell{0.593 \\[-2pt] {\footnotesize (0.187)}}  & \cellcolor{gray!15} \makecell{0.594 \\[-2pt] {\footnotesize (0.197)}}  & \cellcolor{purple!21} \makecell{0.645 \\[-2pt] {\footnotesize (0.218)}}  & \cellcolor{purple!25} \makecell{0.639 \\[-2pt] {\footnotesize (0.195)}}  & \cellcolor{purple!35} \makecell{0.642 \\[-2pt] {\footnotesize (0.180)}}  & \cellcolor{gray!15} \makecell{0.642 \\[-2pt] {\footnotesize (0.198)}} \\[3pt]
te & \cellcolor{purple!20} \makecell{0.704 \\[-2pt] {\footnotesize (0.308)}}  & \cellcolor{purple!21} \makecell{0.665 \\[-2pt] {\footnotesize (0.244)}}  & \cellcolor{purple!23} \makecell{0.673 \\[-2pt] {\footnotesize (0.264)}}  & \cellcolor{gray!15} \makecell{0.681 \\[-2pt] {\footnotesize (0.272)}}  & \cellcolor{purple!31} \makecell{0.608 \\[-2pt] {\footnotesize (0.208)}}  & \cellcolor{purple!14} \makecell{0.590 \\[-2pt] {\footnotesize (0.209)}}  & \cellcolor{purple!14} \makecell{0.586 \\[-2pt] {\footnotesize (0.197)}}  & \cellcolor{gray!15} \makecell{0.595 \\[-2pt] {\footnotesize (0.205)}}  & \cellcolor{purple!20} \makecell{0.641 \\[-2pt] {\footnotesize (0.231)}}  & \cellcolor{purple!15} \makecell{0.623 \\[-2pt] {\footnotesize (0.196)}}  & \cellcolor{purple!27} \makecell{0.637 \\[-2pt] {\footnotesize (0.178)}}  & \cellcolor{gray!15} \makecell{0.634 \\[-2pt] {\footnotesize (0.202)}} \\[3pt]
th & \cellcolor{purple!14} \makecell{0.686 \\[-2pt] {\footnotesize (0.438)}}  & \cellcolor{purple!16} \makecell{0.656 \\[-2pt] {\footnotesize (0.371)}}  & \cellcolor{purple!16} \makecell{0.658 \\[-2pt] {\footnotesize (0.359)}}  & \cellcolor{gray!15} \makecell{0.666 \\[-2pt] {\footnotesize (0.390)}}  & \cellcolor{purple!44} \makecell{0.631 \\[-2pt] {\footnotesize (0.259)}}  & \cellcolor{purple!26} \makecell{0.605 \\[-2pt] {\footnotesize (0.257)}}  & \cellcolor{purple!48} \makecell{0.643 \\[-2pt] {\footnotesize (0.241)}}  & \cellcolor{gray!15} \makecell{0.627 \\[-2pt] {\footnotesize (0.252)}}  & \cellcolor{purple!33} \makecell{0.668 \\[-2pt] {\footnotesize (0.356)}}  & \cellcolor{purple!43} \makecell{0.668 \\[-2pt] {\footnotesize (0.269)}}  & \cellcolor{purple!46} \makecell{0.667 \\[-2pt] {\footnotesize (0.231)}}  & \cellcolor{gray!15} \makecell{0.668 \\[-2pt] {\footnotesize (0.285)}} \\[3pt]
tr & \cellcolor{purple!37} \makecell{0.728 \\[-2pt] {\footnotesize (0.468)}}  & \cellcolor{purple!33} \makecell{0.690 \\[-2pt] {\footnotesize (0.344)}}  & \cellcolor{purple!31} \makecell{0.685 \\[-2pt] {\footnotesize (0.341)}}  & \cellcolor{gray!15} \makecell{0.701 \\[-2pt] {\footnotesize (0.384)}}  & \cellcolor{purple!27} \makecell{0.601 \\[-2pt] {\footnotesize (0.235)}}  & \cellcolor{purple!35} \makecell{0.616 \\[-2pt] {\footnotesize (0.241)}}  & \cellcolor{purple!37} \makecell{0.621 \\[-2pt] {\footnotesize (0.213)}}  & \cellcolor{gray!15} \makecell{0.613 \\[-2pt] {\footnotesize (0.230)}}  & \cellcolor{purple!35} \makecell{0.670 \\[-2pt] {\footnotesize (0.322)}}  & \cellcolor{purple!31} \makecell{0.660 \\[-2pt] {\footnotesize (0.251)}}  & \cellcolor{purple!39} \makecell{0.650 \\[-2pt] {\footnotesize (0.214)}}  & \cellcolor{gray!15} \makecell{0.660 \\[-2pt] {\footnotesize (0.262)}} \\[3pt]
uk & \cellcolor{purple!39} \makecell{0.733 \\[-2pt] {\footnotesize (0.550)}}  & \cellcolor{purple!39} \makecell{0.695 \\[-2pt] {\footnotesize (0.361)}}  & \cellcolor{purple!39} \makecell{0.695 \\[-2pt] {\footnotesize (0.360)}}  & \cellcolor{gray!15} \makecell{0.708 \\[-2pt] {\footnotesize (0.424)}}  & \cellcolor{purple!25} \makecell{0.595 \\[-2pt] {\footnotesize (0.252)}}  & \cellcolor{purple!31} \makecell{0.611 \\[-2pt] {\footnotesize (0.238)}}  & \cellcolor{purple!31} \makecell{0.610 \\[-2pt] {\footnotesize (0.215)}}  & \cellcolor{gray!15} \makecell{0.605 \\[-2pt] {\footnotesize (0.235)}}  & \cellcolor{purple!31} \makecell{0.664 \\[-2pt] {\footnotesize (0.406)}}  & \cellcolor{purple!39} \makecell{0.663 \\[-2pt] {\footnotesize (0.266)}}  & \cellcolor{purple!37} \makecell{0.647 \\[-2pt] {\footnotesize (0.202)}}  & \cellcolor{gray!15} \makecell{0.658 \\[-2pt] {\footnotesize (0.291)}} \\[3pt]
vi & \cellcolor{purple!21} \makecell{0.710 \\[-2pt] {\footnotesize (0.476)}}  & \cellcolor{purple!20} \makecell{0.661 \\[-2pt] {\footnotesize (0.297)}}  & \cellcolor{purple!12} \makecell{0.655 \\[-2pt] {\footnotesize (0.294)}}  & \cellcolor{gray!15} \makecell{0.675 \\[-2pt] {\footnotesize (0.356)}}  & \cellcolor{purple!16} \makecell{0.573 \\[-2pt] {\footnotesize (0.246)}}  & \cellcolor{purple!21} \makecell{0.599 \\[-2pt] {\footnotesize (0.239)}}  & \cellcolor{purple!22} \makecell{0.597 \\[-2pt] {\footnotesize (0.216)}}  & \cellcolor{gray!15} \makecell{0.590 \\[-2pt] {\footnotesize (0.234)}}  & \cellcolor{purple!39} \makecell{0.684 \\[-2pt] {\footnotesize (0.467)}}  & \cellcolor{purple!27} \makecell{0.645 \\[-2pt] {\footnotesize (0.269)}}  & \cellcolor{purple!18} \makecell{0.619 \\[-2pt] {\footnotesize (0.217)}}  & \cellcolor{gray!15} \makecell{0.649 \\[-2pt] {\footnotesize (0.318)}} \\[3pt]
Avg & \cellcolor{gray!15} \makecell{0.727 \\[-2pt] {\footnotesize (0.411)}}  & \cellcolor{gray!15} \makecell{0.691 \\[-2pt] {\footnotesize (0.302)}}  & \cellcolor{gray!15} \makecell{0.692 \\[-2pt] {\footnotesize (0.305)}}  & \cellcolor{gray!15} \textbf{\makecell{0.703 \\[-2pt] {\footnotesize (0.339)}}}  & \cellcolor{gray!15} \makecell{0.612 \\[-2pt] {\footnotesize (0.233)}}  & \cellcolor{gray!15} \makecell{0.619 \\[-2pt] {\footnotesize (0.229)}}  & \cellcolor{gray!15} \makecell{0.619 \\[-2pt] {\footnotesize (0.210)}}  & \cellcolor{gray!15} \textbf{\makecell{0.617 \\[-2pt] {\footnotesize (0.224)}}}  & \cellcolor{gray!15} \makecell{0.665 \\[-2pt] {\footnotesize (0.310)}}  & \cellcolor{gray!15} \makecell{0.658 \\[-2pt] {\footnotesize (0.240)}}  & \cellcolor{gray!15} \makecell{0.647 \\[-2pt] {\footnotesize (0.200)}}  & \cellcolor{gray!15} \textbf{\makecell{0.657 \\[-2pt] {\footnotesize (0.250)}}} \\[3pt]
\bottomrule
\end{tabular}%
}
\vspace{-2mm}
\caption{\textbf{Summarization$\rightarrow$Translation direction results}. 24 English$\rightarrow$XX language pairs are evaluated across Few-shot (FS), Zero-Shot (ZS), and Chain-of-Thoughts ZS (CoT) settings using \gemini, \qwen and \gemma LLMs. Both BERTScore (top) and X-Comet-XL (bottom, in parentheses) metrics are reported.}
\label{tab:results_eng2xx}
\end{table*}

To this end, we exploit the terminal numeric identifier embedded in each source and target URL, which provides stable, article-level provenance. Pairing this identifier with its associated language yields a \emph{language-qualified identifier} for each article, which we use to construct two complementary keys. A \emph{directional identifier}, \texttt{\{src\_lang\}\_\{src\_id\}-\{tgt\_lang\}\_\{tgt\_id\}}, preserves the ordered source--target language direction. A \emph{canonical identifier}, obtained by sorting the two language-qualified article identifiers independently of direction, instead identifies the underlying article pair regardless of translation direction: reverse-direction instances such as \texttt{french\_55217075-english\_55209763} and \texttt{english\_55209763-french\_55217075} map to the same canonical key. This canonical key allows us to both deduplicate instances within each direction and, more importantly, to identify the reverse-direction counterpart of each pair, enabling us to construct parallel source$\rightarrow$target and target$\rightarrow$source summarization instances from a single underlying article.

Following text-based curation, each article and summary is synthesized into speech using OmniVoice with a fixed reference speaker \cite{zhu2026omnivoiceomnilingualzeroshottexttospeech}. Among other recent text-to-speech (TTS) models, OmniVoice was selected for its broad multilingual coverage of 600 languages and lowest Character Error Rate (CER) scores. Prior to synthesis, we normalize whitespace and segment each text into sentence-level units using punctuation-aware rules that support Latin, Arabic, and CJK scripts. Consecutive sentences are then grouped into chunks of up to 1{,}200 characters. Each chunk is synthesized independently, resampled to a common sampling rate where necessary, and concatenated in its original order to yield a single long-form recording at 24 kHz. As a final quality-control step, we discard any language whose synthesized speech yields a CER above 25 when transcribed, ensuring that only languages with reliably intelligible synthesis are retained in the final benchmark. The resulting dataset contains approximately 29 hours of audio per language on average (Table~\ref{tab:cer_nisqa}).

\begin{figure*}[t]
    \centering
    \includegraphics[width=0.7\linewidth]{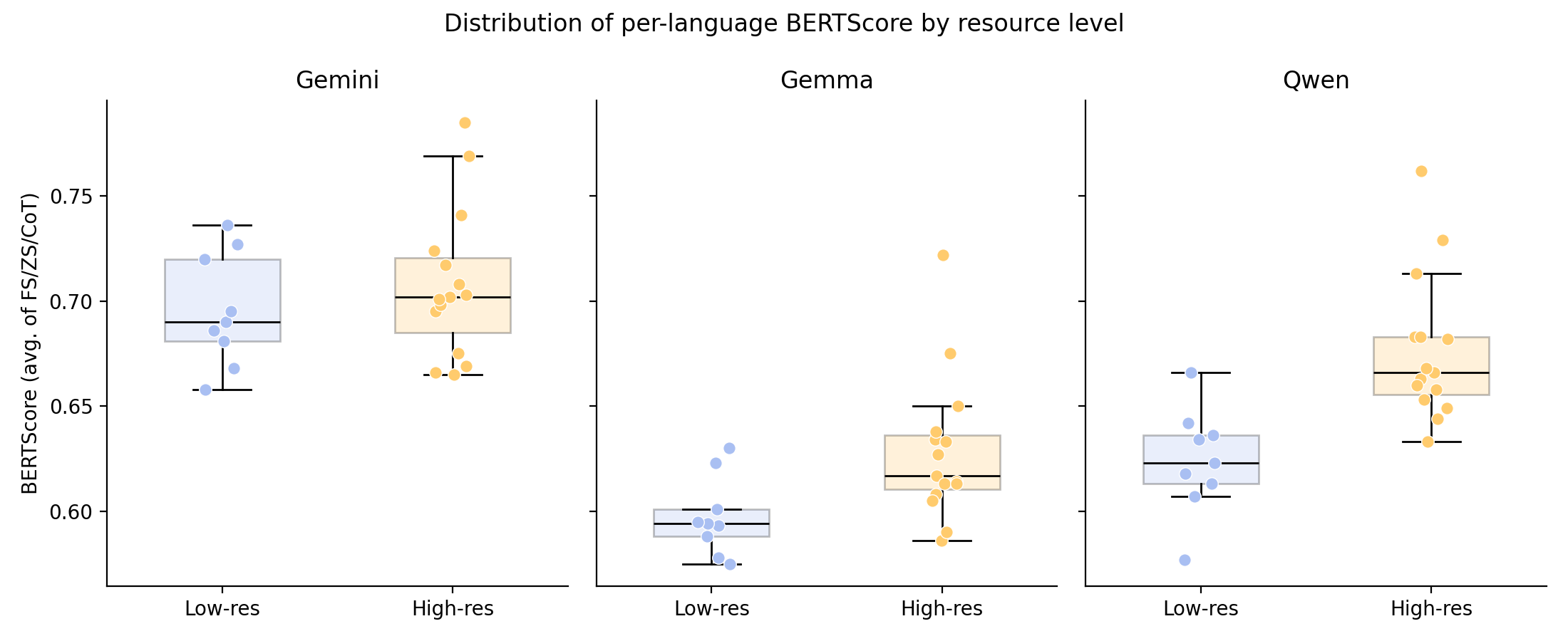}
    \vspace{-3mm}
    \caption{\textbf{Low-resource versus high-resource language comparisons}. BERTScore-F1 results are pooled across FS, ZS and CoT settings for each of our models.}
    \label{fig:low_res_per_model}
\end{figure*}

\vspace{-5pt}
\subsection{Speech Quality and Human Evaluation}
We validate synthesized speech quality at two stages: during generation and after the full dataset is complete. 
During generation, we first verify that every waveform is structurally valid (e.g., non-empty, correctly encoded, and of expected duration). A fluent or native speaker of each language then manually reviews a 20\% sample of the generated files per direction, checking for audio quality, intelligibility, and naturalness.

After full dataset generation, we compute NISQA \citep{nisqa}\footnote{\url{https://github.com/Lightning-AI/torchmetrics/blob/master/src/torchmetrics/audio/nisqa.py}} to automatically assess perceived naturalness, yielding an average score of 4.39 (Table~\ref{tab:cer_nisqa}). To complement this, 30 external annotators separately rate whether the synthesized article and summary audios coherently convey their intended content, using a 1\textasciitilde{}5 Likert scale.\footnote{Due to annotation cost, this external coherence validation is conducted on a subset of 10 of the 24 languages.} As shown in Figure \ref{fig:external_human_quality}, audios receive an average rating of 4.07, consistent with the NISQA results and confirming that the synthesized speech is perceived as natural and intelligible.

\vspace{-3pt}
\section{Experimental Setup}
\vspace{-4pt}
\subsection{Models}
\vspace{-4pt}
Three representative speech-capable large language models (LLMs) across a diverse range of parameter scales are selected to evaluate JSumT: edge-deployable open-weight \gemma, mid-weight \qwen \texttt{(30B)}, and proprietary \gemini. 
To examine the effects of contextual supervision and explicit reasoning, each model is evaluated under three prompting settings: zero-shot, five-shot, and CoT. The complete prompts are provided in Appendix~\ref{sec:prompts}. For all experiments, \gemini is evaluated using its default high-thinking configuration with a maximum generation length of 256 tokens. \gemma and \qwen are evaluated with a batch size of 8 and a maximum generation length of 256 tokens.

\vspace{-6pt}
\paragraph{Zero-Shot (ZS) and Few-Shot (FS) Prompting} For both zero-shot and few-shot evaluation, models are given a source-language spoken news article and tasked with generating a summary in the target language. We adopt a cascaded prompting strategy in which models first summarize the source content and then translate the resulting summary into the target language (or vice versa). Evaluation is performed on 200 held-out examples for each language direction. For few-shot, models are additionally provided with five demonstration examples comprising source-language audio articles and their corresponding target-language summaries. All example samples are selected from a non-overlapping pool and are excluded from the evaluation split.

\vspace{-6pt}
\paragraph{CoT Reasoning} In order to investigate whether structured reasoning improves performance, we additionally evaluate a CoT prompting strategy, as in \citet{chen_cothssum_2025} and \cite{yuan-zhang-2026-understanding}. Motivated by prior work on decomposed reasoning \citep{wang2023element}, models are guided through two implicit reasoning stages before generating the final output. First, the model identifies the key informational content of the spoken document using a 5W1H framework (\emph{who}, \emph{what}, \emph{when}, \emph{where}, \emph{why}, and \emph{how}), extracting salient entities, events, and their temporal, spatial, causal, and procedural relationships. Second, the model synthesizes these elements into a coherent semantic representation, determines the information most central to the document's meaning, and expresses the resulting  one-sentence target-language summary. Details of the CoT prompt is available in Appendix~\ref{sec:prompts}

\begin{figure*}[t]
    \centering
    \begin{subfigure}{0.48\linewidth}
        \resizebox{\linewidth}{!}{\includegraphics{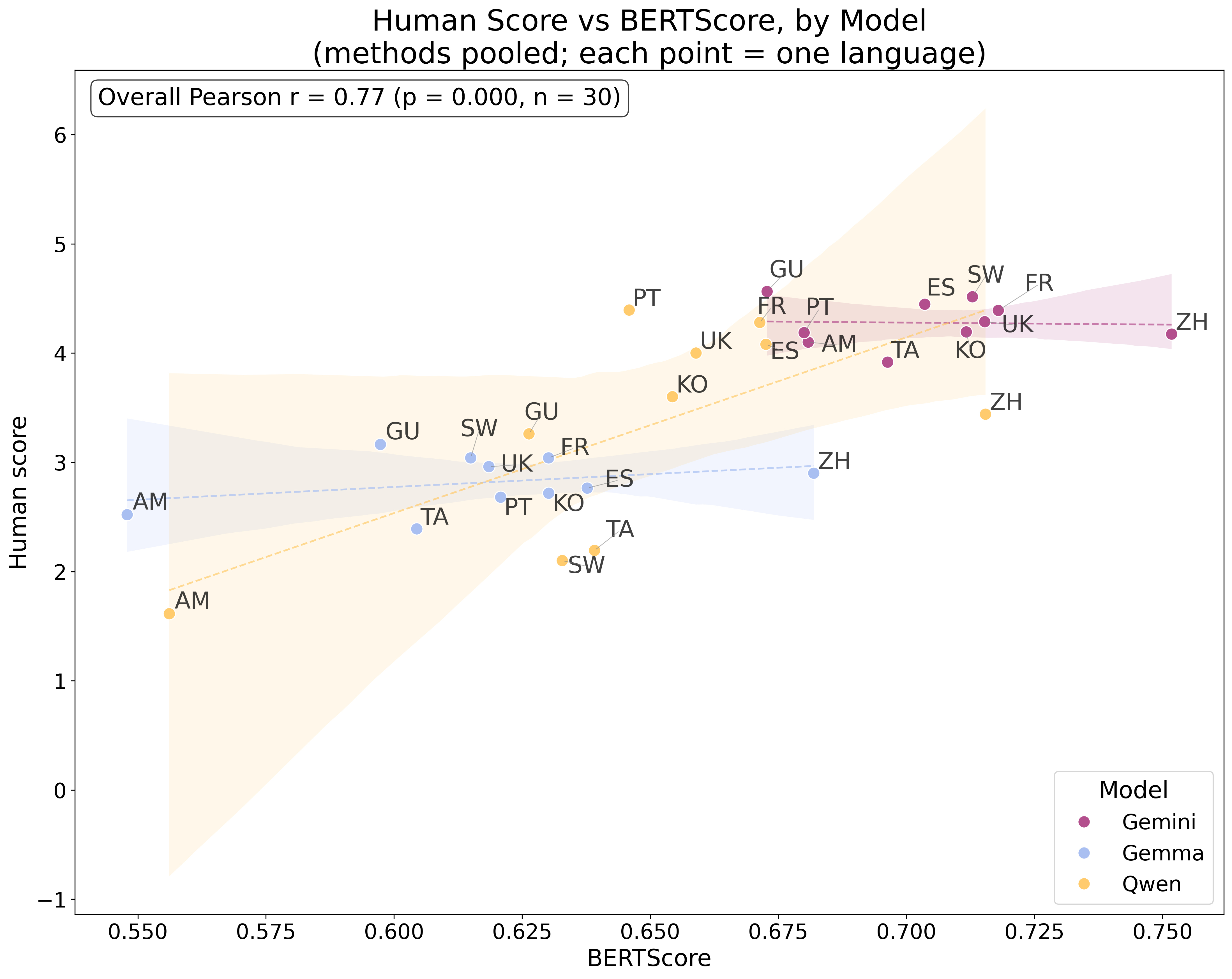}}
        \caption{\textbf{BERTScore F1}}
        \label{fig:human_bertscore}
    \end{subfigure}
    \hfill
    \begin{subfigure}{0.48\linewidth}
        \resizebox{\linewidth}{!}{\includegraphics{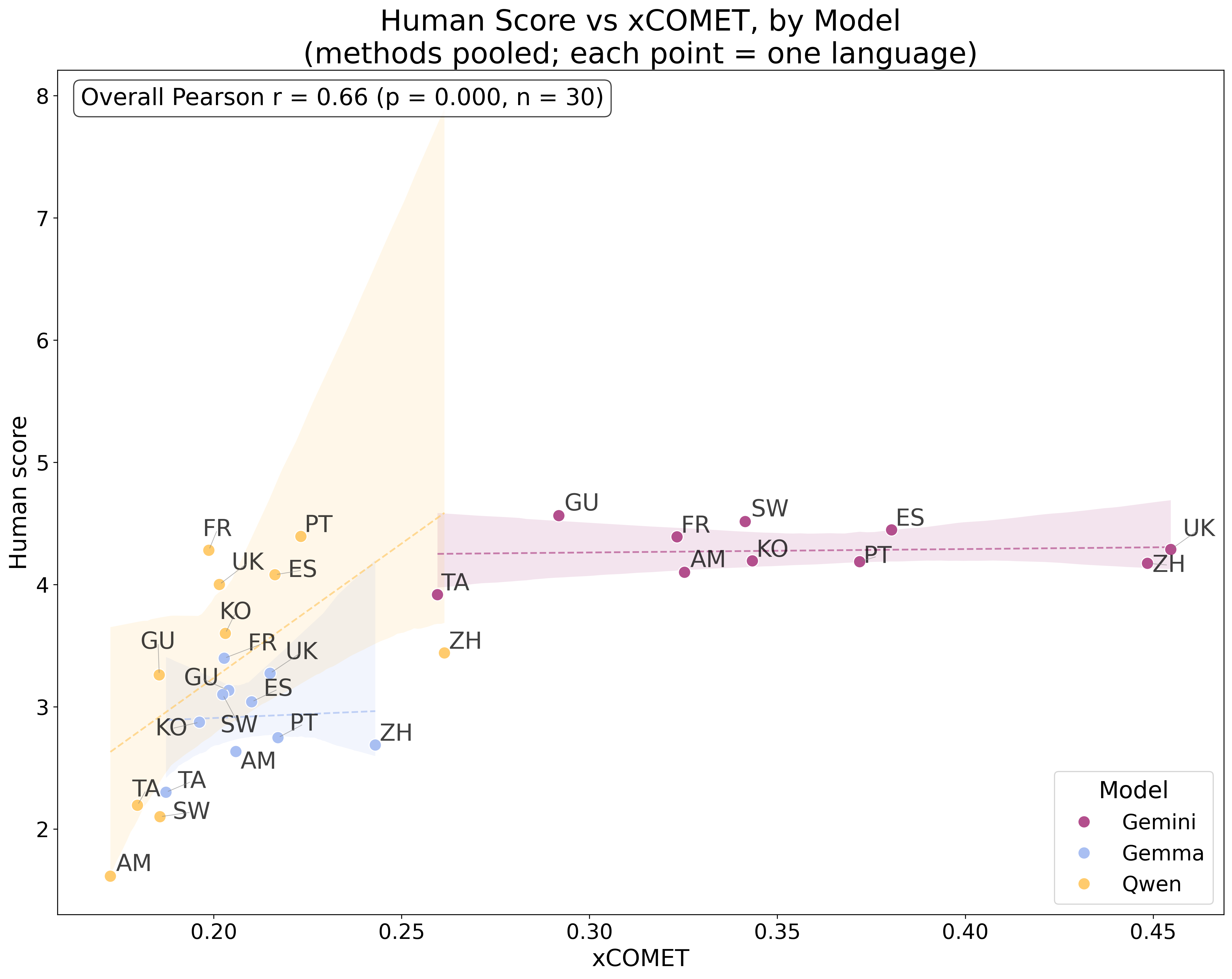}}
        \caption{\textbf{xCOMET}}
        \label{fig:human_xcomet}
    \end{subfigure}
    \vspace{-2mm}
    \caption{\textbf{Correlation between Human evaluation and individual metrics}. Pearson correlation reported.}
    \label{fig:human_metric_alignment}
\end{figure*}

\vspace{-5pt}
\subsection{Evaluation}
\vspace{-1pt}
\noindent \textbf{Objective Metrics: } Since the task involves jointly summarizing and translating a source-language article into a target-language summary, evaluation must capture both the fidelity of the summarized content and the quality of the cross-lingual generation. Thus, summarization outputs are assessed using BERTScore-F1 \cite{chhibbar-kalita-2024-automatic, Bertscore_usage}, while xCOMET-XL \citep{10.1162/tacl_a_00683} reflects translation quality.
Specifically, BERTScore-F1 computes token-level semantic similarity between each predicted summary and its reference \cite{zhangbertscore}. It is thus well suited to capturing meaning-level overlap that surface-level lexical metrics tend to miss, which is a property especially important in our joint summarization-translation setting, where outputs must be assessed across both content and language. See Appendix \ref{sec:add_metrics} for details.

Following prior cross-lingual evaluation work \citep{adelani-etal-2026-speech}, we report xCOMET-XL \citep{10.1162/tacl_a_00683} to evaluate translation. While xCOMET-XL is conventionally computed using the source text, reference, and prediction, we adopt its quality-estimation (QE) configuration\footnote{See Appendix \ref{sec:add_metrics}.} by scoring the predicted summary directly against the reference without access to the source article. This isolates the semantic and translation-quality alignment between the model's output and the target-language reference, independent of source-side variation introduced by the summarization step.

\begin{table}[t]
\centering
\resizebox{0.9\linewidth}{!}{%
\begin{tabular}{ l *{2}{C{2.0cm}} }
\toprule
 & Eng$\rightarrow$XX & XX$\rightarrow$Eng \\
\midrule
\gemini FS & \makecell{0.724 \\[-2pt] \textcolor{red!70!black}{{\footnotesize (-0.003)}}}  & \makecell{0.730 \\[-2pt] \textcolor{green!60!black}{{\footnotesize (+0.005)}}} \\[3pt]
\gemini ZS & \makecell{0.686 \\[-2pt] \textcolor{red!70!black}{{\footnotesize (-0.005)}}}  & \makecell{0.692 \\[-2pt] \textcolor{red!70!black}{{\footnotesize (-0.001)}}} \\[3pt]
\gemini CoT & 0.692  & 0.693 \\[3pt]
\midrule
\gemma FS & \makecell{0.580 \\[-2pt] \textcolor{red!70!black}{{\footnotesize (-0.032)}}}  & \makecell{0.621 \\[-2pt] \textcolor{red!70!black}{{\footnotesize (-0.006)}}} \\[3pt]
\gemma ZS & \makecell{0.600 \\[-2pt] \textcolor{red!70!black}{{\footnotesize (-0.019)}}}  & \makecell{0.630 \\[-2pt] \textcolor{green!60!black}{{\footnotesize (+0.005)}}} \\[3pt]
\gemma CoT & 0.619  & 0.630 \\[3pt]
\midrule
\qwen FS & \makecell{0.668 \\[-2pt] \textcolor{green!60!black}{{\footnotesize (+0.002)}}}  & \makecell{0.687 \\[-2pt] \textcolor{green!60!black}{{\footnotesize (+0.000)}}} \\[3pt]
\qwen ZS & \makecell{0.211 \\[-2pt] \textcolor{red!70!black}{{\footnotesize (-0.448)}}}  & \makecell{0.529 \\[-2pt] \textcolor{red!70!black}{{\footnotesize (-0.071)}}} \\[3pt]
\qwen CoT & 0.647  & 0.661 \\[3pt]
\midrule
\textbf{Avg} & \makecell{0.578 \\[-2pt] \textcolor{red!70!black}{{\footnotesize (-0.081)}}}  & \makecell{0.648 \\[-2pt] \textcolor{red!70!black}{{\footnotesize (-0.012)}}} \\[3pt]
\bottomrule
\end{tabular}%
}
\vspace{-1mm}
\caption{\textbf{Translation$\rightarrow$Summarization results averaged across all languages}. Values in parentheses show the difference relative to the opposite Summarization$\rightarrow$Translation direction. (\textcolor{green!60!black}{Green} = Translation$\rightarrow$Summarization is better, \textcolor{red!70!black}{red} = lower.)}
\label{tab:trans_sum_vs_sum_trans}
\end{table}

\vspace{-5pt}
\paragraph{Subjective Metrics}
\vspace{-5pt}
To assess multilingual summarization quality, we conduct human evaluations spanning a diverse range of languages: 5 high-resource (French, Spanish, Portuguese, Korean, and Chinese) and 5 mid/low-resource (Amharic, Gujarati, Tamil, Swahili, and Ukrainian), in both directions (XX$\rightarrow$English, English$\rightarrow$XX). For each language, three bilingual annotators independently complete a survey in which they listen to 50 audio articles (25 per direction), each paired with its source-text article and a set of anonymized, model-wise candidate summaries. Annotators first listen to the full audio article, then rate each candidate summary on quality on a 1-to-5 scale. In the survey, a high quality summary is defined as accurate, engaging and informative (Appendix~\ref{sec:Human_eval_protocol}).

\begin{figure}[t]
\centering
\resizebox{0.95\linewidth}{!}{%
\includegraphics{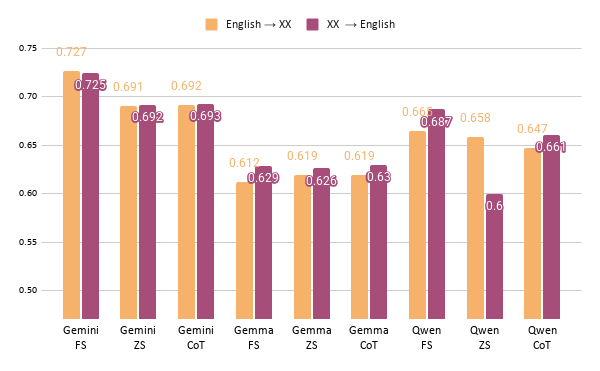}%
}
\vspace{-3mm}
\caption{\textbf{Influence of language direction on performance in the Summarization$\rightarrow$Translation setting.}}
\label{fig:lang_dir}
\end{figure}

\vspace{-2pt}
\section{Results}
\vspace{-6pt}

Table~\ref{tab:results_eng2xx} reveals systematic differences across both models and prompting strategies. \gemini attains the strongest average BERTScore of 0.703, followed by \qwen (0.657) and \gemma (0.617). With respect to prompting methodology, FS prompting proves particularly effective for \gemini, reaching 0.727 and outperforming its ZS and CoT variants by approximately 0.036. \qwen exhibits a similar, though more modest, advantage under FS prompting, whereas \gemma remains comparatively insensitive to prompting strategy altogether. This model-level disparity is further reflected at the language level: both \gemma and \qwen exhibit markedly greater performance variation between low- (e.g. Amharic, Gujarati, Kyrgyz) and high-resource languages (e.g. Chinese, French, Spanish) than \gemini, which performs comparably across resource settings (Figure~\ref{fig:low_res_per_model}). Nonetheless, average performance consistently improves with language resource availability for all models.

These trends hold irrespective of the evaluation metric used: both BERTScore and xCOMET consistently rank \gemini above \qwen, which in turn outperforms \gemma. For \gemini and \qwen specifically, both metrics further agree that FS prompting outperforms CoT, which in turn outperforms ZS. Crucially, these automatic metrics correlate strongly with human judgment, lending further credibility to the observed trends: the overall Pearson correlation for BertScore and xCOMET is 0.77 and 0.66, respectively (Figures~\ref{fig:human_bertscore}, \ref{fig:human_xcomet}).

\begin{figure}[t]
    \centering
    \includegraphics[width=1\linewidth]{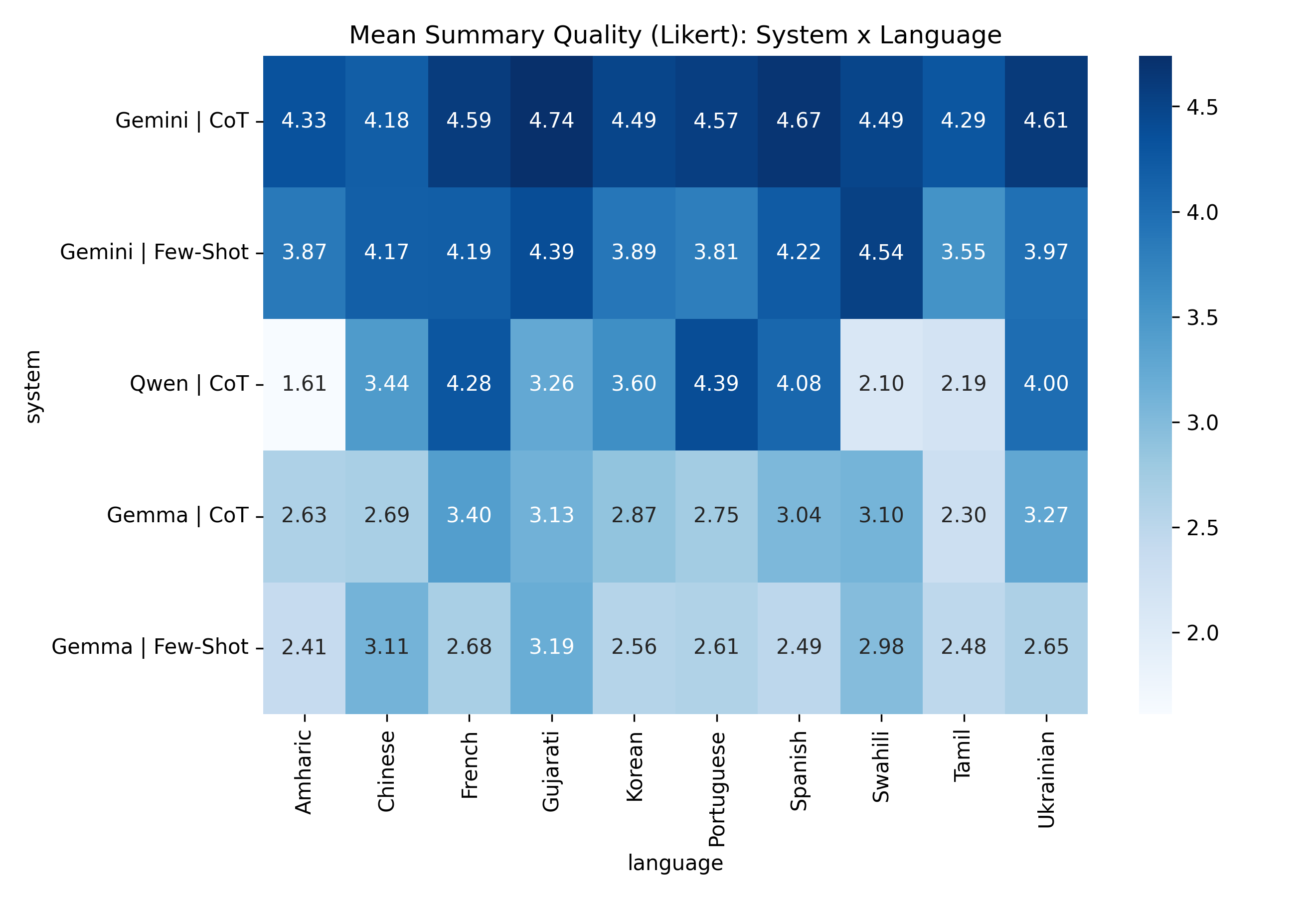}
    \vspace{-5mm}
    \caption{\textbf{Human evaluation for summary quality}. Quality is evaluated on a Likert 1-5 scale for each model, setting (CoT and Few-Shot) and language.}
    \label{fig:heatmap_human_eval}
\end{figure}

\vspace{-5pt}
\paragraph{Effect of Task Direction}
Table~\ref{tab:results_eng2xx} prompts models to summarize and translate jointly in a single pass. To isolate the effect of task ordering, we instead prompt models to first translate the full audio article into the target language, and only then generate a one-sentence summary of that translation. As shown in Table~\ref{tab:trans_sum_vs_sum_trans}, this reordering produces only a minimal drop in the overall average performance across all models, but the effect is asymmetric across language directions: English$\rightarrow$XX is affected substantially more (-0.081) than XX$\rightarrow$English (-0.012). We attribute this asymmetry to instruction-following failures specific to the translate-first ordering: after translating a long audio document, models more frequently fail to complete the remaining instruction, in some cases omitting the summary entirely or hallucinating toward the end of the translation. This effect is most severe for \qwen under ZS prompting (English$\rightarrow$XX: -0.448 vs. XX$\rightarrow$English: -0.071), suggesting that translate-then-summarize pipelines are less capable prompting regimes.

\begin{figure}[t]
    \centering
    \includegraphics[width=0.85\linewidth]{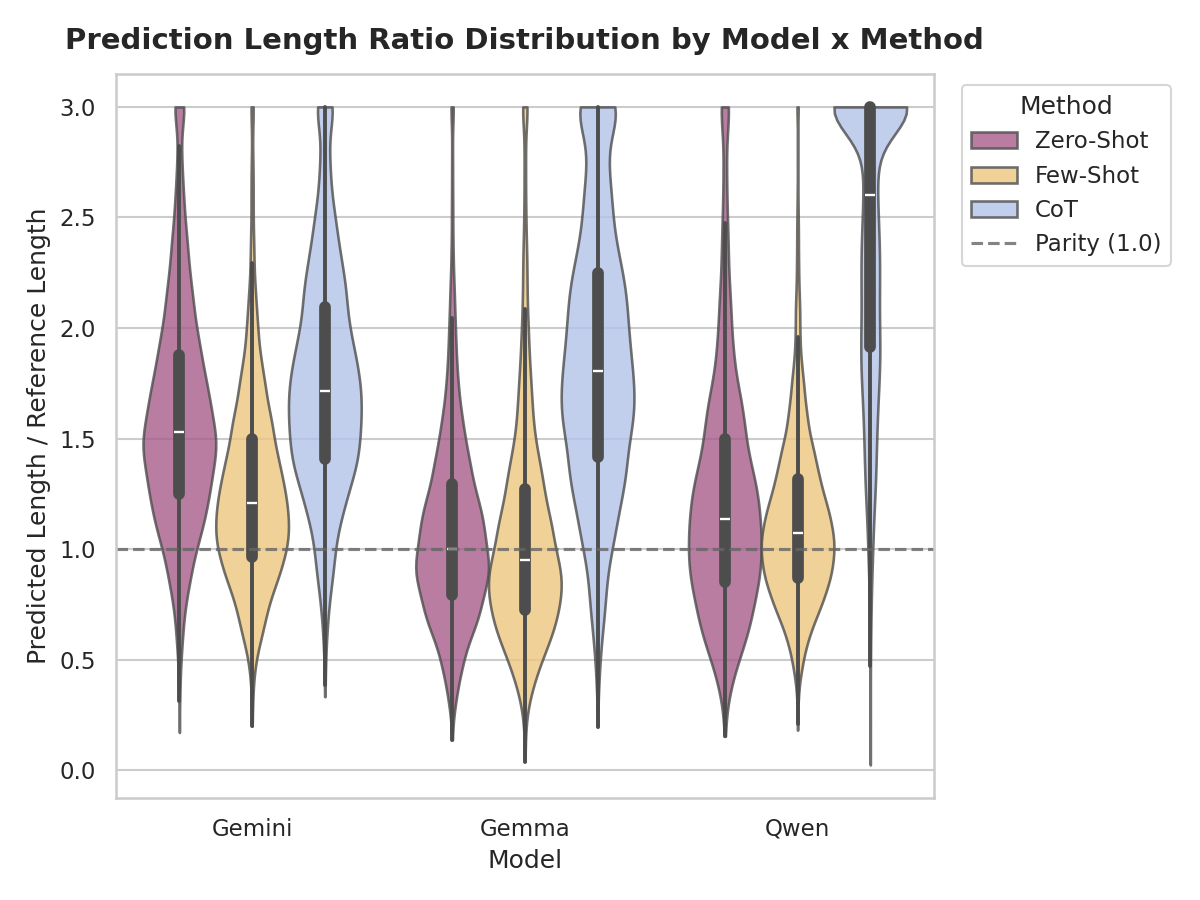}
    \caption{\textbf{Sentence length analysis between reference and predicted summaries}. We focus on the Summarization$\rightarrow$Translation direction. Results are pooled across methods for each model.}
    \label{fig:sentence_length_analysis}
\end{figure}

\vspace{-5pt}
\paragraph{Effect of Language Direction.} 
Independent of task ordering, we further find that the direction of language itself affects performance. As seen in Table~\ref{tab:trans_sum_vs_sum_trans} and Figure~\ref{fig:lang_dir}, generating an English summary from non-English speech (XX$\rightarrow$Eng) almost always outperforms generating a non-English summary from English speech (Eng$\rightarrow$XX) across all models and both task orderings (Summarization$\leftrightarrow$Translation). Specifically, while language direction has little effect on \gemini regardless of task ordering (with only a marginally larger gap under Translation$\rightarrow$Summarization), \gemma and \qwen are considerably more sensitive to language direction when the task is ordered as Translation$\rightarrow$Summarization. Averaged across prompting methods, the Eng$\rightarrow$XX/XX$\rightarrow$Eng relative difference is 0.012 under Summarization$\rightarrow$Translation versus 0.027 under Translation$\rightarrow$Summarization for \gemma, and 0.031 versus 0.0117 for \qwen.

\begin{figure*}[t]
    \centering
    \includegraphics[width=0.9\linewidth]{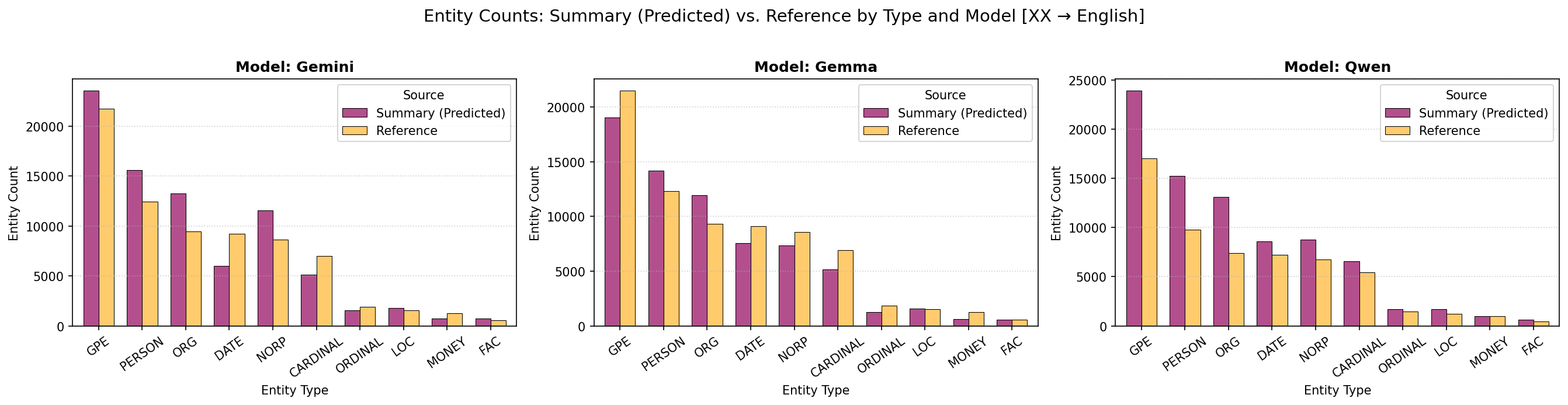}
    \vspace{-3mm}
    \caption{\textbf{Named entities count between reference and predicted summaries.} Entities predicted by \texttt{Spacy}.}
    \label{fig:ner_by_type}
\end{figure*}

This pattern can be attributed to the compounding effect of sustained non-English generation: Translation$\rightarrow$Summarization models must first translate the entire long-form document into the target language before condensing it, requiring long, sustained generation in the non-English target and increasing the opportunity for language-specific errors to accumulate before summarization even begins. In contrast, for Summarization$\rightarrow$Translation, the model first condenses the content in English, its likely dominant pretraining language, and only afterward performs the comparatively shorter task of translating a brief summary, limiting the extent of non-English generation and constraining error accumulation.

\begin{figure}[t]
    \centering
    \includegraphics[width=0.85\linewidth]{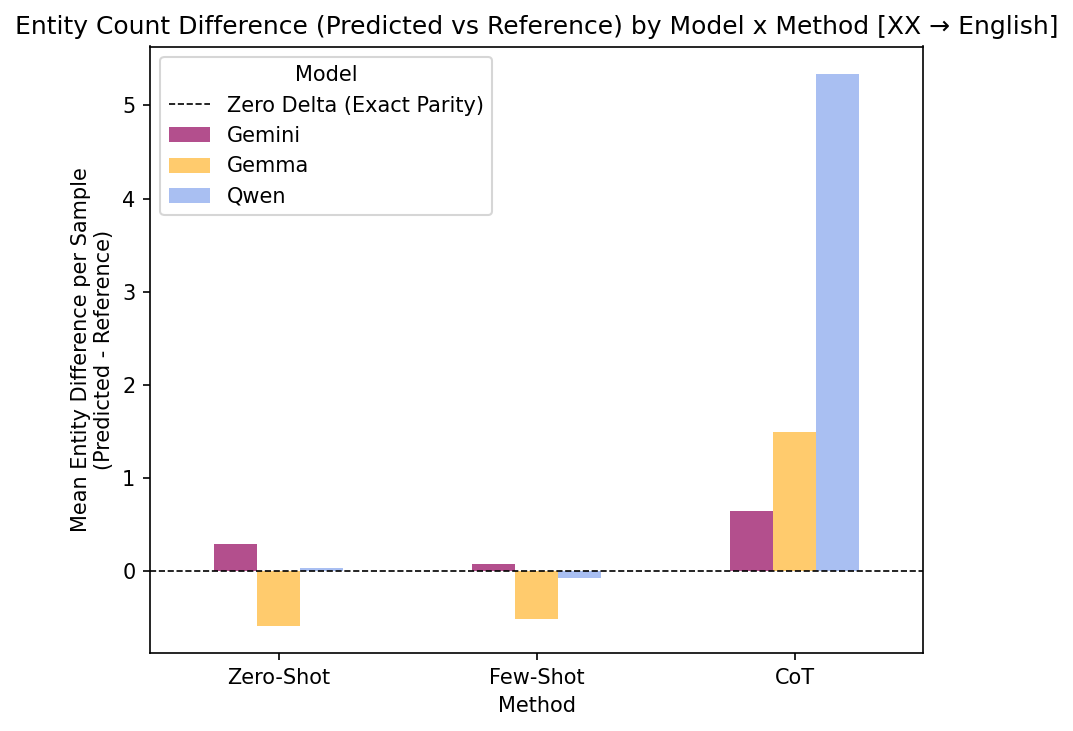}
    \vspace{-3mm}
    \caption{\textbf{NER average count difference between reference and predicted summary for each model.}}
    \label{fig:ner_count_diff}
\end{figure}

\section{Discussion}
\vspace{-5pt}
In order to further analyze what characterizes higher-quality summaries, we conduct several complementary analyses. We first perform human annotation across the generated summaries of different models (Figure~\ref{fig:heatmap_human_eval}, Appendix~\ref{sec:Human_eval_protocol}), which reveals a consistent ranking of \gemini, \qwen, then \gemma, and aligns with all previous metric-based results.
To better understand what drives this ranking, we conduct a macro-level analysis of summary length (Figure~\ref{fig:sentence_length_analysis}). We find that \gemini and \qwen consistently generate longer summaries than the reference ground truth, while \gemma produces shorter summaries than the reference across all prompting methods except CoT, where its output is only marginally longer than the reference. This pattern suggests a possible association between summary length and perceived quality, with longer summaries corresponding to higher-ranked models.

To test this hypothesis more directly, we move beyond macro-level length statistics to a micro-level analysis of summary content, examining the specific entities each model includes through named entity recognition (NER) using SpaCy (en\_core\_web\_sm) on the English output and reference only, for consistency. As shown in Figure \ref{fig:ner_count_diff}, this finer-grained analysis reinforces the length-based pattern: \gemini and \qwen include more named entities than the reference across the ZS and CoT settings and a similar amont in the FS setting, while \gemma includes a similar amount of named entities as the reference in the ZS and FS settings, and more in CoT. Entity category analysis (Figure~\ref{fig:ner_by_type}) reveals that this trend holds nearly uniformly across entity types (e.g., Geopolitical Entity (GPE), individual names (PERSON), Organization (ORG), and NORP (Nationalities, Other groups, Religious groups, Political groups)); \gemini and \qwen surface more entities than the reference in almost every category, while \gemma demonstrates fewer.

\vspace{-6pt}
\section{Conclusion}
\vspace{-6pt}
In this paper, we have introduced JSumT, a task requiring models to compress the salient content of a long spoken document into a summary rendered in a target language, and present \textsc{VoxSumm}, the first multilingual and cross-lingual benchmark for long-form speech summarization spanning 24 languages. 
Our evaluation reveals substantial variation: \gemini performs best, few-shot prompting benefits stronger models, and generating English summaries from non-English speech is typically better than the reverse. Moreover, translating before summarizing amplifies instruction-following failures, making summarize-then-translate the more robust pipeline. Although high-resource languages outperform low-resource ones on average, this pattern breaks down at the individual-language level (e.g., Korean and Portuguese perform worse than lower-resourced Kyrgyz and Thai). Overall, these findings establish JSumT as a challenging task, and we hope \textsc{VoxSumm} and these findings provide a foundation for building more robust and inclusive multilingual and cross-lingual speech summarization systems.

\section{Limitations}
\textsc{VoxSumm} is derived from the CrossSum dataset, whose cross-lingual pairs are identified automatically using semantic similarity between summaries. As such, this may reflect imperfect alignment, however, we mitigate this by retaining only complete instances, reconstructing bidirectional pairs conservatively, and applying quality-control checks during dataset construction.
Additionally, the speech in \textsc{VoxSumm} is generated from professionally written news text using a multilingual TTS system rather than collected from naturally occurring broadcasts. Despite this, CER, NISQA, and human assessments indicate that the generated audio is intelligible and of high perceptual quality, making it suitable for controlled comparison across languages.

\section{Ethical considerations}
This project involved collecting annotated data from 10 languages (out of 24) to verify the quality of the generated speech data. Additional annotations were collected for validating LLM-as-a-Judge evaluation. All participants  were fairly compensated for their contributions at $\$9$ per hour, on Upwork annotation platform.

\section{Acknowledgement}
This research was supported in part by the Natural Sciences and Engineering Research Council (NSERC) of Canada and in part by the AI2050 program at Schmidt Sciences. This work was partially supported through LLM API credits provided by Google's Gemini Academic Program Award and the OpenAI Researcher Access Award. Finally, we are grateful for the support from IVADO and the Canada First Research Excellence Fund.

\bibliography{custom}

@inproceedings{clifton-etal-2020-100000,
    title = "100,000 Podcasts: A Spoken {E}nglish Document Corpus",
    author = "Clifton, Ann  and
      Reddy, Sravana  and
      Yu, Yongze  and
      Pappu, Aasish  and
      Rezapour, Rezvaneh  and
      Bonab, Hamed  and
      Eskevich, Maria  and
      Jones, Gareth  and
      Karlgren, Jussi  and
      Carterette, Ben  and
      Jones, Rosie",
    editor = "Scott, Donia  and
      Bel, Nuria  and
      Zong, Chengqing",
    booktitle = "Proceedings of the 28th International Conference on Computational Linguistics",
    month = dec,
    year = "2020",
    address = "Barcelona, Spain (Online)",
    publisher = "International Committee on Computational Linguistics",
    url = "https://aclanthology.org/2020.coling-main.519/",
    doi = "10.18653/v1/2020.coling-main.519",
    pages = "5903--5917"
}

@inproceedings{retkowski-etal-2025-summarizing,
    title = "Summarizing Speech: A Comprehensive Survey",
    author = {Retkowski, Fabian  and
      Z{\"u}fle, Maike  and
      Sudmann, Andreas  and
      Pfau, Dinah  and
      Watanabe, Shinji  and
      Niehues, Jan  and
      Waibel, Alexander},
    editor = "Christodoulopoulos, Christos  and
      Chakraborty, Tanmoy  and
      Rose, Carolyn  and
      Peng, Violet",
    booktitle = "Proceedings of the 2025 Conference on Empirical Methods in Natural Language Processing",
    month = nov,
    year = "2025",
    address = "Suzhou, China",
    publisher = "Association for Computational Linguistics",
    url = "https://aclanthology.org/2025.emnlp-main.1388/",
    doi = "10.18653/v1/2025.emnlp-main.1388",
    pages = "27275--27306",
    ISBN = "979-8-89176-332-6"
}

@inproceedings{sharma-etal-2024-speech,
    title = "Speech vs. Transcript: Does It Matter for Human Annotators in Speech Summarization?",
    author = "Sharma, Roshan  and
      Shon, Suwon  and
      Lindsey, Mark  and
      Dhamyal, Hira  and
      Raj, Bhiksha",
    editor = "Ku, Lun-Wei  and
      Martins, Andre  and
      Srikumar, Vivek",
    booktitle = "Proceedings of the 62nd Annual Meeting of the Association for Computational Linguistics (Volume 1: Long Papers)",
    month = aug,
    year = "2024",
    address = "Bangkok, Thailand",
    publisher = "Association for Computational Linguistics",
    url = "https://aclanthology.org/2024.acl-long.790/",
    doi = "10.18653/v1/2024.acl-long.790",
    pages = "14779--14797"
}

@INPROCEEDINGS{kano-etal-2023-long,
  author={Kano, Takatomo and Ogawa, Atsunori and Delcroix, Marc and Sharma, Roshan and Matsuura, Kohei and Watanabe, Shinji},
  booktitle={ICASSP 2023 - 2023 IEEE International Conference on Acoustics, Speech and Signal Processing (ICASSP)}, 
  title={Speech Summarization of Long Spoken Document: Improving Memory Efficiency of Speech/Text Encoders}, 
  year={2023},
  volume={},
  number={},
  pages={1-5},
  doi={10.1109/ICASSP49357.2023.10095019}}

@inproceedings{sharma-etal-2024-r,
    title = "{R}-{BASS} : Relevance-aided Block-wise Adaptation for Speech Summarization",
    author = "Sharma, Roshan  and
      Sharma, Ruchira  and
      Dhamyal, Hira  and
      Singh, Rita  and
      Raj, Bhiksha",
    editor = "Duh, Kevin  and
      Gomez, Helena  and
      Bethard, Steven",
    booktitle = "Findings of the Association for Computational Linguistics: NAACL 2024",
    month = jun,
    year = "2024",
    address = "Mexico City, Mexico",
    publisher = "Association for Computational Linguistics",
    url = "https://aclanthology.org/2024.findings-naacl.54/",
    doi = "10.18653/v1/2024.findings-naacl.54",
    pages = "848--857"
}

@inproceedings{hasan-etal-2021-xl,
    title = "{XL}-Sum: Large-Scale Multilingual Abstractive Summarization for 44 Languages",
    author = "Hasan, Tahmid  and
      Bhattacharjee, Abhik  and
      Islam, Md. Saiful  and
      Mubasshir, Kazi  and
      Li, Yuan-Fang  and
      Kang, Yong-Bin  and
      Rahman, M. Sohel  and
      Shahriyar, Rifat",
    editor = "Zong, Chengqing  and
      Xia, Fei  and
      Li, Wenjie  and
      Navigli, Roberto",
    booktitle = "Findings of the Association for Computational Linguistics: ACL-IJCNLP 2021",
    month = aug,
    year = "2021",
    address = "Online",
    publisher = "Association for Computational Linguistics",
    url = "https://aclanthology.org/2021.findings-acl.413/",
    doi = "10.18653/v1/2021.findings-acl.413",
    pages = "4693--4703"
}

@inproceedings{bhattacharjee-etal-2023-crosssum,
    title = "{C}ross{S}um: Beyond {E}nglish-Centric Cross-Lingual Summarization for 1,500+ Language Pairs",
    author = "Bhattacharjee, Abhik  and
      Hasan, Tahmid  and
      Ahmad, Wasi Uddin  and
      Li, Yuan-Fang  and
      Kang, Yong-Bin  and
      Shahriyar, Rifat",
    editor = "Rogers, Anna  and
      Boyd-Graber, Jordan  and
      Okazaki, Naoaki",
    booktitle = "Proceedings of the 61st Annual Meeting of the Association for Computational Linguistics (Volume 1: Long Papers)",
    month = jul,
    year = "2023",
    address = "Toronto, Canada",
    publisher = "Association for Computational Linguistics",
    url = "https://aclanthology.org/2023.acl-long.143/",
    doi = "10.18653/v1/2023.acl-long.143",
    pages = "2541--2564"
}

@inproceedings{di-gangi-etal-2019-mustc,
    title = "{M}u{ST}-{C}: a {M}ultilingual {S}peech {T}ranslation {C}orpus",
    author = "Di Gangi, Mattia A.  and
      Cattoni, Roldano  and
      Bentivogli, Luisa  and
      Negri, Matteo  and
      Turchi, Marco",
    editor = "Burstein, Jill  and
      Doran, Christy  and
      Solorio, Thamar",
    booktitle = "Proceedings of the 2019 Conference of the North {A}merican Chapter of the Association for Computational Linguistics: Human Language Technologies, Volume 1 (Long and Short Papers)",
    month = jun,
    year = "2019",
    address = "Minneapolis, Minnesota",
    publisher = "Association for Computational Linguistics",
    url = "https://aclanthology.org/N19-1202/",
    doi = "10.18653/v1/N19-1202",
    pages = "2012--2017"
}

@inproceedings{wang-etal-2021-covost,
      title={{CoVoST 2 and Massively Multilingual Speech-to-Text Translation}}, 
      author={Changhan Wang and Anne Wu and Juan Pino},
      year={2021},
      booktitle = {Proceedings of Interspeech 2021}
}

@inproceedings{ICLR2024_f7b77476,
    title={{BooookScore: A systematic exploration of book-length summarization in the era of LLMs}},
    author={Yapei Chang and Kyle Lo and Tanya Goyal and Mohit Iyyer},
    booktitle={The Twelfth International Conference on Learning Representations},
    year={2024},
    url={https://arxiv.org/pdf/2310.00785.pdf}
}

@inproceedings{rehbein-etal-2020-improving,
    title = "Improving Sentence Boundary Detection for Spoken Language Transcripts",
    author = "Rehbein, Ines  and
      Ruppenhofer, Josef  and
      Schmidt, Thomas",
    editor = "Calzolari, Nicoletta  and
      B{\'e}chet, Fr{\'e}d{\'e}ric  and
      Blache, Philippe  and
      Choukri, Khalid  and
      Cieri, Christopher  and
      Declerck, Thierry  and
      Goggi, Sara  and
      Isahara, Hitoshi  and
      Maegaard, Bente  and
      Mariani, Joseph  and
      Mazo, H{\'e}l{\`e}ne  and
      Moreno, Asuncion  and
      Odijk, Jan  and
      Piperidis, Stelios",
    booktitle = "Proceedings of the Twelfth Language Resources and Evaluation Conference",
    month = may,
    year = "2020",
    address = "Marseille, France",
    publisher = "European Language Resources Association",
    url = "https://aclanthology.org/2020.lrec-1.878/",
    pages = "7102--7111",
    language = "eng",
    ISBN = "979-10-95546-34-4"
}

@inproceedings{zechner-waibel-2000-diasumm,
    title = "{DIASUMM}: Flexible Summarization of Spontaneous Dialogues in Unrestricted Domains",
    author = "Zechner, Klaus  and
      Waibel, Alex",
    booktitle = "{COLING} 2000 Volume 2: The 18th International Conference on Computational Linguistics",
    year = "2000",
    url = "https://aclanthology.org/C00-2140/"
}

@misc{barrault2023seamless,
      title={Seamless: Multilingual Expressive and Streaming Speech Translation}, 
      author={Seamless Communication and Loïc Barrault and Yu-An Chung and Mariano Coria Meglioli and David Dale and Ning Dong and Mark Duppenthaler and Paul-Ambroise Duquenne and Brian Ellis and Hady Elsahar and Justin Haaheim and John Hoffman and Min-Jae Hwang and Hirofumi Inaguma and Christopher Klaiber and Ilia Kulikov and Pengwei Li and Daniel Licht and Jean Maillard and Ruslan Mavlyutov and Alice Rakotoarison and Kaushik Ram Sadagopan and Abinesh Ramakrishnan and Tuan Tran and Guillaume Wenzek and Yilin Yang and Ethan Ye and Ivan Evtimov and Pierre Fernandez and Cynthia Gao and Prangthip Hansanti and Elahe Kalbassi and Amanda Kallet and Artyom Kozhevnikov and Gabriel Mejia Gonzalez and Robin San Roman and Christophe Touret and Corinne Wong and Carleigh Wood and Bokai Yu and Pierre Andrews and Can Balioglu and Peng-Jen Chen and Marta R. Costa-jussà and Maha Elbayad and Hongyu Gong and Francisco Guzmán and Kevin Heffernan and Somya Jain and Justine Kao and Ann Lee and Xutai Ma and Alex Mourachko and Benjamin Peloquin and Juan Pino and Sravya Popuri and Christophe Ropers and Safiyyah Saleem and Holger Schwenk and Anna Sun and Paden Tomasello and Changhan Wang and Jeff Wang and Skyler Wang and Mary Williamson},
      year={2023},
      eprint={2312.05187},
      archivePrefix={arXiv},
      primaryClass={cs.CL},
      url={https://arxiv.org/abs/2312.05187}, 
}

@misc{qian2023polyvoice,
      title={PolyVoice: Language Models for Speech to Speech Translation}, 
      author={Qianqian Dong and Zhiying Huang and Qiao Tian and Chen Xu and Tom Ko and Yunlong Zhao and Siyuan Feng and Tang Li and Kexin Wang and Xuxin Cheng and Fengpeng Yue and Ye Bai and Xi Chen and Lu Lu and Zejun Ma and Yuping Wang and Mingxuan Wang and Yuxuan Wang},
      year={2023},
      eprint={2306.02982},
      archivePrefix={arXiv},
      primaryClass={cs.CL},
      url={https://arxiv.org/abs/2306.02982}, 
}

@misc{alastruey2026omnilingual,
      title={Omnilingual MT: Machine Translation for 1,600 Languages}, 
      author={Omnilingual MT Team and Belen Alastruey and Niyati Bafna and Andrea Caciolai and Kevin Heffernan and Artyom Kozhevnikov and Christophe Ropers and Eduardo Sánchez and Charles-Eric Saint-James and Ioannis Tsiamas and Xiang "Tony" Cao and Chierh Cheng and Joe Chuang and Paul-Ambroise Duquenne and Mark Duppenthaler and Nate Ekberg and Cynthia Gao and Pere Lluís Huguet Cabot and João Maria Janeiro and Jean Maillard and Gabriel Mejia Gonzalez and Holger Schwenk and Edan Toledo and Arina Turkatenko and Albert Ventayol-Boada and Rashel Moritz and Alexandre Mourachko and Surya Parimi and Mary Williamson and Shireen Yates and David Dale and Marta R. Costa-jussà},
      year={2026},
      eprint={2603.16309},
      archivePrefix={arXiv},
      primaryClass={cs.CL},
      url={https://arxiv.org/abs/2603.16309}, 
}

@inproceedings{pu2025empowering,
author = {Pu, Yu and Liu, Xiaoqian and Zhang, Guangyu and Yan, Zheng and Zhang, Wei-Qiang and Chen, Xie},
year = {2025},
month = {08},
pages = {26-30},
title = {Empowering Large Language Models for End-to-End Speech Translation Leveraging Synthetic Data},
doi = {10.21437/Interspeech.2025-2341}
}

@inproceedings{nisqa, 
   title={NISQA: A Deep CNN-Self-Attention Model for Multidimensional Speech Quality Prediction with Crowdsourced Datasets},
   url={http://dx.doi.org/10.21437/Interspeech.2021-299},
   DOI={10.21437/interspeech.2021-299},
   booktitle={Interspeech 2021},
   publisher={ISCA},
   author={Mittag, Gabriel and Naderi, Babak and Chehadi, Assmaa and Möller, Sebastian},
   year={2021},
   month=Aug, pages={2127–2131},
   collection={interspeech_2021} }

@misc{zhu2026omnivoiceomnilingualzeroshottexttospeech,
      title={OmniVoice: Towards Omnilingual Zero-Shot Text-to-Speech with Diffusion Language Models}, 
      author={Han Zhu and Lingxuan Ye and Wei Kang and Zengwei Yao and Liyong Guo and Fangjun Kuang and Zhifeng Han and Weiji Zhuang and Long Lin and Daniel Povey},
      year={2026},
      eprint={2604.00688},
      archivePrefix={arXiv},
      primaryClass={cs.CL},
      url={https://arxiv.org/abs/2604.00688}, 
}

@inproceedings{chhibbar-kalita-2024-automatic,
    title = "Automatic Summarization of Long Documents",
    author = "Chhibbar, Naman  and
      Kalita, Jugal",
    editor = "Lalitha Devi, Sobha  and
      Arora, Karunesh",
    booktitle = "Proceedings of the 21st International Conference on Natural Language Processing (ICON)",
    month = dec,
    year = "2024",
    address = "AU-KBC Research Centre, Chennai, India",
    publisher = "NLP Association of India (NLPAI)",
    url = "https://aclanthology.org/2024.icon-1.72/",
    pages = "607--615"
}

@ARTICLE{Bertscore_usage,
  author={Abdulreda Kadhim, Estabraq and Feizi-Derakhshi, Mohammad-Reza and Aghdasi, Hadi S.},
  journal={IEEE Access}, 
  title={Advanced Text Summarization Model Incorporating NLP Techniques and Feature-Based Scoring}, 
  year={2025},
  volume={13},
  number={},
  pages={19302-19319},
  doi={10.1109/ACCESS.2025.3528830}}

@inproceedings{adelani-etal-2026-speech,
    title = "Speech Translation and Metrics in 2026: Findings of the {IWSLT} Campaign",
    author = {Adelani, David Ifeoluwa  and
      Agostinelli, Victor  and
      Anastasopoulos, Antonios  and
      Bentivogli, Luisa  and
      Bojar, Ond{\v{r}}ej  and
      Brati{\`e}res, S{\'e}bastien  and
      Carpuat, Marine  and
      Carraro, Fabr{\'i}cio  and
      Cattoni, Roldano  and
      Cettolo, Mauro  and
      Chen, Lizhong  and
      Federico, Marcello  and
      Gaido, Marco  and
      Gupta, Mahendra  and
      Han, HyoJung  and
      Hatami, Ali  and
      Howe, Lewis C.  and
      Javorsk{\'y}, D{\'a}vid  and
      Jeon, Yejin  and
      Kasztelnik, Marek  and
      Laurent, Antoine  and
      Liu, Danni  and
      Luu, Nam  and
      Ma, Min  and
      Mach{\'a}{\v{c}}ek, Dominik  and
      Maltais, Marie  and
      Matusov, Evgeny  and
      McCrae, John  and
      Meng, Chutong  and
      Maurya, Chandresh Kumar  and
      Mohammadamini, Mohammad  and
      Moslem, Yasmin  and
      Murray, Kenton  and
      Nakamura, Satoshi  and
      Negri, Matteo  and
      Niehues, Jan  and
      Ojha, Atul Kr.  and
      Ortega, John E.  and
      Ouyang, Siqi  and
      Papi, Sara  and
      Pol{\'a}k, Peter  and
      Retkowski, Fabian  and
      S{\'a}nchez, Stephanny  and
      Savoldi, Beatrice  and
      Sikasote, Claytone  and
      Sperber, Matthias  and
      St{\"u}ker, Sebastian  and
      Sudoh, Katsuhito  and
      Tahon, Marie  and
      Turchi, Marco  and
      Waibel, Alexander  and
      Wilken, Patrick  and
      Zevallos, Rodolfo Joel  and
      Zouhar, Vilem  and
      Z{\"u}fle, Maike},
    editor = "Salesky, Elizabeth  and
      Anastasopoulos, Antonios  and
      Negri, Matteo  and
      Federico, Marcello",
    booktitle = "Proceedings of the 23rd International Conference on Spoken Language Translation ({IWSLT} 2026)",
    month = jul,
    year = "2026",
    address = "San Diego, USA (in-person and online)",
    publisher = "Association for Computational Linguistics",
    url = "https://aclanthology.org/2026.iwslt-1.39/",
    doi = "10.18653/v1/2026.iwslt-1.39",
    pages = "336--422",
    ISBN = "979-8-89176-411-8"
}

@inproceedings{hermann-etal-2015-teaching,
author = {Hermann, Karl Moritz and Ko\v{c}isk\'{y}, Tom\'{a}\v{s} and Grefenstette, Edward and Espeholt, Lasse and Kay, Will and Suleyman, Mustafa and Blunsom, Phil},
title = {Teaching machines to read and comprehend},
year = {2015},
publisher = {MIT Press},
address = {Cambridge, MA, USA},
booktitle = {Proceedings of the 29th International Conference on Neural Information Processing Systems - Volume 1},
pages = {1693–1701},
numpages = {9},
location = {Montreal, Canada},
series = {NIPS'15}
}

@inproceedings{narayan-etal-2018-dont,
    title = "Don{'}t Give Me the Details, Just the Summary! Topic-Aware Convolutional Neural Networks for Extreme Summarization",
    author = "Narayan, Shashi  and
      Cohen, Shay B.  and
      Lapata, Mirella",
    editor = "Riloff, Ellen  and
      Chiang, David  and
      Hockenmaier, Julia  and
      Tsujii, Jun{'}ichi",
    booktitle = "Proceedings of the 2018 Conference on Empirical Methods in Natural Language Processing",
    month = oct # "-" # nov,
    year = "2018",
    address = "Brussels, Belgium",
    publisher = "Association for Computational Linguistics",
    url = "https://aclanthology.org/D18-1206/",
    doi = "10.18653/v1/D18-1206",
    pages = "1797--1807"
}

@inproceedings{scialom-etal-2020-mlsum,
    title = "{MLSUM}: The Multilingual Summarization Corpus",
    author = "Scialom, Thomas  and
      Dray, Paul-Alexis  and
      Lamprier, Sylvain  and
      Piwowarski, Benjamin  and
      Staiano, Jacopo",
    editor = "Webber, Bonnie  and
      Cohn, Trevor  and
      He, Yulan  and
      Liu, Yang",
    booktitle = "Proceedings of the 2020 Conference on Empirical Methods in Natural Language Processing (EMNLP)",
    month = nov,
    year = "2020",
    address = "Online",
    publisher = "Association for Computational Linguistics",
    url = "https://aclanthology.org/2020.emnlp-main.647/",
    doi = "10.18653/v1/2020.emnlp-main.647",
    pages = "8051--8067"
}

@inproceedings{ladhak-etal-2020-wikilingua,
    title = "{W}iki{L}ingua: A New Benchmark Dataset for Cross-Lingual Abstractive Summarization",
    author = "Ladhak, Faisal  and
      Durmus, Esin  and
      Cardie, Claire  and
      McKeown, Kathleen",
    editor = "Cohn, Trevor  and
      He, Yulan  and
      Liu, Yang",
    booktitle = "Findings of the Association for Computational Linguistics: EMNLP 2020",
    month = nov,
    year = "2020",
    address = "Online",
    publisher = "Association for Computational Linguistics",
    url = "https://aclanthology.org/2020.findings-emnlp.360/",
    doi = "10.18653/v1/2020.findings-emnlp.360",
    pages = "4034--4048"
}

@inproceedings{palaskar-etal-2019-multimodal,
    title = "Multimodal Abstractive Summarization for How2 Videos",
    author = "Palaskar, Shruti  and
      Libovick{\'y}, Jind{\v{r}}ich  and
      Gella, Spandana  and
      Metze, Florian",
    editor = "Korhonen, Anna  and
      Traum, David  and
      M{\`a}rquez, Llu{\'i}s",
    booktitle = "Proceedings of the 57th Annual Meeting of the Association for Computational Linguistics",
    month = jul,
    year = "2019",
    address = "Florence, Italy",
    publisher = "Association for Computational Linguistics",
    url = "https://aclanthology.org/P19-1659/",
    doi = "10.18653/v1/P19-1659",
    pages = "6587--6596"
}

@inproceedings{zhong-etal-2021-qmsum,
    title = "{QMS}um: A New Benchmark for Query-based Multi-domain Meeting Summarization",
    author = "Zhong, Ming  and
      Yin, Da  and
      Yu, Tao  and
      Zaidi, Ahmad  and
      Mutuma, Mutethia  and
      Jha, Rahul  and
      Awadallah, Ahmed Hassan  and
      Celikyilmaz, Asli  and
      Liu, Yang  and
      Qiu, Xipeng  and
      Radev, Dragomir",
    editor = "Toutanova, Kristina  and
      Rumshisky, Anna  and
      Zettlemoyer, Luke  and
      Hakkani-Tur, Dilek  and
      Beltagy, Iz  and
      Bethard, Steven  and
      Cotterell, Ryan  and
      Chakraborty, Tanmoy  and
      Zhou, Yichao",
    booktitle = "Proceedings of the 2021 Conference of the North American Chapter of the Association for Computational Linguistics: Human Language Technologies",
    month = jun,
    year = "2021",
    address = "Online",
    publisher = "Association for Computational Linguistics",
    url = "https://aclanthology.org/2021.naacl-main.472/",
    doi = "10.18653/v1/2021.naacl-main.472",
    pages = "5905--5921"
}

@article{zhangbertscore,
  title={BERTScore: Evaluating Text Generation with BERT},
  author={Tianyi Zhang and Varsha Kishore and Felix Wu and Kilian Q. Weinberger and Yoav Artzi},
  journal={ArXiv},
  year={2019},
  volume={abs/1904.09675},
  url={https://api.semanticscholar.org/CorpusID:127986044}
}

@inproceedings{wang2023element,
    title = "Element-aware Summarization with Large Language Models: Expert-aligned Evaluation and Chain-of-Thought Method",
    author = "Wang, Yiming  and
      Zhang, Zhuosheng  and
      Wang, Rui",
    editor = "Rogers, Anna  and
      Boyd-Graber, Jordan  and
      Okazaki, Naoaki",
    booktitle = "Proceedings of the 61st Annual Meeting of the Association for Computational Linguistics (Volume 1: Long Papers)",
    month = jul,
    year = "2023",
    address = "Toronto, Canada",
    publisher = "Association for Computational Linguistics",
    url = "https://aclanthology.org/2023.acl-long.482/",
    doi = "10.18653/v1/2023.acl-long.482",
    pages = "8640--8665"
}

@article{10.1162/tacl_a_00683,
    title = "x{COMET}: Transparent Machine Translation Evaluation through Fine-grained Error Detection",
    author = "Guerreiro, Nuno M.  and
      Rei, Ricardo  and
      van Stigt, Daan  and
      Coheur, Luisa  and
      Colombo, Pierre  and
      Martins, Andr{\'e} F. T.",
    journal = "Transactions of the Association for Computational Linguistics",
    volume = "12",
    year = "2024",
    address = "Cambridge, MA",
    publisher = "MIT Press",
    url = "https://aclanthology.org/2024.tacl-1.54/",
    doi = "10.1162/tacl_a_00683",
    pages = "979--995"
}

@inproceedings{liu-etal-2023-g,
    title = "{G}-Eval: {NLG} Evaluation using Gpt-4 with Better Human Alignment",
    author = "Liu, Yang  and
      Iter, Dan  and
      Xu, Yichong  and
      Wang, Shuohang  and
      Xu, Ruochen  and
      Zhu, Chenguang",
    editor = "Bouamor, Houda  and
      Pino, Juan  and
      Bali, Kalika",
    booktitle = "Proceedings of the 2023 Conference on Empirical Methods in Natural Language Processing",
    month = dec,
    year = "2023",
    address = "Singapore",
    publisher = "Association for Computational Linguistics",
    url = "https://aclanthology.org/2023.emnlp-main.153/",
    doi = "10.18653/v1/2023.emnlp-main.153",
    pages = "2511--2522"
}

@article{ZHANG2025128369,
author = {Zhang, Yanyue and Lai, Yilong and Wang, Zhenglin and Zhou, Deyu},
title = {DimSum: Disentangling representation with automatically generated multi-category summary templates for fine-grained opinion summarization},
year = {2025},
issue_date = {Sep 2025},
publisher = {Pergamon Press, Inc.},
address = {USA},
volume = {290},
number = {C},
issn = {0957-4174},
url = {https://doi.org/10.1016/j.eswa.2025.128369},
doi = {10.1016/j.eswa.2025.128369},
journal = {Expert Syst. Appl.},
month = sep,
numpages = {12}
}

@misc{yuan-zhang-2026-understanding,
      title={Understanding LLM Reasoning for Abstractive Summarization}, 
      author={Haohan Yuan and Haopeng Zhang},
      year={2026},
      eprint={2512.03503},
      archivePrefix={arXiv},
      primaryClass={cs.CL},
      url={https://arxiv.org/abs/2512.03503}, 
}

@article{AURIEMMACITARELLA2025102571,
title = {Assessing the effectiveness of ROUGE as unbiased metric in Extractive vs. Abstractive summarization techniques},
journal = {Journal of Computational Science},
volume = {87},
pages = {102571},
year = {2025},
issn = {1877-7503},
doi = {https://doi.org/10.1016/j.jocs.2025.102571},
url = {https://www.sciencedirect.com/science/article/pii/S1877750325000481},
author = {Alessia {Auriemma Citarella} and Marcello Barbella and Madalina G. Ciobanu and Fabiola {De Marco} and Luigi {Di Biasi} and Genoveffa Tortora}
}

@article{chen_cothssum_2025,
author = {Chen, Xiaoyong and Chen, Zhiqiang and Cheng, Shi},
year = {2025},
month = {05},
pages = {},
title = {CoTHSSum: Structured long-document summarization via chain-of-thought reasoning and hierarchical segmentation},
volume = {37},
journal = {Journal of King Saud University Computer and Information Sciences},
doi = {10.1007/s44443-025-00041-2}
}

@article{conneau2023fleurs,
  title={FLEURS: FEW-Shot Learning Evaluation of Universal Representations of Speech},
  author={Alexis Conneau and Min Ma and Simran Khanuja and Yu Zhang and Vera Axelrod and Siddharth Dalmia and Jason Riesa and Clara Rivera and Ankur Bapna},
  journal={2022 IEEE Spoken Language Technology Workshop (SLT)},
  year={2022},
  pages={798-805},
  url={https://api.semanticscholar.org/CorpusID:249062909}
}

\appendix

\section{Additional Experimental Setup}
\label{sec:results_additional}

\subsection{Metrics}
\label{sec:add_metrics}

To account for missing summaries, a common issue for some systems and lower-resourced language, we weight each system's BERTScore, xCOMET, and ROUGE-L scores accordingly. Specifically, we penalize missing summaries by scaling the score by the proportion of valid generated samples, leaving scores unaffected when a system generates summaries for the entire test set.

\noindent \textbf{xCOMET} Previously, the quality-estimation (QE) configuration of xCOMET-XL was adopted, which scores the predicted summary directly against the reference without access to the source article. We additionally evaluate the machine translation aspect of our tasks through the MT evaluation setting of xCOMET-XL (i.e., src + ref) by using the full source language article as the source text, the reference summary and the predicted summary \cite{10.1162/tacl_a_00683}. As shown in Table~\ref{tab:xComet_MT_scores}, the resulting model rankings are consistent across both the MT and QE configurations.

\begin{table}[t]
\centering
\label{tab:xcomet_scores_mt_setting}
\resizebox{0.9\linewidth}{!}{
\begin{tabular}{l ccc}
\toprule
\textbf{Target Language} & \textbf{\gemini} & \textbf{\gemma} & \textbf{\qwen} \\
\midrule
\midrule
Amharic         & 0.270 & 0.205 & 0.184 \\
Arabic          & 0.281 & 0.217 & 0.228 \\
Bengali         & 0.245 & 0.202 & 0.192 \\
Chinese (Simp.) & 0.327 & 0.241 & 0.275 \\
French          & 0.259 & 0.206 & 0.225 \\
Gujarati        & 0.254 & 0.200 & 0.179 \\
Hindi           & 0.259 & 0.210 & 0.210 \\
Indonesian      & 0.304 & 0.226 & 0.250 \\
Japanese        & 0.298 & 0.211 & 0.238 \\
Korean          & 0.274 & 0.210 & 0.227 \\
Kyrgyz          & 0.261 & 0.195 & 0.183 \\
Persian         & 0.304 & 0.215 & 0.229 \\
Portuguese      & 0.279 & 0.215 & 0.237 \\
Punjabi         & 0.249 & 0.201 & 0.195 \\
Russian         & 0.320 & 0.235 & 0.261 \\
Sinhala         & 0.261 & 0.193 & 0.177 \\
Spanish         & 0.287 & 0.211 & 0.245 \\
Swahili         & 0.267 & 0.202 & 0.193 \\
Tamil           & 0.237 & 0.194 & 0.188 \\
Telugu          & 0.248 & 0.199 & 0.190 \\
Thai            & 0.306 & 0.230 & 0.246 \\
Turkish         & 0.300 & 0.216 & 0.232 \\
Ukrainian       & 0.319 & 0.218 & 0.241 \\
Vietnamese      & 0.285 & 0.215 & 0.258 \\
\midrule
\textbf{Avg}    & 0.279 & 0.211 & 0.220 \\
\bottomrule
\bottomrule
\end{tabular}
}
\caption{\textbf{xCOMET scores from the MT evaluation} setting for our Table \ref{tab:results_eng2xx} results, averaged across FS, ZS, and CoT prompting methods, with the full source text article, reference summary, and predicted summary as inputs in the English$\rightarrow$XX direction.}
\label{tab:xComet_MT_scores}
\end{table}

\paragraph{ROUGE-L}
Given that BERTScore and xCOMET exhibit the highest correlation with human judgment, we have prioritized these two metrics as our primary evaluation measures. Nonetheless, given that ROUGE (Recall-Oriented Understudy for Gisting Evaluation) remains among the most widely used metrics in automatic summarization evaluation \cite{AURIEMMACITARELLA2025102571}, we additionally report ROUGE-L results in Table~\ref{tab:results_eng2xx_rouge_bertsrc}. As shown, ROUGE-L yields the same overall ranking as BERTScore, with \gemini performing best, followed by \qwen and then \gemma, further corroborating our findings.

\noindent  \textbf{G-Eval}
As done in current literature \citep{ZHANG2025128369, yuan-zhang-2026-understanding}, we use \textbf{Gemini 3.5 Flash} to perform G-Eval for a more explicit look at specific summarization criteria \citep{liu-etal-2023-g}. As \gemini is multimodal, we input the source audio and the source text for each evaluated summary.  We evaluate on the four different quality aspects presented in \citet{liu-etal-2023-g}: Coherence, Consistency, Fluency and Relevance. All G-Eval prompts are available in Appendix \ref{sec:prompts}. We corroborate these results with human evaluation, which is further detailed in Appendix \ref{sec:Human_eval_protocol}.

\begin{table}[t]
\centering
\resizebox{\linewidth}{!}{%
\begin{tabular}{l cc cc}
\toprule
\multirow{2}{*}{\textbf{Language}} & \multicolumn{2}{c}{\textbf{X-Comet-XL}} & \multicolumn{2}{c}{\textbf{BERTScore F1 (ref)}} \\
\cmidrule(lr){2-3} \cmidrule(lr){4-5}
 & Trans$\rightarrow$Sum & Sum$\rightarrow$Trans & Trans.$\rightarrow$Sum. & Sum$\rightarrow$Trans \\
\midrule
Amharic & \textbf{0.246} & 0.228 & 0.711 & \textbf{0.717} \\
Arabic & \textbf{0.272} & 0.241 & \textbf{0.756} & 0.753 \\
Bengali & \textbf{0.214} & 0.205 & 0.666 & \textbf{0.667} \\
Chinese & \textbf{0.328} & 0.322 & 0.774 & \textbf{0.774} \\
French & \textbf{0.246} & 0.223 & \textbf{0.708} & 0.707 \\
Gujarati & \textbf{0.237} & 0.212 & \textbf{0.646} & 0.637 \\
Hindi & \textbf{0.234} & 0.215 & \textbf{0.652} & 0.648 \\
Indonesian & \textbf{0.273} & 0.270 & 0.674 & \textbf{0.679} \\
Japanese & \textbf{0.288} & 0.276 & 0.731 & \textbf{0.734} \\
Korean & \textbf{0.258} & 0.233 & 0.677 & \textbf{0.678} \\
Kyrgyz & \textbf{0.252} & 0.224 & \textbf{0.706} & 0.703 \\
Persian & \textbf{0.283} & 0.270 & 0.682 & \textbf{0.689} \\
Portuguese & \textbf{0.290} & 0.274 & 0.657 & \textbf{0.658} \\
Punjabi & \textbf{0.227} & 0.204 & 0.648 & \textbf{0.649} \\
Russian & \textbf{0.339} & 0.300 & 0.701 & \textbf{0.703} \\
Sinhala & \textbf{0.218} & 0.206 & 0.702 & \textbf{0.705} \\
Spanish & \textbf{0.259} & 0.254 & 0.682 & \textbf{0.688} \\
Swahili & \textbf{0.221} & 0.219 & 0.665 & \textbf{0.674} \\
Tamil & \textbf{0.201} & 0.196 & 0.675 & \textbf{0.676} \\
Telugu & \textbf{0.232} & 0.208 & 0.668 & \textbf{0.671} \\
Thai & \textbf{0.327} & 0.311 & 0.619 & \textbf{0.670} \\
Turkish & \textbf{0.279} & 0.242 & \textbf{0.680} & 0.679 \\
Ukrainian & \textbf{0.319} & 0.284 & \textbf{0.692} & 0.692 \\
Vietnamese & \textbf{0.278} & 0.249 & 0.651 & \textbf{0.654} \\
\midrule
\textbf{Average} & \textbf{0.263} & 0.244 & 0.684 & \textbf{0.688} \\
\bottomrule
\end{tabular}%
}
\caption{\textbf{Comparison of the two cascaded pipeline orderings }on Eng$\rightarrow$XX (\gemini). \textit{Trans$\rightarrow$Sum} translates the English source first and then summarizes in the target language; \textit{Sum$\rightarrow$Trans} summarizes the English source first and then translates the summary into the target language. Both are evaluated against target-language references with xCOMET and BERTScore-F1. Best score per language and metric is in \textbf{bold}.}
\label{tab:cascade_compare_eng2xx_xcomet}
\end{table}

\begin{table}[t]
\centering
\small
\resizebox{0.8\linewidth}{!}{%
\begin{tabular}{lccc}
\toprule
\textbf{Language} & \textbf{FS} & \textbf{FS Text Only} & \textbf{$\Delta$} \\
\midrule
Amharic & 0.589 & 0.638 & \color{green!60!black}{+0.049} \\
Arabic & 0.745 & 0.758 & \color{green!60!black}{+0.014} \\
Bengali & 0.646 & 0.654 & \color{green!60!black}{+0.008} \\
Chinese & 0.771 & 0.780 & \color{green!60!black}{+0.010} \\
French & 0.701 & 0.716 & \color{green!60!black}{+0.015} \\
Gujarati & 0.620 & 0.639 & \color{green!60!black}{+0.020} \\
Hindi & 0.656 & 0.660 & \color{green!60!black}{+0.004} \\
Indonesian & 0.687 & 0.699 & \color{green!60!black}{+0.012} \\
Japanese & 0.718 & 0.739 & \color{green!60!black}{+0.021} \\
Korean & 0.690 & 0.710 & \color{green!60!black}{+0.020} \\
Kyrgyz & 0.678 & 0.677 & \color{red!70!black}{-0.001} \\
Persian & 0.655 & 0.673 & \color{green!60!black}{+0.018 }\\
Portuguese & 0.649 & 0.660 & \color{green!60!black}{+0.011} \\
Punjabi & 0.626 & 0.631 & \color{green!60!black}{+0.005} \\
Russian & 0.707 & 0.714 & \color{green!60!black}{+0.007} \\
Sinhala & 0.570 & 0.600 & \color{green!60!black}{+0.030} \\
Spanish & 0.697 & 0.701 & \color{green!60!black}{+0.003} \\
Swahili & 0.596 & 0.634 & \color{green!60!black}{+0.038} \\
Tamil & 0.645 & 0.657 & \color{green!60!black}{+0.012} \\
Telugu & 0.641 & 0.653 & \color{green!60!black}{+0.012} \\
Thai & 0.668 & 0.702 & \color{green!60!black}{+0.034} \\
Turkish & 0.670 & 0.693 & \color{green!60!black}{+0.023} \\
Ukrainian & 0.664 & 0.694 & \color{green!60!black}{+0.030 }\\
Vietnamese & 0.684 & 0.689 & \color{green!60!black}{+0.004} \\
\midrule
\textbf{Average} & \textbf{0.666} & \textbf{0.682} & \textbf{+0.017} \\
\bottomrule
\end{tabular}%
}
\caption{\textbf{Influence of input modality }on performance. BERTScore results between \qwen FS text and audio article inputs.}
\label{tab:bertscore_comp_text_audio}
\end{table}

\subsection{Experimental Details and Results}
While \gemini and \gemma officially support all 24 languages included in our benchmark, \qwen lacks official support for some of the low-resource languages: Amharic, Bengali, Gujarati, Hindi, Kyrgyz, Persian, Punjabi, Sinhala, Swahili, Tamil, Telugu, Thai, Ukrainian and Vietnamese. However, we do not exclude them for our evaluation of \qwen, as preliminary analysis showed that it still performed relatively well, particularly in CoT and FS settings. 

\paragraph{Cascading Pipeline} To further assess the performance of actual cascaded systems, we construct the following pipeline: Omnilingual-ASR (LLM-1B) first transcribes the source audio, after which \gemini summarizes the resulting transcript; the summary is then passed to a separate instance of \gemini, which translates it into the target language. We apply the same cascaded procedure in reverse for the Translation$\rightarrow$Summarization setting. As shown in Table~\ref{tab:cascade_compare_eng2xx_xcomet}, this cascaded pipeline degrades Gemini's performance relative to its corresponding end-to-end systems in Table~\ref{tab:results_eng2xx}, regardless of whether summarization or translation is performed first. This degradation underscores the substantial impact of error propagation across the successive ASR, translation, and summarization stages inherent to cascaded architectures.

\paragraph{Modality Ablation}
In order to isolate the impact of acoustic encoding on JSumT performance, we evaluate \qwen under a text-only FS setting across all languages, substituting the synthesized audio input with the original source-language article text (Table \ref{tab:bertscore_comp_text_audio}). We observe a consistent improvement across all languages under text input, with an average BERTScore-F1 gain of +0.017 relative to the audio-based setting. This indicates that JSumT's difficulty is not confined to cross-lingual generation alone, but also stems from information loss introduced during acoustic encoding. 

Furthermore, in order to compare the difficulty of speech summarization versus speech JSumT, we evaluate \qwen in a FS English-to-English summarization setting. We observe an improvement over the average \qwen FS Summarization$\rightarrow$Translation (0.665) in the summarization-only task (0.700), which is a \textcolor{green!60!black}{0.035} gain in BERTScore-F1 performance.

\begin{table*}[t]
\centering
\resizebox{\textwidth}{!}{%
\begin{tabular}{ l *{12}{C{1.4cm}} }
\toprule
\multirow{2}{*}{\textbf{Pair}} & \multicolumn{4}{c}{\textbf{\gemini}} & \multicolumn{4}{c}{\textbf{\gemma}} & \multicolumn{4}{c}{\textbf{\qwen}} \\
\cmidrule(lr){2-5} \cmidrule(lr){6-9} \cmidrule(lr){10-13}
 & FS & ZS & CoT & Avg & FS & ZS & CoT & Avg & FS & ZS & CoT & Avg \\
\midrule
am & \cellcolor{purple!50} \makecell{0.780 \\[-2pt] {\footnotesize (0.083)}}  & \cellcolor{purple!48} \makecell{0.716 \\[-2pt] {\footnotesize (0.058)}}  & \cellcolor{purple!46} \makecell{0.713 \\[-2pt] {\footnotesize (0.047)}}  & \cellcolor{gray!15} \makecell{0.736 \\[-2pt] {\footnotesize (0.062)}}  & \cellcolor{purple!39} \makecell{0.624 \\[-2pt] {\footnotesize (0.017)}}  & \cellcolor{purple!41} \makecell{0.623 \\[-2pt] {\footnotesize (0.013)}}  & \cellcolor{purple!41} \makecell{0.623 \\[-2pt] {\footnotesize (0.021)}}  & \cellcolor{gray!15} \makecell{0.623 \\[-2pt] {\footnotesize (0.017)}}  & \cellcolor{purple!12} \makecell{0.589 \\[-2pt] {\footnotesize (0.005)}}  & \cellcolor{purple!23} \makecell{0.637 \\[-2pt] {\footnotesize (0.033)}}  & \cellcolor{purple!21} \makecell{0.627 \\[-2pt] {\footnotesize (0.034)}}  & \cellcolor{gray!15} \makecell{0.618 \\[-2pt] {\footnotesize (0.024)}} \\[3pt]
ar & \cellcolor{purple!52} \makecell{0.792 \\[-2pt] {\footnotesize (0.051)}}  & \cellcolor{purple!52} \makecell{0.756 \\[-2pt] {\footnotesize (0.047)}}  & \cellcolor{purple!52} \makecell{0.759 \\[-2pt] {\footnotesize (0.047)}}  & \cellcolor{gray!15} \makecell{0.769 \\[-2pt] {\footnotesize (0.048)}}  & \cellcolor{purple!52} \makecell{0.658 \\[-2pt] {\footnotesize (0.003)}}  & \cellcolor{purple!52} \makecell{0.685 \\[-2pt] {\footnotesize (0.018)}}  & \cellcolor{purple!52} \makecell{0.682 \\[-2pt] {\footnotesize (0.029)}}  & \cellcolor{gray!15} \makecell{0.675 \\[-2pt] {\footnotesize (0.017)}}  & \cellcolor{purple!52} \makecell{0.745 \\[-2pt] {\footnotesize (0.030)}}  & \cellcolor{purple!52} \makecell{0.729 \\[-2pt] {\footnotesize (0.032)}}  & \cellcolor{purple!50} \makecell{0.714 \\[-2pt] {\footnotesize (0.025)}}  & \cellcolor{gray!15} \makecell{0.729 \\[-2pt] {\footnotesize (0.029)}} \\[3pt]
bn & \cellcolor{purple!23} \makecell{0.716 \\[-2pt] {\footnotesize (0.000)}}  & \cellcolor{purple!23} \makecell{0.672 \\[-2pt] {\footnotesize (0.000)}}  & \cellcolor{purple!21} \makecell{0.671 \\[-2pt] {\footnotesize (0.000)}}  & \cellcolor{gray!15} \makecell{0.686 \\[-2pt] {\footnotesize (0.000)}}  & \cellcolor{purple!10} \makecell{0.560 \\[-2pt] {\footnotesize (0.000)}}  & \cellcolor{purple!29} \makecell{0.608 \\[-2pt] {\footnotesize (0.000)}}  & \cellcolor{purple!22} \makecell{0.597 \\[-2pt] {\footnotesize (0.000)}}  & \cellcolor{gray!15} \makecell{0.588 \\[-2pt] {\footnotesize (0.000)}}  & \cellcolor{purple!23} \makecell{0.646 \\[-2pt] {\footnotesize (0.000)}}  & \cellcolor{purple!21} \makecell{0.634 \\[-2pt] {\footnotesize (0.000)}}  & \cellcolor{purple!23} \makecell{0.627 \\[-2pt] {\footnotesize (0.000)}}  & \cellcolor{gray!15} \makecell{0.636 \\[-2pt] {\footnotesize (0.000)}} \\[3pt]
zh & \cellcolor{purple!54} \makecell{0.802 \\[-2pt] {\footnotesize (0.070)}}  & \cellcolor{purple!54} \makecell{0.778 \\[-2pt] {\footnotesize (0.066)}}  & \cellcolor{purple!54} \makecell{0.775 \\[-2pt] {\footnotesize (0.074)}}  & \cellcolor{gray!15} \makecell{0.785 \\[-2pt] {\footnotesize (0.070)}}  & \cellcolor{purple!54} \makecell{0.716 \\[-2pt] {\footnotesize (0.038)}}  & \cellcolor{purple!54} \makecell{0.723 \\[-2pt] {\footnotesize (0.040)}}  & \cellcolor{purple!54} \makecell{0.726 \\[-2pt] {\footnotesize (0.049)}}  & \cellcolor{gray!15} \makecell{0.722 \\[-2pt] {\footnotesize (0.042)}}  & \cellcolor{purple!54} \makecell{0.770 \\[-2pt] {\footnotesize (0.047)}}  & \cellcolor{purple!54} \makecell{0.763 \\[-2pt] {\footnotesize (0.059)}}  & \cellcolor{purple!54} \makecell{0.751 \\[-2pt] {\footnotesize (0.058)}}  & \cellcolor{gray!15} \makecell{0.762 \\[-2pt] {\footnotesize (0.055)}} \\[3pt]
fr & \cellcolor{purple!44} \makecell{0.745 \\[-2pt] {\footnotesize (0.275)}}  & \cellcolor{purple!46} \makecell{0.715 \\[-2pt] {\footnotesize (0.220)}}  & \cellcolor{purple!43} \makecell{0.710 \\[-2pt] {\footnotesize (0.212)}}  & \cellcolor{gray!15} \makecell{0.724 \\[-2pt] {\footnotesize (0.236)}}  & \cellcolor{purple!41} \makecell{0.625 \\[-2pt] {\footnotesize (0.158)}}  & \cellcolor{purple!48} \makecell{0.641 \\[-2pt] {\footnotesize (0.166)}}  & \cellcolor{purple!43} \makecell{0.635 \\[-2pt] {\footnotesize (0.149)}}  & \cellcolor{gray!15} \makecell{0.634 \\[-2pt] {\footnotesize (0.158)}}  & \cellcolor{purple!46} \makecell{0.701 \\[-2pt] {\footnotesize (0.170)}}  & \cellcolor{purple!48} \makecell{0.684 \\[-2pt] {\footnotesize (0.187)}}  & \cellcolor{purple!43} \makecell{0.664 \\[-2pt] {\footnotesize (0.138)}}  & \cellcolor{gray!15} \makecell{0.683 \\[-2pt] {\footnotesize (0.165)}} \\[3pt]
gu & \cellcolor{purple!10} \makecell{0.679 \\[-2pt] {\footnotesize (0.070)}}  & \cellcolor{purple!10} \makecell{0.644 \\[-2pt] {\footnotesize (0.062)}}  & \cellcolor{purple!10} \makecell{0.651 \\[-2pt] {\footnotesize (0.035)}}  & \cellcolor{gray!15} \makecell{0.658 \\[-2pt] {\footnotesize (0.056)}}  & \cellcolor{purple!12} \makecell{0.568 \\[-2pt] {\footnotesize (0.031)}}  & \cellcolor{purple!10} \makecell{0.575 \\[-2pt] {\footnotesize (0.019)}}  & \cellcolor{purple!12} \makecell{0.581 \\[-2pt] {\footnotesize (0.015)}}  & \cellcolor{gray!15} \makecell{0.575 \\[-2pt] {\footnotesize (0.022)}}  & \cellcolor{purple!16} \makecell{0.620 \\[-2pt] {\footnotesize (0.015)}}  & \cellcolor{purple!12} \makecell{0.612 \\[-2pt] {\footnotesize (0.042)}}  & \cellcolor{purple!14} \makecell{0.606 \\[-2pt] {\footnotesize (0.018)}}  & \cellcolor{gray!15} \makecell{0.613 \\[-2pt] {\footnotesize (0.025)}} \\[3pt]
hi & \cellcolor{purple!18} \makecell{0.694 \\[-2pt] {\footnotesize (0.078)}}  & \cellcolor{purple!14} \makecell{0.656 \\[-2pt] {\footnotesize (0.057)}}  & \cellcolor{purple!14} \makecell{0.657 \\[-2pt] {\footnotesize (0.070)}}  & \cellcolor{gray!15} \makecell{0.669 \\[-2pt] {\footnotesize (0.068)}}  & \cellcolor{purple!18} \makecell{0.578 \\[-2pt] {\footnotesize (0.005)}}  & \cellcolor{purple!16} \makecell{0.594 \\[-2pt] {\footnotesize (0.016)}}  & \cellcolor{purple!16} \makecell{0.587 \\[-2pt] {\footnotesize (0.007)}}  & \cellcolor{gray!15} \makecell{0.586 \\[-2pt] {\footnotesize (0.009)}}  & \cellcolor{purple!29} \makecell{0.656 \\[-2pt] {\footnotesize (0.012)}}  & \cellcolor{purple!15} \makecell{0.623 \\[-2pt] {\footnotesize (0.027)}}  & \cellcolor{purple!20} \makecell{0.622 \\[-2pt] {\footnotesize (0.025)}}  & \cellcolor{gray!15} \makecell{0.633 \\[-2pt] {\footnotesize (0.021)}} \\[3pt]
id & \cellcolor{purple!33} \makecell{0.721 \\[-2pt] {\footnotesize (0.242)}}  & \cellcolor{purple!29} \makecell{0.686 \\[-2pt] {\footnotesize (0.195)}}  & \cellcolor{purple!27} \makecell{0.676 \\[-2pt] {\footnotesize (0.174)}}  & \cellcolor{gray!15} \makecell{0.695 \\[-2pt] {\footnotesize (0.204)}}  & \cellcolor{purple!29} \makecell{0.607 \\[-2pt] {\footnotesize (0.126)}}  & \cellcolor{purple!37} \makecell{0.616 \\[-2pt] {\footnotesize (0.129)}}  & \cellcolor{purple!35} \makecell{0.620 \\[-2pt] {\footnotesize (0.123)}}  & \cellcolor{gray!15} \makecell{0.614 \\[-2pt] {\footnotesize (0.126)}}  & \cellcolor{purple!41} \makecell{0.686 \\[-2pt] {\footnotesize (0.193)}}  & \cellcolor{purple!37} \makecell{0.662 \\[-2pt] {\footnotesize (0.168)}}  & \cellcolor{purple!29} \makecell{0.641 \\[-2pt] {\footnotesize (0.091)}}  & \cellcolor{gray!15} \makecell{0.663 \\[-2pt] {\footnotesize (0.151)}} \\[3pt]
ja & \cellcolor{purple!48} \makecell{0.768 \\[-2pt] {\footnotesize (0.197)}}  & \cellcolor{purple!50} \makecell{0.721 \\[-2pt] {\footnotesize (0.141)}}  & \cellcolor{purple!50} \makecell{0.734 \\[-2pt] {\footnotesize (0.138)}}  & \cellcolor{gray!15} \makecell{0.741 \\[-2pt] {\footnotesize (0.159)}}  & \cellcolor{purple!50} \makecell{0.644 \\[-2pt] {\footnotesize (0.071)}}  & \cellcolor{purple!50} \makecell{0.651 \\[-2pt] {\footnotesize (0.103)}}  & \cellcolor{purple!50} \makecell{0.656 \\[-2pt] {\footnotesize (0.116)}}  & \cellcolor{gray!15} \makecell{0.650 \\[-2pt] {\footnotesize (0.097)}}  & \cellcolor{purple!50} \makecell{0.718 \\[-2pt] {\footnotesize (0.152)}}  & \cellcolor{purple!50} \makecell{0.708 \\[-2pt] {\footnotesize (0.126)}}  & \cellcolor{purple!52} \makecell{0.714 \\[-2pt] {\footnotesize (0.103)}}  & \cellcolor{gray!15} \makecell{0.713 \\[-2pt] {\footnotesize (0.127)}} \\[3pt]
ko & \cellcolor{purple!29} \makecell{0.717 \\[-2pt] {\footnotesize (0.201)}}  & \cellcolor{purple!31} \makecell{0.687 \\[-2pt] {\footnotesize (0.165)}}  & \cellcolor{purple!33} \makecell{0.690 \\[-2pt] {\footnotesize (0.191)}}  & \cellcolor{gray!15} \makecell{0.698 \\[-2pt] {\footnotesize (0.185)}}  & \cellcolor{purple!43} \makecell{0.630 \\[-2pt] {\footnotesize (0.062)}}  & \cellcolor{purple!33} \makecell{0.614 \\[-2pt] {\footnotesize (0.047)}}  & \cellcolor{purple!29} \makecell{0.608 \\[-2pt] {\footnotesize (0.063)}}  & \cellcolor{gray!15} \makecell{0.617 \\[-2pt] {\footnotesize (0.057)}}  & \cellcolor{purple!43} \makecell{0.690 \\[-2pt] {\footnotesize (0.161)}}  & \cellcolor{purple!41} \makecell{0.666 \\[-2pt] {\footnotesize (0.085)}}  & \cellcolor{purple!31} \makecell{0.641 \\[-2pt] {\footnotesize (0.092)}}  & \cellcolor{gray!15} \makecell{0.666 \\[-2pt] {\footnotesize (0.113)}} \\[3pt]
ky & \cellcolor{purple!41} \makecell{0.737 \\[-2pt] {\footnotesize (0.154)}}  & \cellcolor{purple!43} \makecell{0.710 \\[-2pt] {\footnotesize (0.144)}}  & \cellcolor{purple!44} \makecell{0.712 \\[-2pt] {\footnotesize (0.145)}}  & \cellcolor{gray!15} \makecell{0.720 \\[-2pt] {\footnotesize (0.148)}}  & \cellcolor{purple!46} \makecell{0.634 \\[-2pt] {\footnotesize (0.051)}}  & \cellcolor{purple!43} \makecell{0.635 \\[-2pt] {\footnotesize (0.064)}}  & \cellcolor{purple!39} \makecell{0.622 \\[-2pt] {\footnotesize (0.056)}}  & \cellcolor{gray!15} \makecell{0.630 \\[-2pt] {\footnotesize (0.057)}}  & \cellcolor{purple!37} \makecell{0.678 \\[-2pt] {\footnotesize (0.076)}}  & \cellcolor{purple!33} \makecell{0.660 \\[-2pt] {\footnotesize (0.061)}}  & \cellcolor{purple!41} \makecell{0.660 \\[-2pt] {\footnotesize (0.049)}}  & \cellcolor{gray!15} \makecell{0.666 \\[-2pt] {\footnotesize (0.062)}} \\[3pt]
fa & \cellcolor{purple!35} \makecell{0.724 \\[-2pt] {\footnotesize (0.000)}}  & \cellcolor{purple!35} \makecell{0.691 \\[-2pt] {\footnotesize (0.000)}}  & \cellcolor{purple!35} \makecell{0.694 \\[-2pt] {\footnotesize (0.000)}}  & \cellcolor{gray!15} \makecell{0.703 \\[-2pt] {\footnotesize (0.000)}}  & \cellcolor{purple!33} \makecell{0.609 \\[-2pt] {\footnotesize (0.000)}}  & \cellcolor{purple!39} \makecell{0.618 \\[-2pt] {\footnotesize (0.005)}}  & \cellcolor{purple!33} \makecell{0.612 \\[-2pt] {\footnotesize (0.000)}}  & \cellcolor{gray!15} \makecell{0.613 \\[-2pt] {\footnotesize (0.002)}}  & \cellcolor{purple!27} \makecell{0.655 \\[-2pt] {\footnotesize (0.000)}}  & \cellcolor{purple!35} \makecell{0.662 \\[-2pt] {\footnotesize (0.000)}}  & \cellcolor{purple!33} \makecell{0.642 \\[-2pt] {\footnotesize (0.000)}}  & \cellcolor{gray!15} \makecell{0.653 \\[-2pt] {\footnotesize (0.000)}} \\[3pt]
pt & \cellcolor{purple!12} \makecell{0.679 \\[-2pt] {\footnotesize (0.225)}}  & \cellcolor{purple!18} \makecell{0.658 \\[-2pt] {\footnotesize (0.196)}}  & \cellcolor{purple!18} \makecell{0.658 \\[-2pt] {\footnotesize (0.189)}}  & \cellcolor{gray!15} \makecell{0.665 \\[-2pt] {\footnotesize (0.203)}}  & \cellcolor{purple!35} \makecell{0.613 \\[-2pt] {\footnotesize (0.154)}}  & \cellcolor{purple!26} \makecell{0.605 \\[-2pt] {\footnotesize (0.160)}}  & \cellcolor{purple!27} \makecell{0.607 \\[-2pt] {\footnotesize (0.153)}}  & \cellcolor{gray!15} \makecell{0.608 \\[-2pt] {\footnotesize (0.156)}}  & \cellcolor{purple!25} \makecell{0.649 \\[-2pt] {\footnotesize (0.185)}}  & \cellcolor{purple!29} \makecell{0.647 \\[-2pt] {\footnotesize (0.185)}}  & \cellcolor{purple!25} \makecell{0.636 \\[-2pt] {\footnotesize (0.142)}}  & \cellcolor{gray!15} \makecell{0.644 \\[-2pt] {\footnotesize (0.171)}} \\[3pt]
pa & \cellcolor{purple!16} \makecell{0.686 \\[-2pt] {\footnotesize (0.051)}}  & \cellcolor{purple!12} \makecell{0.652 \\[-2pt] {\footnotesize (0.041)}}  & \cellcolor{purple!20} \makecell{0.665 \\[-2pt] {\footnotesize (0.037)}}  & \cellcolor{gray!15} \makecell{0.668 \\[-2pt] {\footnotesize (0.043)}}  & \cellcolor{purple!14} \makecell{0.573 \\[-2pt] {\footnotesize (0.010)}}  & \cellcolor{purple!12} \makecell{0.585 \\[-2pt] {\footnotesize (0.013)}}  & \cellcolor{purple!10} \makecell{0.577 \\[-2pt] {\footnotesize (0.010)}}  & \cellcolor{gray!15} \makecell{0.578 \\[-2pt] {\footnotesize (0.011)}}  & \cellcolor{purple!18} \makecell{0.626 \\[-2pt] {\footnotesize (0.015)}}  & \cellcolor{purple!19} \makecell{0.629 \\[-2pt] {\footnotesize (0.033)}}  & \cellcolor{purple!16} \makecell{0.613 \\[-2pt] {\footnotesize (0.024)}}  & \cellcolor{gray!15} \makecell{0.623 \\[-2pt] {\footnotesize (0.024)}} \\[3pt]
ru & \cellcolor{purple!43} \makecell{0.743 \\[-2pt] {\footnotesize (0.159)}}  & \cellcolor{purple!41} \makecell{0.705 \\[-2pt] {\footnotesize (0.119)}}  & \cellcolor{purple!41} \makecell{0.702 \\[-2pt] {\footnotesize (0.129)}}  & \cellcolor{gray!15} \makecell{0.717 \\[-2pt] {\footnotesize (0.136)}}  & \cellcolor{purple!37} \makecell{0.624 \\[-2pt] {\footnotesize (0.037)}}  & \cellcolor{purple!46} \makecell{0.638 \\[-2pt] {\footnotesize (0.059)}}  & \cellcolor{purple!44} \makecell{0.636 \\[-2pt] {\footnotesize (0.033)}}  & \cellcolor{gray!15} \makecell{0.633 \\[-2pt] {\footnotesize (0.043)}}  & \cellcolor{purple!48} \makecell{0.707 \\[-2pt] {\footnotesize (0.112)}}  & \cellcolor{purple!44} \makecell{0.677 \\[-2pt] {\footnotesize (0.092)}}  & \cellcolor{purple!44} \makecell{0.665 \\[-2pt] {\footnotesize (0.069)}}  & \cellcolor{gray!15} \makecell{0.683 \\[-2pt] {\footnotesize (0.091)}} \\[3pt]
si & \cellcolor{purple!46} \makecell{0.754 \\[-2pt] {\footnotesize (0.035)}}  & \cellcolor{purple!44} \makecell{0.711 \\[-2pt] {\footnotesize (0.042)}}  & \cellcolor{purple!48} \makecell{0.717 \\[-2pt] {\footnotesize (0.059)}}  & \cellcolor{gray!15} \makecell{0.727 \\[-2pt] {\footnotesize (0.045)}}  & \cellcolor{purple!23} \makecell{0.595 \\[-2pt] {\footnotesize (0.025)}}  & \cellcolor{purple!23} \makecell{0.603 \\[-2pt] {\footnotesize (0.012)}}  & \cellcolor{purple!25} \makecell{0.604 \\[-2pt] {\footnotesize (0.018)}}  & \cellcolor{gray!15} \makecell{0.601 \\[-2pt] {\footnotesize (0.019)}}  & \cellcolor{purple!10} \makecell{0.570 \\[-2pt] {\footnotesize (0.005)}}  & \cellcolor{purple!10} \makecell{0.592 \\[-2pt] {\footnotesize (0.008)}}  & \cellcolor{purple!10} \makecell{0.568 \\[-2pt] {\footnotesize (0.011)}}  & \cellcolor{gray!15} \makecell{0.577 \\[-2pt] {\footnotesize (0.008)}} \\[3pt]
es & \cellcolor{purple!25} \makecell{0.716 \\[-2pt] {\footnotesize (0.242)}}  & \cellcolor{purple!37} \makecell{0.694 \\[-2pt] {\footnotesize (0.214)}}  & \cellcolor{purple!37} \makecell{0.695 \\[-2pt] {\footnotesize (0.201)}}  & \cellcolor{gray!15} \makecell{0.702 \\[-2pt] {\footnotesize (0.219)}}  & \cellcolor{purple!48} \makecell{0.640 \\[-2pt] {\footnotesize (0.169)}}  & \cellcolor{purple!44} \makecell{0.635 \\[-2pt] {\footnotesize (0.170)}}  & \cellcolor{purple!46} \makecell{0.639 \\[-2pt] {\footnotesize (0.161)}}  & \cellcolor{gray!15} \makecell{0.638 \\[-2pt] {\footnotesize (0.167)}}  & \cellcolor{purple!44} \makecell{0.697 \\[-2pt] {\footnotesize (0.221)}}  & \cellcolor{purple!46} \makecell{0.679 \\[-2pt] {\footnotesize (0.192)}}  & \cellcolor{purple!48} \makecell{0.669 \\[-2pt] {\footnotesize (0.149)}}  & \cellcolor{gray!15} \makecell{0.682 \\[-2pt] {\footnotesize (0.188)}} \\[3pt]
sw & \cellcolor{purple!31} \makecell{0.718 \\[-2pt] {\footnotesize (0.270)}}  & \cellcolor{purple!25} \makecell{0.678 \\[-2pt] {\footnotesize (0.211)}}  & \cellcolor{purple!25} \makecell{0.675 \\[-2pt] {\footnotesize (0.206)}}  & \cellcolor{gray!15} \makecell{0.690 \\[-2pt] {\footnotesize (0.229)}}  & \cellcolor{purple!20} \makecell{0.588 \\[-2pt] {\footnotesize (0.144)}}  & \cellcolor{purple!18} \makecell{0.595 \\[-2pt] {\footnotesize (0.147)}}  & \cellcolor{purple!20} \makecell{0.597 \\[-2pt] {\footnotesize (0.141)}}  & \cellcolor{gray!15} \makecell{0.593 \\[-2pt] {\footnotesize (0.144)}}  & \cellcolor{purple!14} \makecell{0.596 \\[-2pt] {\footnotesize (0.145)}}  & \cellcolor{purple!19} \makecell{0.629 \\[-2pt] {\footnotesize (0.160)}}  & \cellcolor{purple!12} \makecell{0.596 \\[-2pt] {\footnotesize (0.084)}}  & \cellcolor{gray!15} \makecell{0.607 \\[-2pt] {\footnotesize (0.130)}} \\[3pt]
ta & \cellcolor{purple!27} \makecell{0.717 \\[-2pt] {\footnotesize (0.066)}}  & \cellcolor{purple!27} \makecell{0.683 \\[-2pt] {\footnotesize (0.067)}}  & \cellcolor{purple!29} \makecell{0.685 \\[-2pt] {\footnotesize (0.064)}}  & \cellcolor{gray!15} \makecell{0.695 \\[-2pt] {\footnotesize (0.066)}}  & \cellcolor{purple!21} \makecell{0.590 \\[-2pt] {\footnotesize (0.008)}}  & \cellcolor{purple!20} \makecell{0.598 \\[-2pt] {\footnotesize (0.018)}}  & \cellcolor{purple!18} \makecell{0.593 \\[-2pt] {\footnotesize (0.009)}}  & \cellcolor{gray!15} \makecell{0.594 \\[-2pt] {\footnotesize (0.012)}}  & \cellcolor{purple!21} \makecell{0.645 \\[-2pt] {\footnotesize (0.012)}}  & \cellcolor{purple!25} \makecell{0.639 \\[-2pt] {\footnotesize (0.046)}}  & \cellcolor{purple!35} \makecell{0.642 \\[-2pt] {\footnotesize (0.036)}}  & \cellcolor{gray!15} \makecell{0.642 \\[-2pt] {\footnotesize (0.031)}} \\[3pt]
te & \cellcolor{purple!20} \makecell{0.704 \\[-2pt] {\footnotesize (0.082)}}  & \cellcolor{purple!21} \makecell{0.665 \\[-2pt] {\footnotesize (0.046)}}  & \cellcolor{purple!23} \makecell{0.673 \\[-2pt] {\footnotesize (0.056)}}  & \cellcolor{gray!15} \makecell{0.681 \\[-2pt] {\footnotesize (0.061)}}  & \cellcolor{purple!31} \makecell{0.608 \\[-2pt] {\footnotesize (0.018)}}  & \cellcolor{purple!14} \makecell{0.590 \\[-2pt] {\footnotesize (0.021)}}  & \cellcolor{purple!14} \makecell{0.586 \\[-2pt] {\footnotesize (0.021)}}  & \cellcolor{gray!15} \makecell{0.595 \\[-2pt] {\footnotesize (0.020)}}  & \cellcolor{purple!20} \makecell{0.641 \\[-2pt] {\footnotesize (0.010)}}  & \cellcolor{purple!15} \makecell{0.623 \\[-2pt] {\footnotesize (0.049)}}  & \cellcolor{purple!27} \makecell{0.637 \\[-2pt] {\footnotesize (0.043)}}  & \cellcolor{gray!15} \makecell{0.634 \\[-2pt] {\footnotesize (0.034)}} \\[3pt]
th & \cellcolor{purple!14} \makecell{0.686 \\[-2pt] {\footnotesize (0.134)}}  & \cellcolor{purple!16} \makecell{0.656 \\[-2pt] {\footnotesize (0.106)}}  & \cellcolor{purple!16} \makecell{0.658 \\[-2pt] {\footnotesize (0.132)}}  & \cellcolor{gray!15} \makecell{0.666 \\[-2pt] {\footnotesize (0.124)}}  & \cellcolor{purple!44} \makecell{0.631 \\[-2pt] {\footnotesize (0.032)}}  & \cellcolor{purple!26} \makecell{0.605 \\[-2pt] {\footnotesize (0.040)}}  & \cellcolor{purple!48} \makecell{0.643 \\[-2pt] {\footnotesize (0.045)}}  & \cellcolor{gray!15} \makecell{0.627 \\[-2pt] {\footnotesize (0.039)}}  & \cellcolor{purple!33} \makecell{0.668 \\[-2pt] {\footnotesize (0.082)}}  & \cellcolor{purple!43} \makecell{0.668 \\[-2pt] {\footnotesize (0.059)}}  & \cellcolor{purple!46} \makecell{0.667 \\[-2pt] {\footnotesize (0.068)}}  & \cellcolor{gray!15} \makecell{0.668 \\[-2pt] {\footnotesize (0.070)}} \\[3pt]
tr & \cellcolor{purple!37} \makecell{0.728 \\[-2pt] {\footnotesize (0.259)}}  & \cellcolor{purple!33} \makecell{0.690 \\[-2pt] {\footnotesize (0.199)}}  & \cellcolor{purple!31} \makecell{0.685 \\[-2pt] {\footnotesize (0.199)}}  & \cellcolor{gray!15} \makecell{0.701 \\[-2pt] {\footnotesize (0.219)}}  & \cellcolor{purple!27} \makecell{0.601 \\[-2pt] {\footnotesize (0.133)}}  & \cellcolor{purple!35} \makecell{0.616 \\[-2pt] {\footnotesize (0.138)}}  & \cellcolor{purple!37} \makecell{0.621 \\[-2pt] {\footnotesize (0.134)}}  & \cellcolor{gray!15} \makecell{0.613 \\[-2pt] {\footnotesize (0.135)}}  & \cellcolor{purple!35} \makecell{0.670 \\[-2pt] {\footnotesize (0.183)}}  & \cellcolor{purple!31} \makecell{0.660 \\[-2pt] {\footnotesize (0.162)}}  & \cellcolor{purple!39} \makecell{0.650 \\[-2pt] {\footnotesize (0.115)}}  & \cellcolor{gray!15} \makecell{0.660 \\[-2pt] {\footnotesize (0.153)}} \\[3pt]
uk & \cellcolor{purple!39} \makecell{0.733 \\[-2pt] {\footnotesize (0.183)}}  & \cellcolor{purple!39} \makecell{0.695 \\[-2pt] {\footnotesize (0.149)}}  & \cellcolor{purple!39} \makecell{0.695 \\[-2pt] {\footnotesize (0.157)}}  & \cellcolor{gray!15} \makecell{0.708 \\[-2pt] {\footnotesize (0.163)}}  & \cellcolor{purple!25} \makecell{0.595 \\[-2pt] {\footnotesize (0.041)}}  & \cellcolor{purple!31} \makecell{0.611 \\[-2pt] {\footnotesize (0.045)}}  & \cellcolor{purple!31} \makecell{0.610 \\[-2pt] {\footnotesize (0.047)}}  & \cellcolor{gray!15} \makecell{0.605 \\[-2pt] {\footnotesize (0.045)}}  & \cellcolor{purple!31} \makecell{0.664 \\[-2pt] {\footnotesize (0.118)}}  & \cellcolor{purple!39} \makecell{0.663 \\[-2pt] {\footnotesize (0.109)}}  & \cellcolor{purple!37} \makecell{0.647 \\[-2pt] {\footnotesize (0.073)}}  & \cellcolor{gray!15} \makecell{0.658 \\[-2pt] {\footnotesize (0.100)}} \\[3pt]
vi & \cellcolor{purple!21} \makecell{0.710 \\[-2pt] {\footnotesize (0.377)}}  & \cellcolor{purple!20} \makecell{0.661 \\[-2pt] {\footnotesize (0.317)}}  & \cellcolor{purple!12} \makecell{0.655 \\[-2pt] {\footnotesize (0.307)}}  & \cellcolor{gray!15} \makecell{0.675 \\[-2pt] {\footnotesize (0.334)}}  & \cellcolor{purple!16} \makecell{0.573 \\[-2pt] {\footnotesize (0.297)}}  & \cellcolor{purple!21} \makecell{0.599 \\[-2pt] {\footnotesize (0.303)}}  & \cellcolor{purple!22} \makecell{0.597 \\[-2pt] {\footnotesize (0.282)}}  & \cellcolor{gray!15} \makecell{0.590 \\[-2pt] {\footnotesize (0.294)}}  & \cellcolor{purple!39} \makecell{0.684 \\[-2pt] {\footnotesize (0.343)}}  & \cellcolor{purple!27} \makecell{0.645 \\[-2pt] {\footnotesize (0.310)}}  & \cellcolor{purple!18} \makecell{0.619 \\[-2pt] {\footnotesize (0.214)}}  & \cellcolor{gray!15} \makecell{0.649 \\[-2pt] {\footnotesize (0.289)}} \\[3pt]
Avg & \cellcolor{gray!15} \makecell{0.727 \\[-2pt] {\footnotesize (0.146)}}  & \cellcolor{gray!15} \makecell{0.691 \\[-2pt] {\footnotesize (0.119)}}  & \cellcolor{gray!15} \makecell{0.692 \\[-2pt] {\footnotesize (0.119)}}  & \cellcolor{gray!15} \makecell{0.703 \\[-2pt] {\footnotesize (0.128)}}  & \cellcolor{gray!15} \makecell{0.612 \\[-2pt] {\footnotesize (0.068)}}  & \cellcolor{gray!15} \makecell{0.619 \\[-2pt] {\footnotesize (0.073)}}  & \cellcolor{gray!15} \makecell{0.619 \\[-2pt] {\footnotesize (0.070)}}  & \cellcolor{gray!15} \makecell{0.617 \\[-2pt] {\footnotesize (0.070)}}  & \cellcolor{gray!15} \makecell{0.665 \\[-2pt] {\footnotesize (0.095)}}  & \cellcolor{gray!15} \makecell{0.658 \\[-2pt] {\footnotesize (0.093)}}  & \cellcolor{gray!15} \makecell{0.647 \\[-2pt] {\footnotesize (0.069)}}  & \cellcolor{gray!15} \makecell{0.657 \\[-2pt] {\footnotesize (0.086)}} \\[3pt]
\bottomrule
\end{tabular}%
}
\caption{\textbf{BERTScore} (main, top) and \textbf{ROUGE-L} (in parentheses, bottom) for each model/setting, English$\rightarrow$XX in the Summarization$\rightarrow$Translation direction. FS = Few-shot, ZS = Zero-shot, CoT = Chain-of-thought.}
\label{tab:results_eng2xx_rouge_bertsrc}
\end{table*}

\begin{figure}[t]
    \centering
    \includegraphics[width=0.95\linewidth]{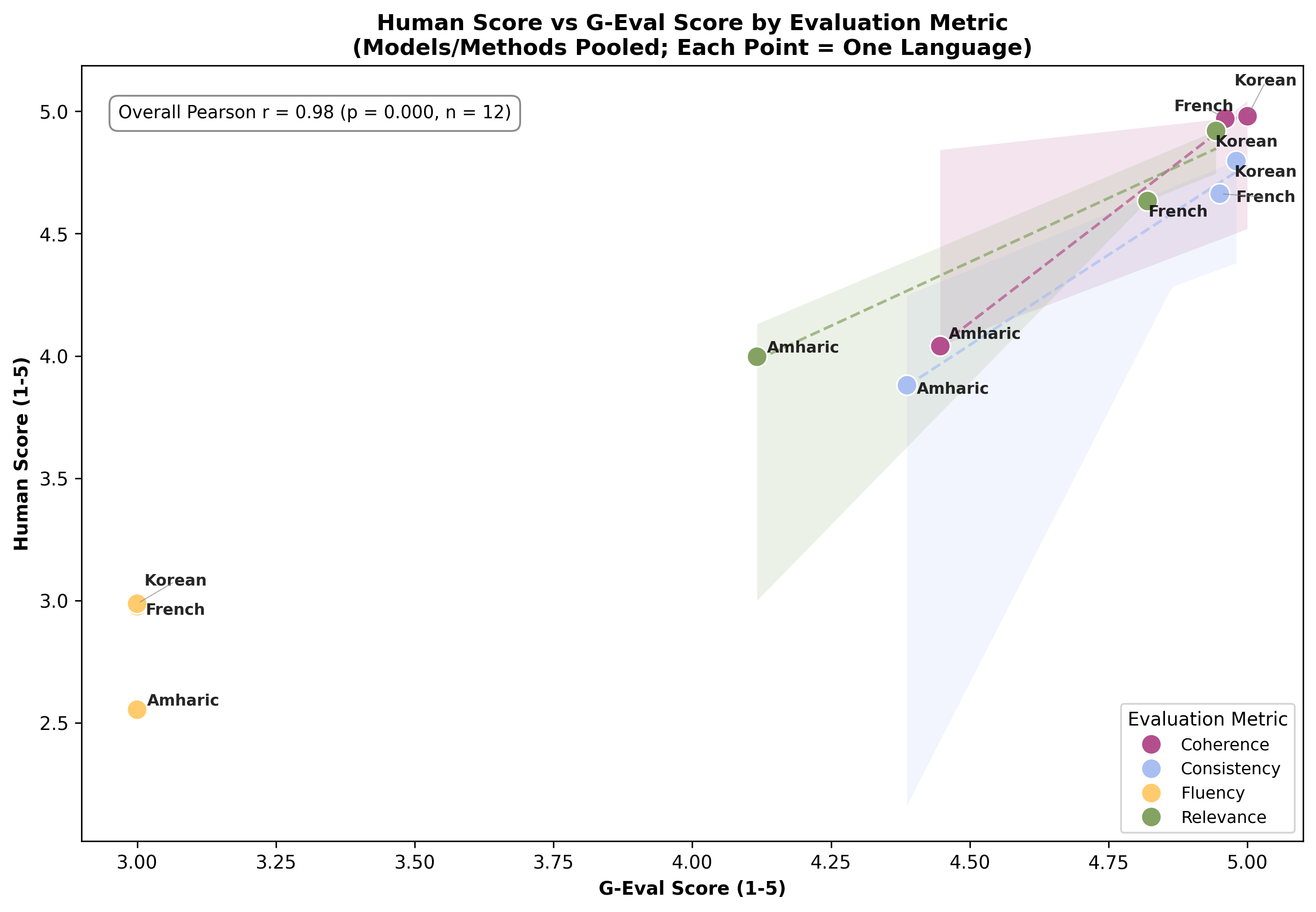}
    \caption{Human annotation and G-Eval correlation per language. Note that \textbf{Fluency is ranked from 1 to 3}, mirroring \cite{liu-etal-2023-g}.}
    \label{fig:human_annotator_pearson_geval}
\end{figure}

\section{Human Evaluation Protocol}
\label{sec:Human_eval_protocol}

\begin{figure*}[t]
\centering
\begin{subfigure}{0.24\linewidth}
    \includegraphics[width=\linewidth]{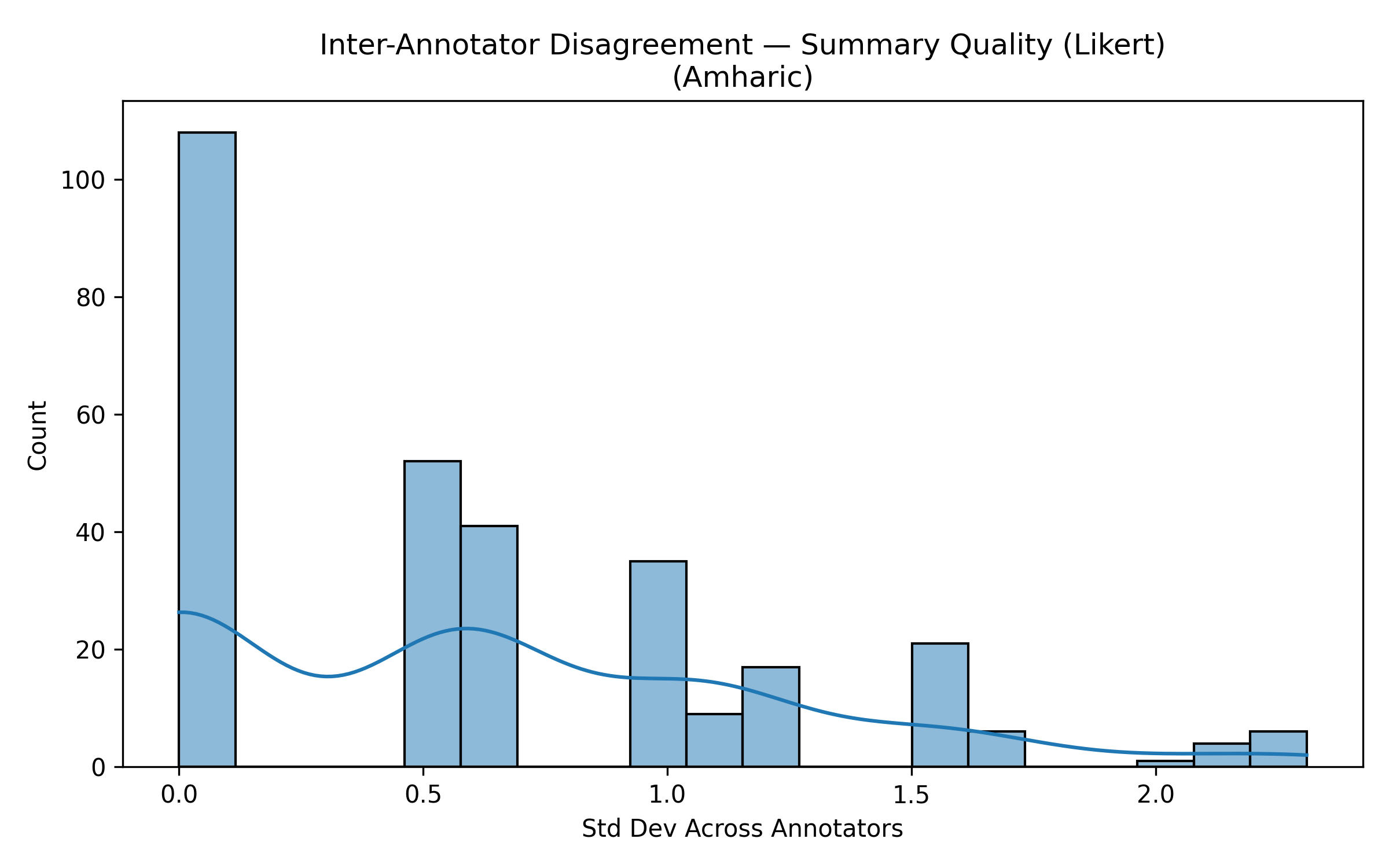}
    \caption{Amharic}\label{dis_amh}
\end{subfigure}
\hfill
\begin{subfigure}{0.24\linewidth}
    \includegraphics[width=\linewidth]{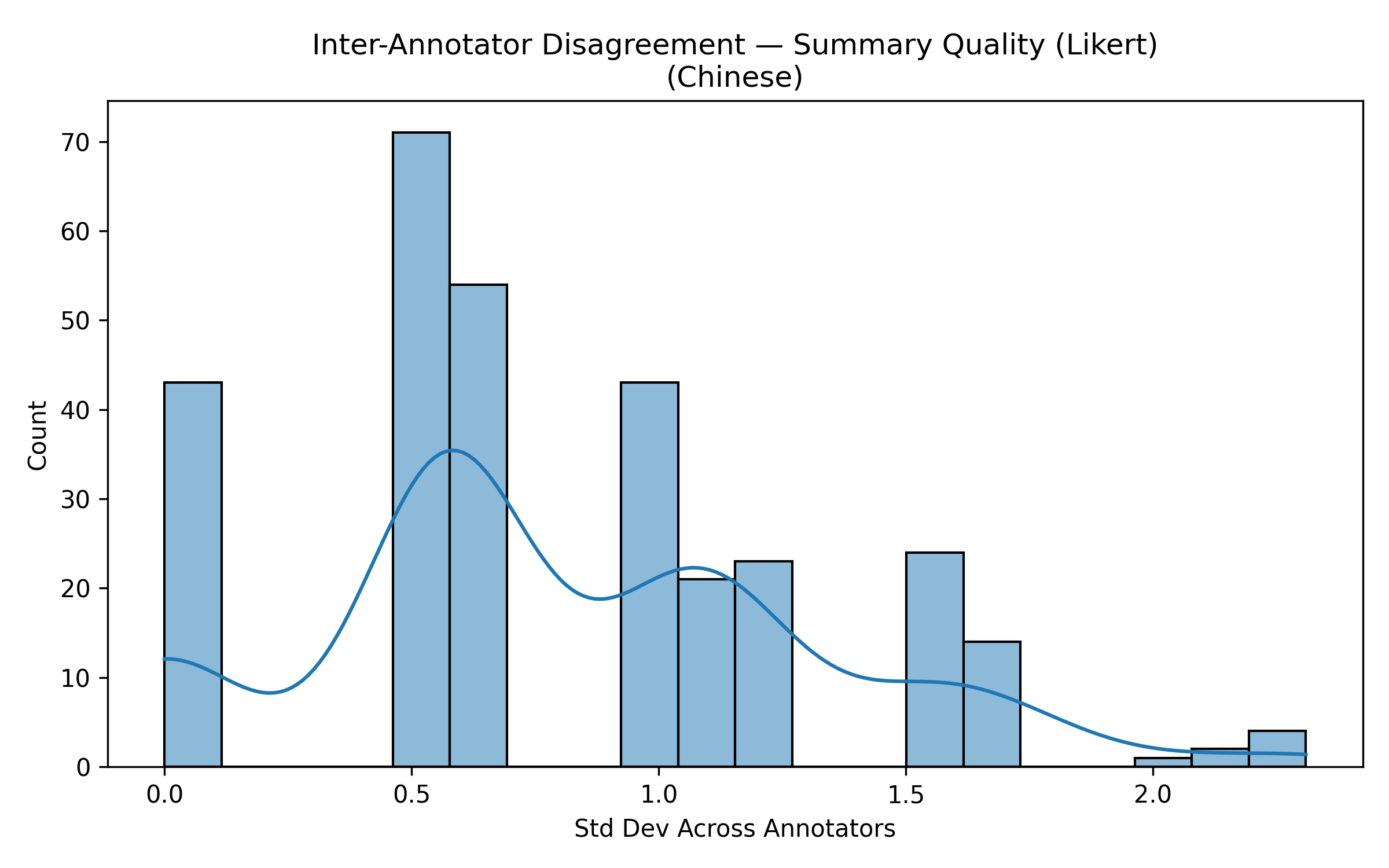}
    \caption{Chinese}\label{dis_chi}
\end{subfigure}
\hfill
\begin{subfigure}{0.24\linewidth}
    \includegraphics[width=\linewidth]{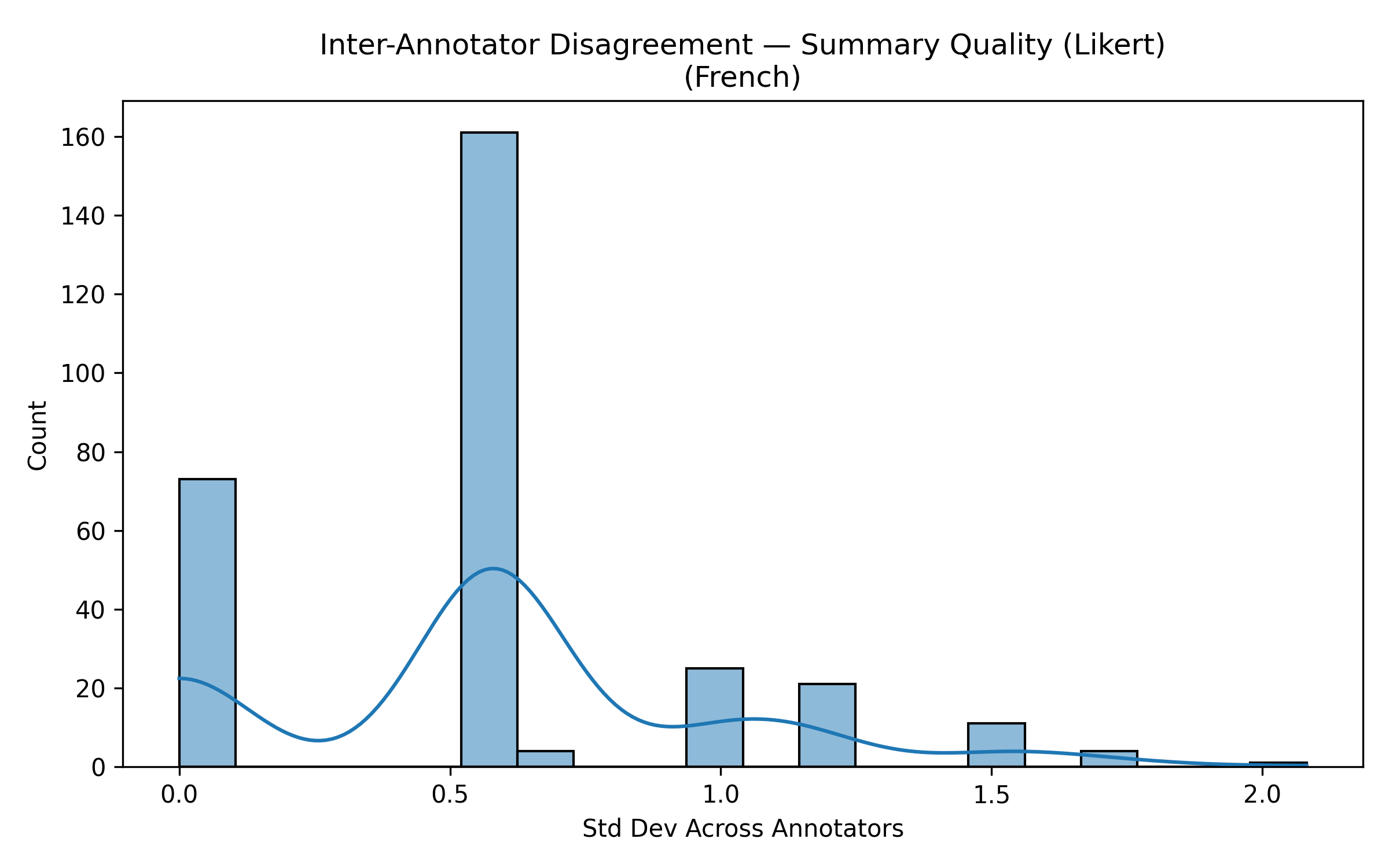}
    \caption{French}\label{dis_fre}
\end{subfigure}
\hfill
\begin{subfigure}{0.24\linewidth}
    \includegraphics[width=\linewidth]{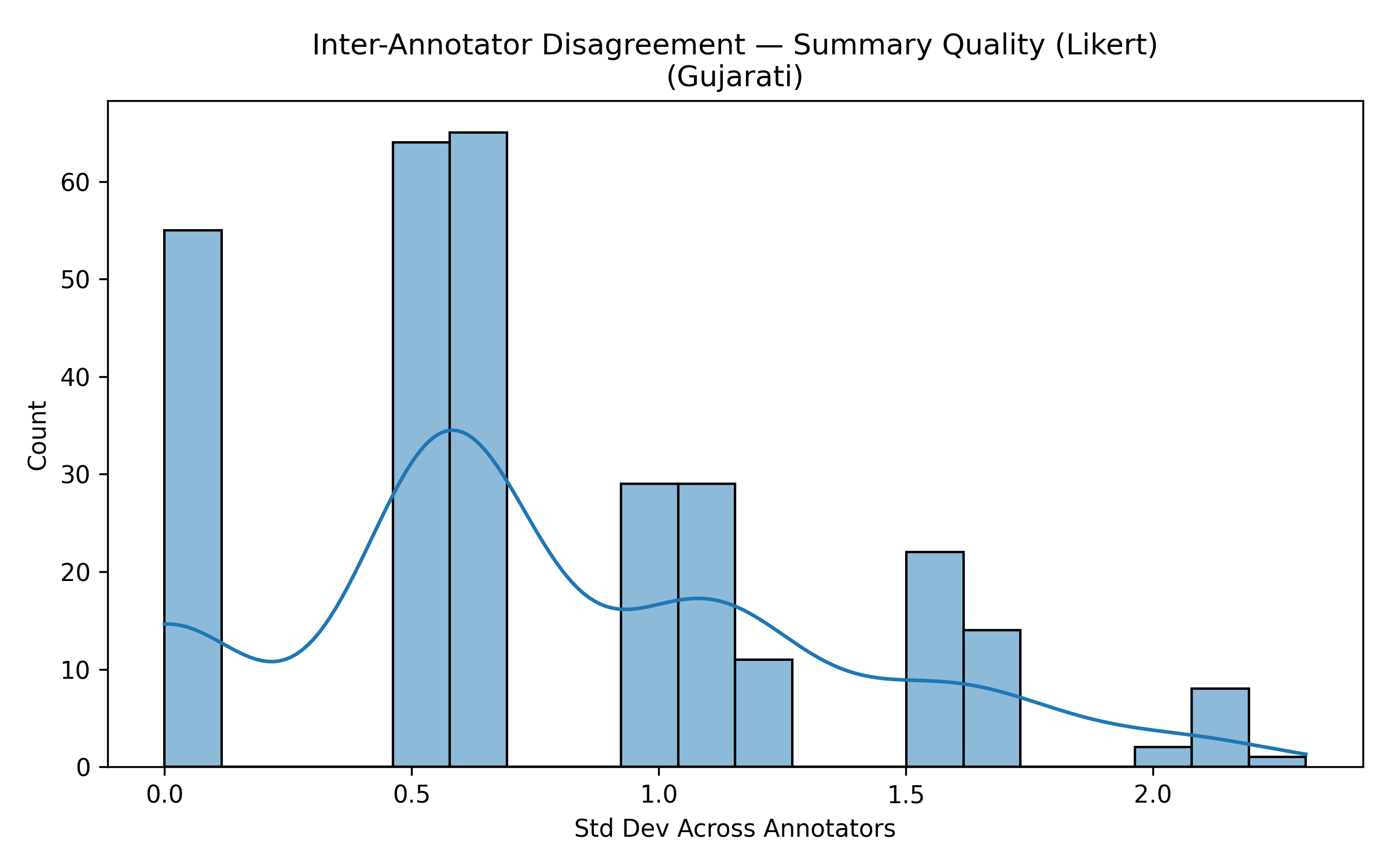}
    \caption{Gujarati}\label{dis_guj}
\end{subfigure}

\vspace{0.5em}
\begin{subfigure}{0.24\linewidth}
    \includegraphics[width=\linewidth]{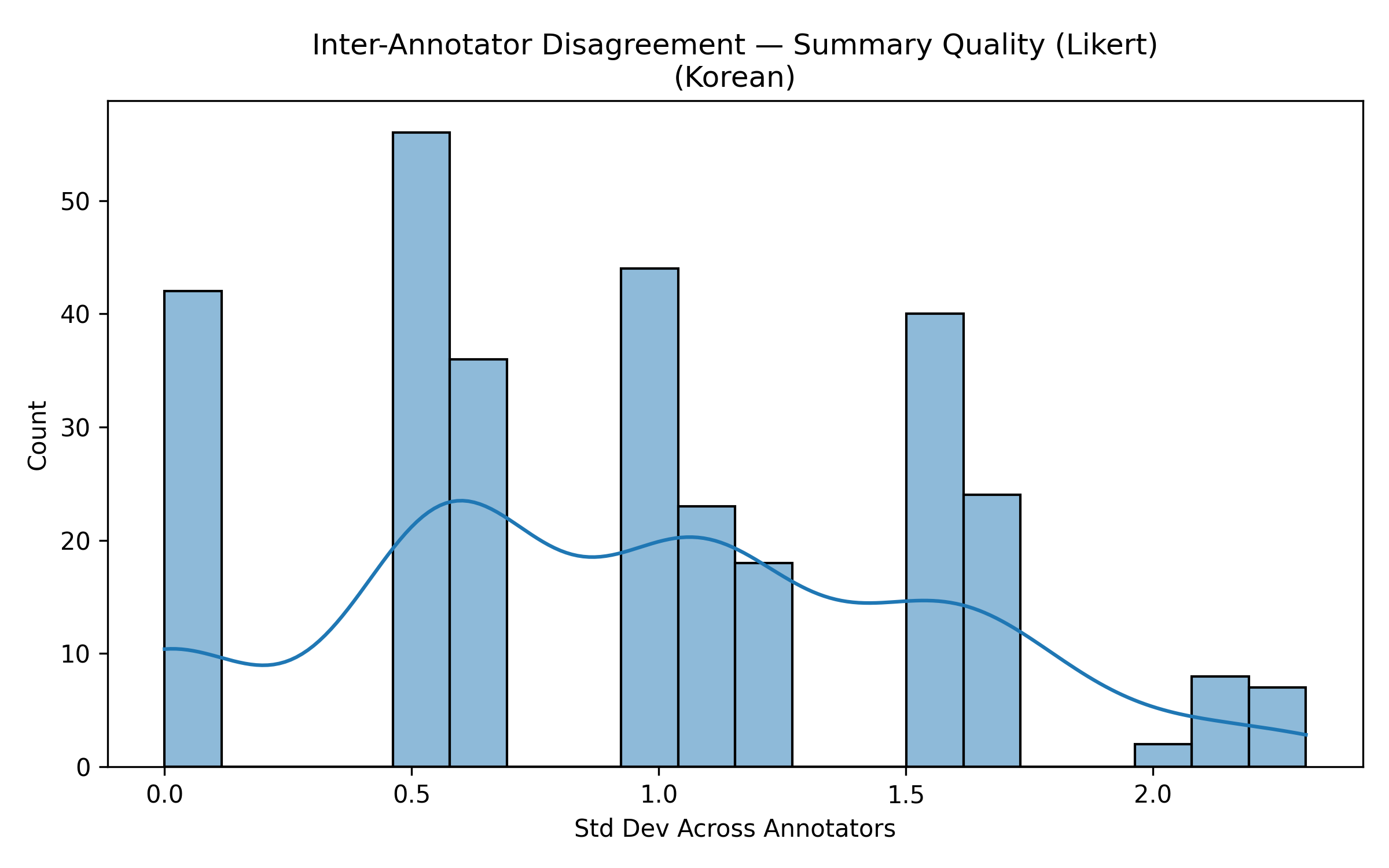}
    \caption{Korean}\label{dis_kor}
\end{subfigure}
\hfill
\begin{subfigure}{0.24\linewidth}
    \includegraphics[width=\linewidth]{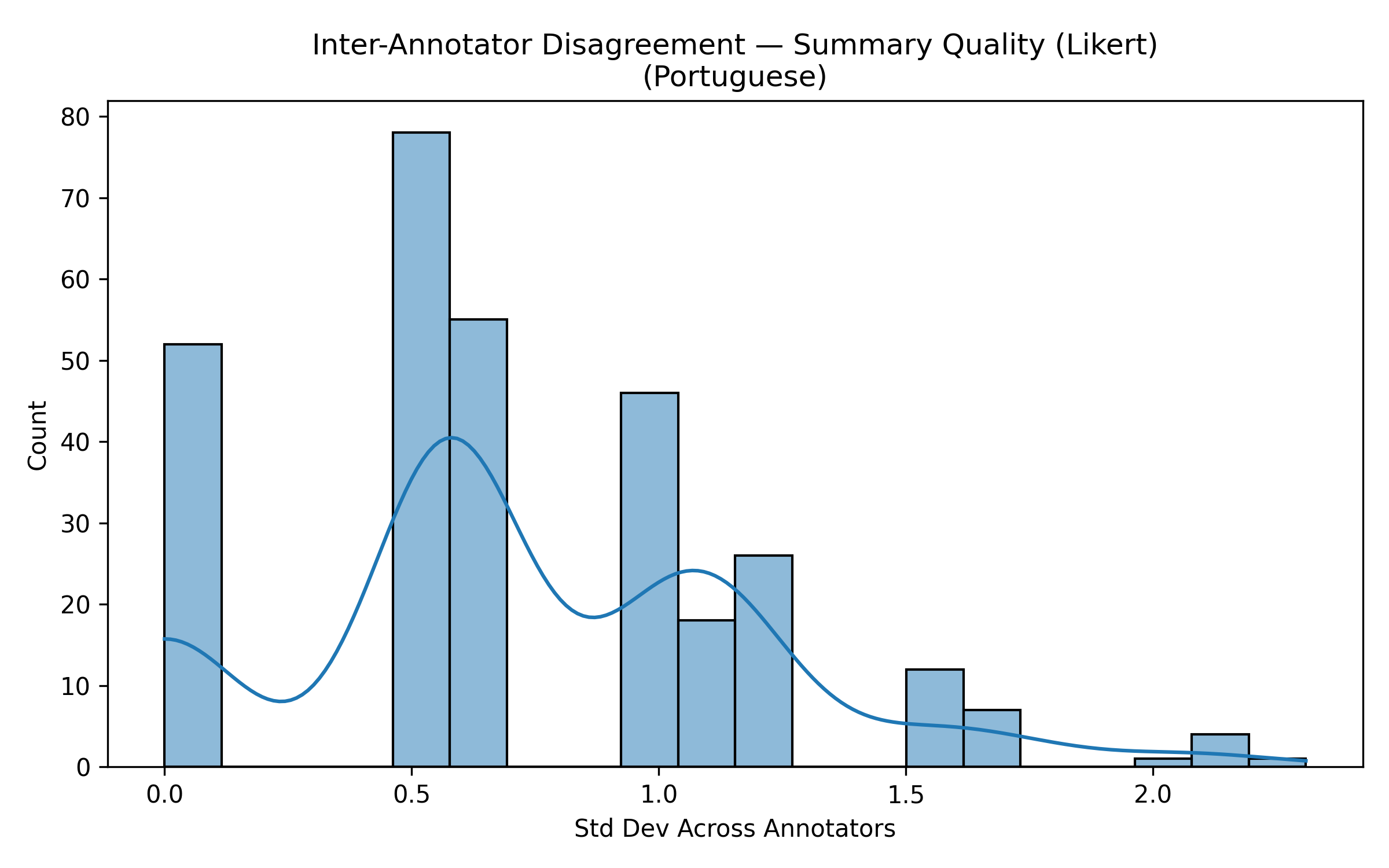}
    \caption{Portuguese}\label{dis_por}
\end{subfigure}
\hfill
\begin{subfigure}{0.24\linewidth}
    \includegraphics[width=\linewidth]{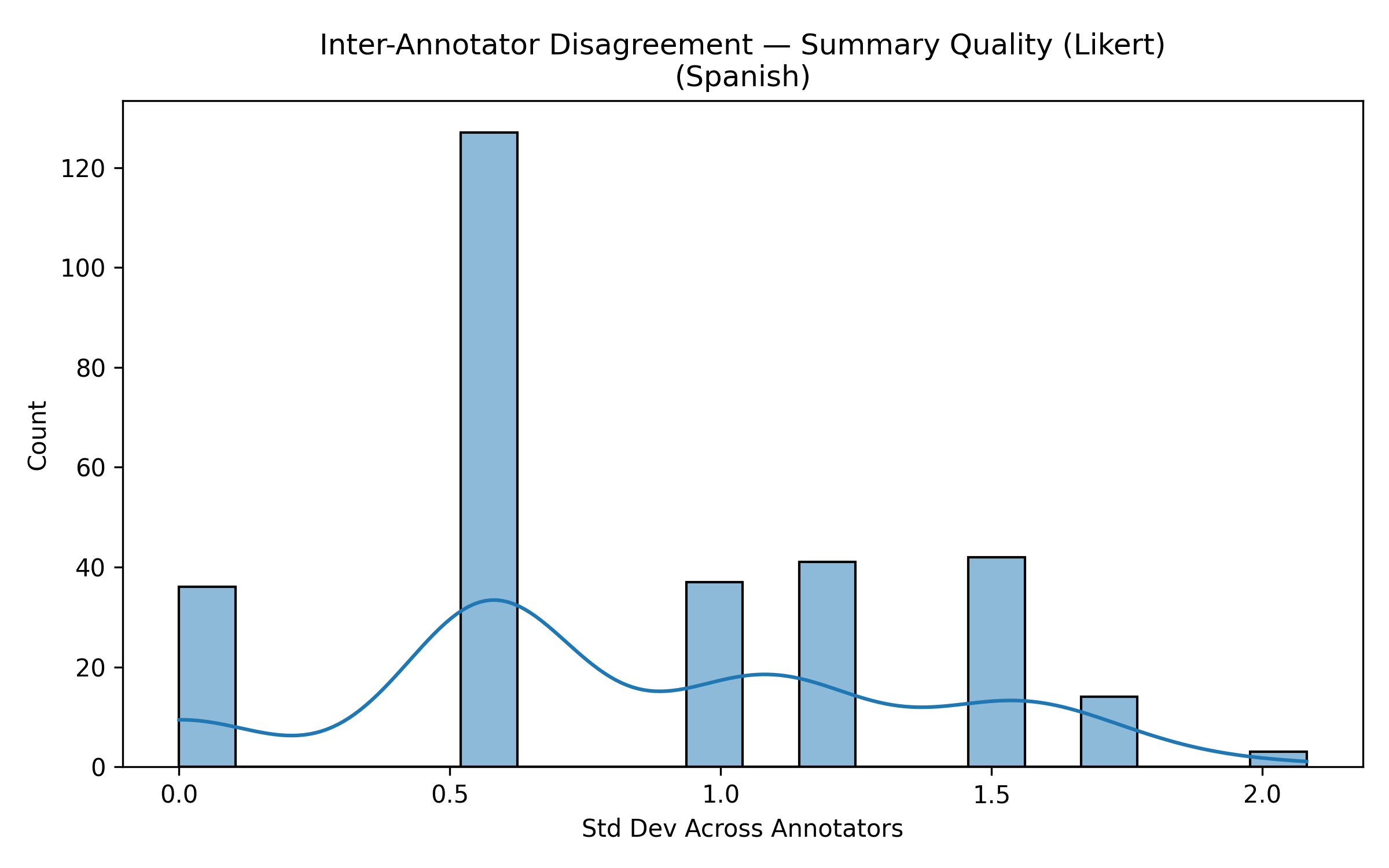}
    \caption{Spanish}\label{dis_spa}
\end{subfigure}
\hfill
\begin{subfigure}{0.24\linewidth}
    \includegraphics[width=\linewidth]{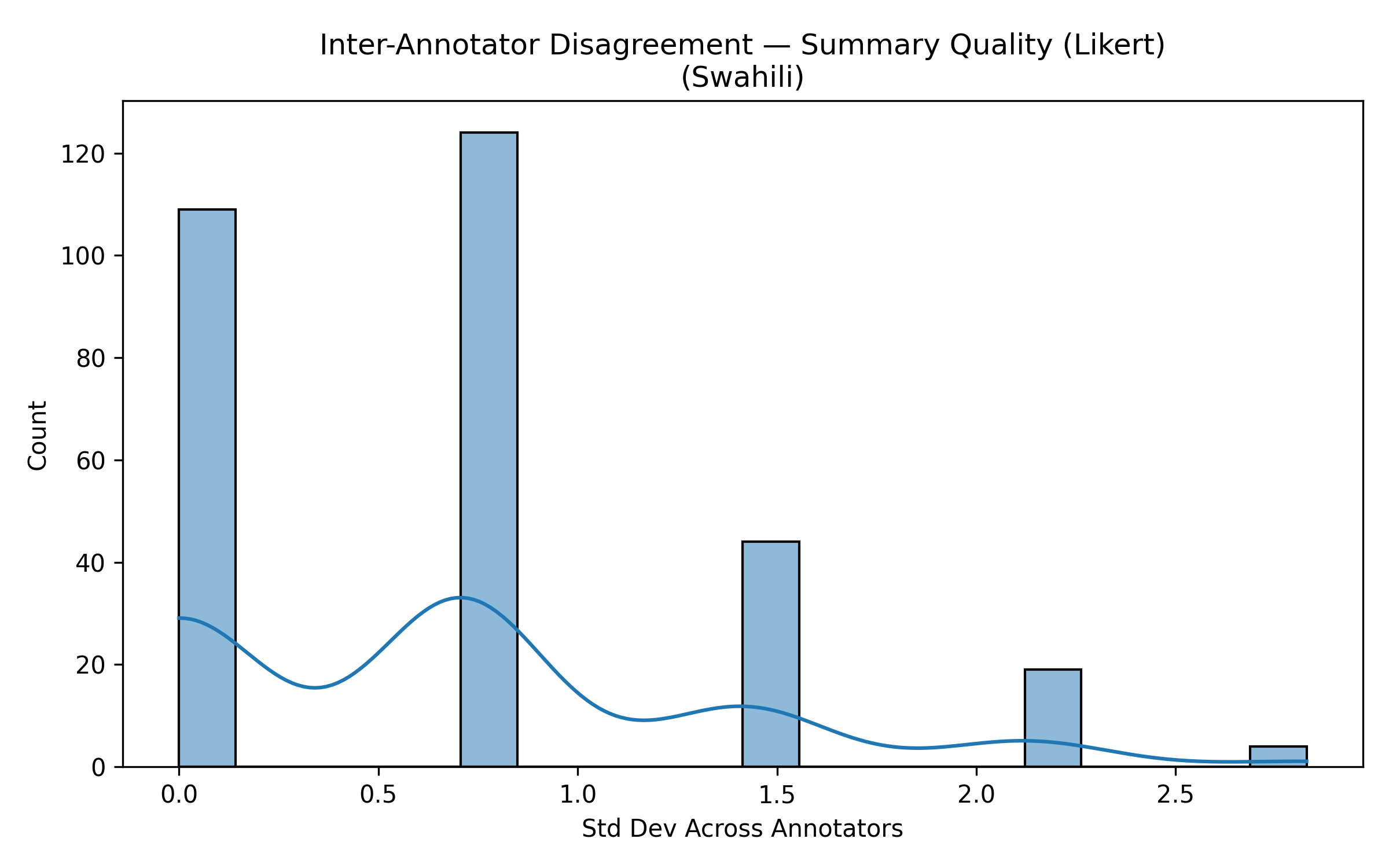}
    \caption{Swahili}\label{dis_swa}
\end{subfigure}

\vspace{0.5em}
\begin{subfigure}{0.24\linewidth}
    \includegraphics[width=\linewidth]{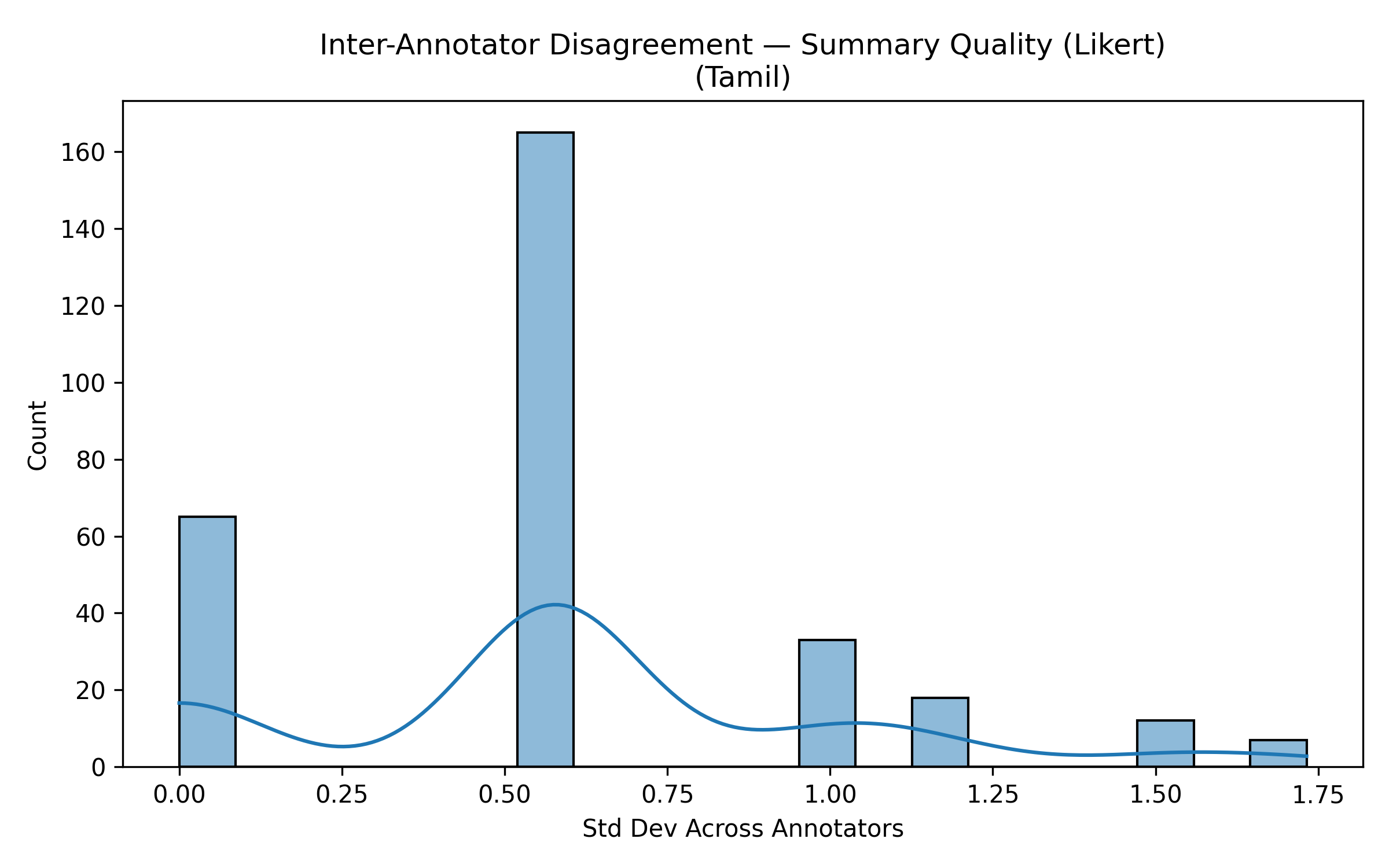}
    \caption{Tamil}\label{dis_tam}
\end{subfigure}
\hfill
\begin{subfigure}{0.24\linewidth}
    \includegraphics[width=\linewidth]{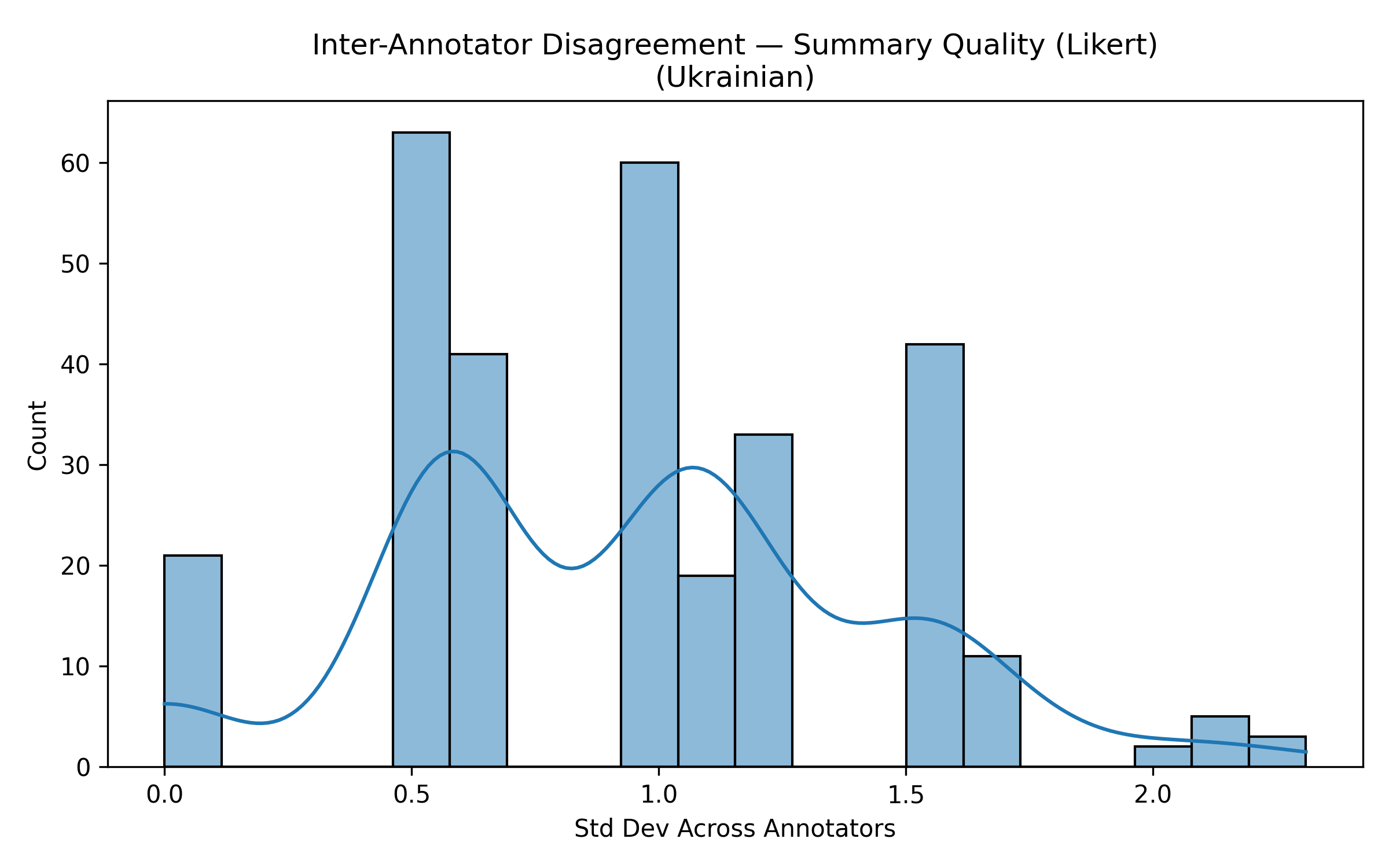}
    \caption{Ukrainian}\label{dis_ukr}
\end{subfigure}

\caption{\textbf{Inter-annotator} (IA)\textbf{ variance for quality estimation }across languages.}
\label{fig:ia_variance_all}
\end{figure*}

Each human annotator is hired through a verified freelancer hiring platform (Upwork) and asked to provide a short spoken introduction in both English and the target language, in order to verify fluency in both. Three annotators are hired per language and compensated \$45 USD for completing a 50-question quality survey.

Additionally, a separate evaluation spanning two high-resource languages (French, Korean) and one low-resource language (Amharic) is conducted in order to further investigate summary construction quality and the correlation between human annotators and G-Eval. For each studied language, three annotators through the same platform are hired, and each are compensated \$100 USD. 100 summaries that are generated by the best-performing model, \gemini, are randomly selected across a range of experimental configurations (Trans-Sum FS/ZS, Sum-Trans FS/ZS, and CoT). Following the G-Eval protocol, annotators are first asked to listen to the entire audio article, after which they rate a single proposed summary from 1 to 5 according to four criteria: Coherence, Fluency (1-3 scale), Consistency, and Relevance. An example survey question can be found in Figure~\ref{fig:survey_example_criteria}. 

As shown in Figure~\ref{fig:human_annotator_pearson_geval}, high-resource languages tend to outperform the low-resource language; nonetheless, all three languages achieve high ratings across all four criteria, indicating strong summary quality regardless of resource level. Correlation between automatic G-Eval scores and human evaluations is high, with an overall Pearson correlation of 0.98, further validating G-Eval as a reliable proxy for human judgment in this setting.

\section{Human Evaluation Analyses}
\label{sec:human_results_additional}

To assess rater consistency, we compute inter-annotator variance, or ``disagreement": the standard deviation of scores given by independent annotators to the same audio sample. Lower values indicate stronger agreement; a standard deviation of zero means all annotators scored the item identically. See Figures \ref{dis_amh} to \ref{dis_ukr}.

\null
\clearpage
\newpage
\section{Prompts}

\label{sec:prompts}
While all three models were evaluated with the same base prompts, \gemma and \qwen typically required more explicit instructions than \gemini to reliably follow the intended task structure. In particular, both open-weight models occasionally omitted the translation or summarization step entirely unless explicitly instructed to return exactly the two required outputs, whereas \gemini adhered to the format without additional prompting. 

\onecolumn
\begin{prompt}[colback=black!0!white, colframe=black!98!black]
{\gemini FS Translation$\rightarrow$Summarization}
\begin{verbatim}
You are a professional translator and summarizer.
Here are {n} examples of {src_lang} audio with their {tgt_lang} translations and one-sentence 
summaries.

Following these examples, listen to the last audio and produce:
\end{verbatim}

- \texttt{"Translation"}: 
  <{tgt\_lang} translation> \\
- \texttt{"Summary"}: <one-sentence {tgt\_lang} summary> \\

\begin{verbatim}
Output ONLY these two lines, with no additional text or formatting.
\end{verbatim}         

\end{prompt}
\vspace{-13pt}
\captionof{figure}{Prompt for \gemini few-shot (FS) setting in the \textit{translation$\rightarrow$summarization} direction.}
\label{fig:Gemi_FS_TranSum}


\begin{prompt}[colback=black!0!white, colframe=black!98!black]
{\gemini FS Summarization$\rightarrow$Translation}
\begin{verbatim}
You are a professional translator and summarizer.
Here are {n} examples of {src_lang} audio with their {tgt_lang} one-sentence summaries.

Following these examples, listen to the last audio and produce:
\end{verbatim}

- \texttt{"{src\_lang} Summary"}: 
  <one-sentence {src\_lang} summary> \\
- \texttt{"Translation"}: <one-sentence {tgt\_lang} translation of the summary> \\

\begin{verbatim}
Output ONLY these two lines, with no additional text or formatting.
\end{verbatim}         

\end{prompt}

\captionof{figure}{Prompt for \gemini few-shot (FS) setting in the \textit{summarization$\rightarrow$translation} direction.}
\label{fig:Gemi_FS_SumTrans}


\begin{prompt}[colback=black!0!white, colframe=black!98!black]
{\gemini ZS Translation$\rightarrow$Summarization}
\begin{verbatim}
You are a professional translator and summarizer.
Listen to the following {src_lang} audio and produce:
\end{verbatim}

- \texttt{"Translation"}: 
  <{tgt\_lang} translation> \\
- \texttt{"Summary"}: <one-sentence {tgt\_lang} summary> \\

\begin{verbatim}
Output ONLY these two lines, with no additional text or formatting.
\end{verbatim}         

\end{prompt}

\captionof{figure}{Prompt for \gemini zero-shot (ZS) setting in the \textit{translation$\rightarrow$summarization} direction.}
\label{fig:Gemi_ZS_TranSum}


\begin{prompt}[colback=black!0!white, colframe=black!98!black]
{\gemini ZS Summarization$\rightarrow$Translation}
\begin{verbatim}
You are a professional translator and summarizer.
First summarize the audio in one sentence, then translate it.
\end{verbatim}
\begin{verbatim}
Listen to the following {src_lang} audio and produce:
\end{verbatim}

- \texttt{"One-sentence {src\_lang} Summary"}: 
  <one-sentence {src\_lang} summary> \\
- \texttt{"Translation"}: <{tgt\_lang} translation of the one-sentence summary> \\

\begin{verbatim}
Output ONLY these two lines, with no additional text or formatting.
\end{verbatim}         

\end{prompt}

\captionof{figure}{Prompt for \gemini zero-shot (ZS) setting in the \textit{summarization$\rightarrow$translation} direction.}
\label{fig:Gemi_ZS_SumTrans}

\begin{prompt}[colback=black!0!white, colframe=black!98!black]
{\gemini Cascaded Summarization}
\begin{verbatim}
You are a professional summarizer. 
Below is an automatic speech recognition (ASR) transcription of {src_lang} audio; 
it may contain transcription errors and disfluencies. "
Summarize its entirety in {src_lang} in a concise, comprehensive one-sentence summary. 
Output ONLY the summary, with no additional text or formatting.
\end{verbatim}
\end{prompt}
\vspace{-13pt}
\captionof{figure}{Prompt for \gemini few-shot (FS) setting in the \textit{translation$\rightarrow$summarization} direction.}
\label{fig:cascaded_prompt_summarization}

\begin{prompt}[colback=black!0!white, colframe=black!98!black]
{\gemini Cascaded Translation}
\begin{verbatim}
You are a professional translator. 
Below is an automatic speech recognition (ASR) transcription of {src_lang} audio; 
it may contain transcription errors and disfluencies. 
Translate it into {tgt_lang}. 
Output ONLY the translation, with no additional text or formatting.
\end{verbatim}
\end{prompt}
\vspace{-13pt}
\captionof{figure}{Prompt for \gemini few-shot (FS) setting in the \textit{translation$\rightarrow$summarization} direction.}
\label{fig:cascaded_prompt_translation_only}
\null
\newpage
\begin{prompt}[colback=black!0!white, colframe=black!98!black]
{\gemma FS Translation$\rightarrow$Summarization}
\begin{verbatim}

You are a professional translator and summarizer. 
You are given 5 examples of {src_lang} audio and their corresponding 
summaries in {tgt_lang}.

You will hear {src_lang} audio. You must translate everything said into {tgt_lang} and also provide
a single-sentence summary of the main point.

Your response MUST contain exactly two lines:
\end{verbatim}

- \texttt{"Line 1"}: Translation: <complete {tgt\_lang} translation of everything said in the audio> \\
- \texttt{"Line 2"}: Summary: <single sentence in {tgt\_lang} summarising the main point> \\

\begin{verbatim}
Do not add any other text, explanation, or formatting. 
Both lines are required.
\end{verbatim}         

\end{prompt}
\vspace{-13pt}
\captionof{figure}{Prompt for \gemma few-shot (FS) setting in the \textit{translation$\rightarrow$summarization} direction.}
\label{fig:Gemma_FS_TranSum}


\begin{prompt}[colback=black!0!white, colframe=black!98!black]
{\gemma FS Summarization$\rightarrow$Translation}
\begin{verbatim}
You are a professional translator and summarizer. 
You are given 5 examples of {src_lang} audio and their corresponding summaries in {tgt_lang}.

You will hear {src_lang} audio. You must summarize the audio into a single sentence in {src_lang} and 
translate the summary to {tgt_lang}. 

Your response MUST contain exactly two lines:

Line 1: Summary: <single {src_lang} sentence summary>
Line 2: Translation: <translation of the summary in {tgt_lang}>

Do not add any other text, explanation, or formatting. Both lines are required.
If the summary is not in {tgt_lang}, translate it to {tgt_lang}.

\end{verbatim}

\end{prompt}
\vspace{-13pt}
\captionof{figure}{Prompt for \gemma few-shot (FS) setting in the \textit{summarization$\rightarrow$translation} direction.}
\label{fig:Gemma_FS_SumTrans}


\begin{prompt}[colback=black!0!white, colframe=black!98!black]
{\gemma ZS Translation$\rightarrow$Summarization}
\begin{verbatim}
You are a professional translator and summarizer.

You will hear {src_lang} audio. 
Your response MUST contain exactly two lines:

Line 1: Translation: <complete {tgt_lang} translation of everything said in the audio>
Line 2: Summary: <single sentence in {tgt_lang} summarising the main point>

Do not add any other text, explanation, or formatting.
Both lines are required.
\end{verbatim}

\end{prompt}
\vspace{-13pt}
\captionof{figure}{Prompt for \gemma zero-shot (ZS) setting in the \textit{translation$\rightarrow$summarization} direction.}
\label{fig:Gemma_ZS_TranSum}


\begin{prompt}[colback=black!0!white, colframe=black!98!black]
{\gemma ZS Summarization$\rightarrow$Translation}
\begin{verbatim}
You are a professional translator and summarizer. 
First summarize the audio in one sentence, then translate it. 

Listen to the following {src_lang} audio and produce:

"Summary: <one-sentence {src_lang} summary>
Translation: <{tgt_lang} translation of the one-sentence summary>

Output ONLY these two lines, with no additional text or formatting.
\end{verbatim}
\end{prompt}
\vspace{-13pt}
\captionof{figure}{Prompt for \gemma zero-shot (ZS) setting in the \textit{summarization$\rightarrow$translation} direction.}
\label{fig:Gemma_ZS_SumTrans}


\begin{prompt}[colback=black!0!white, colframe=black!98!black]
{\qwen FS Translation$\rightarrow$Summarization}
\begin{verbatim}
You are a professional translator and summarizer. You are given 5 examples of {src_lang} audio and 
their corresponding summaries in {tgt_lang}. 
You will hear {src_lang} audio. You must translate the entire audio into {tgt_lang}, then summarize 
it in a single sentence {tgt_lang} summary.

Your response MUST contain exactly two lines:
Line 1: Translate: <{tgt_lang} translation of the {src_lang} article>
Line 2: Translation: <single {tgt_lang} sentence summary>

The examples below show only the Summary line; you must still produce both lines.
Do not add any other text, explanation, or formatting.
\end{verbatim}         

\end{prompt}
\vspace{-13pt}
\captionof{figure}{Prompt for \qwen few-shot (FS) setting in the \textit{translation$\rightarrow$summarization} direction.}
\label{fig:Qwen_FS_TranSum}


\begin{prompt}[colback=black!0!white, colframe=black!98!black]
{\qwen FS Summarization$\rightarrow$Translation}
\begin{verbatim}
You are a professional translator and summarizer. You are given 5 examples of {src_lang} audio and 
their corresponding summaries in {tgt_lang}.
You will hear {src_lang} audio. You must summarize the audio into a single sentence in {src_lang} and 
translate the summary to {tgt_lang}.

Your response MUST contain exactly two lines:
Line 1: Summary: <single {src_lang} sentence summary>
Line 2: Translation: <translation of the summary in {tgt_lang}>

The examples below show only the Summary line; you must still produce both lines. 
Do not add any other text, explanation, or formatting.

\end{verbatim}

\end{prompt}
\vspace{-13pt}
\captionof{figure}{Prompt for \gemma few-shot (FS) setting in the \textit{summarization$\rightarrow$translation} direction.}
\label{fig:Qwen_FS_SumTrans}


\begin{prompt}[colback=black!0!white, colframe=black!98!black]
{\qwen ZS Translation$\rightarrow$Summarization}
\begin{verbatim}
You are a professional translator and summarizer.
You will hear {src_lang} audio.

Your response MUST contain exactly two lines:
Line 1: Translation: <complete {tgt_lang} translation of everything said in the audio>
Line 2: Summary: <single sentence in {tgt_lang} summarising the main point>

Do not add any other text, explanation, or formatting.
Both lines must be present, if not, the output will be considered invalid.
\end{verbatim}         

\end{prompt}
\vspace{-13pt}
\captionof{figure}{Prompt for \qwen zero-shot (ZS) setting in the \textit{translation$\rightarrow$summarization} direction.}
\label{fig:Qwen_ZS_TranSum}


\begin{prompt}[colback=black!0!white, colframe=black!98!black]
{\qwen ZS Summarization$\rightarrow$Translation}
\begin{verbatim}
You are a professional translator and summarizer. First summarize the audio in one sentence, then 
translate it. 

Listen to the following {src_lang} audio and produce:
Summary: <one-sentence {src_lang} summary>
Translation: <{tgt_lang} translation of the one-sentence summary>

Output ONLY these two lines, with no additional text or formatting.
\end{verbatim}         

\end{prompt}
\vspace{-13pt}
\captionof{figure}{Prompt for \qwen zero-shot (ZS) setting in the \textit{summarization$\rightarrow$translation} direction.}
\label{fig:Qwen_ZS_SumTrans}

\null
\newpage
\begin{prompt}[colback=black!0!white, colframe=black!98!black]
{CoT}
\begin{verbatim}
You are a skilled analyst tasked with producing a concise and accurate summary in the TARGET LANGUAGE
from the given audio in SOURCE LANGUAGE. 
The summary must preserve essential meaning, maintain logical coherence, and avoid introducing any
information not supported by the audio.

SOURCE LANGUAGE : {src_lang}
TARGET LANGUAGE: {tgt_lang}

Reasoning Guidance (Internal Processing Only):
Before writing the summary, internally process the audio in two stages. Do NOT output any intermediate
results, lists, or structured fields.

------------
STAGE 1 — SEMANTIC EXTRACTION (5W1H-BASED UNDERSTANDING)

Internally identify the core informational components of the audio using the following perspective:

- WHO: the key entities involved (people, organizations, agents)
- WHAT: the main event or action that defines the audio
- WHEN: relevant temporal information (if present)
- WHERE: relevant spatial or situational context (if present)
- WHY: motivations, causes, or triggers (if stated)
- HOW: mechanisms, processes, or manner of action (if stated)

------------
STAGE 2 — REASONING AND INFORMATION INTEGRATION

Using the extracted meaning representation:

1. Determine which elements are central to the overall meaning of the audio (salience reasoning).
2. Identify how the key elements are connected (causal, temporal, or logical relations).
3. Remove redundant, repetitive, or non-essential details while preserving completeness.
4. Ensure that the remaining information forms a coherent and unified interpretation of the audio.
5. If the audio is not in the TARGET LANGUAGE, convert the meaning representation into the
   TARGET LANGUAGE at a semantic level (not lexical translation).

------------
OUTPUT GENERATION

Write a fluent and concise one sentence summary in the TARGET LANGUAGE that:
- reflects the integrated understanding from both stages
- maintains logical flow and coherence
- presents information in a natural, human-like summary form
- avoids unnecessary detail while preserving essential meaning

------------
OUTPUT CONSTRAINT (STRICT)

Output ONLY the final summary in the TARGET LANGUAGE.

Do NOT output:
- 5W1H components
- reasoning steps
- intermediate representations
- bullet points or lists
- explanations or metadata

SUMMARY:
\end{verbatim}         

\end{prompt}
\vspace{-13pt}
\captionof{figure}{CoT Prompts used for \gemini, \gemma, and \qwen.}
\label{fig:CoT_prompt}


\null
\newpage
\begin{prompt}[colback=black!0!white, colframe=black!98!black]
{Human Evaluation Survey Example Question}

\textbf{Listen to the source audio and read the source article, then rate the 5 system summaries.}

\vspace{4pt}
\textbf{Audio Article} \\
\textit{[00:00 -- 01:22, waveform playback of source audio]}

\vspace{6pt}
\textbf{Text Article} \\
\begin{quote}
\small
\foreignlanguage{ukrainian}{"Після війни за свій гендер, яка тривала все моє життя, я вирішив прийняти себе таким, яким я є -- всередині і зовні", -- йдеться на сторінці поп-зірки в Instagram. "Я дуже нервував перед тим, як анонсувати це, оскільки приділяю забагато значення тому,\ldots}
\end{quote}

\vspace{6pt}
\textbf{Overall audio article understanding} \\
\textit{How well did you understand the source audio article?}
\begin{itemize}
    \setlength\itemsep{1pt}
    \item[$\circ$] 1 -- Did not understand
    \item[$\circ$] 2 -- Limited understanding
    \item[$\circ$] 3 -- Partial understanding
    \item[$\circ$] 4 -- Good understanding
    \item[$\bullet$] 5 -- Excellent understanding \hfill \textit{(selected)}
\end{itemize}
\vspace{6pt}
\textit{\textbf{For each summary, provide a 1--5 overall quality rating: accurate, engaging, and informative.}}

\vspace{4pt}
\textbf{Summary 1} \\
\small British pop star Sam Smith has publicly announced their non-binary gender identity, admitting that they feel neither male nor female.
\vspace{2pt}
\textit{Overall quality: accurate, engaging, and informative}
\begin{itemize}
    \setlength\itemsep{1pt}
    \item[$\circ$] 1 -- Very Poor
    \item[$\circ$] 2 -- Poor
    \item[$\circ$] 3 -- Fair
    \item[$\circ$] 4 -- Good
    \item[$\bullet$] 5 -- Excellent \hfill \textit{(selected)}
\end{itemize}

\vspace{4pt}
\textbf{Summary 2} \\
\small British singer Sam Smith has publicly come out as non-binary, asking for acceptance and receiving praise from LGBTQ+ organizations like Stonewall for the positive impact of their visibility on the community.
\vspace{2pt}
\textit{Overall quality: accurate, engaging, and informative}
\begin{itemize}
    \setlength\itemsep{1pt}
    \item[$\circ$] 1 -- Very Poor
    \item[$\circ$] 2 -- Poor
    \item[$\circ$] 3 -- Fair
    \item[$\bullet$] 4 -- Good \hfill \textit{(selected)}
    \item[$\circ$] 5 -- Excellent
\end{itemize}

\begin{center}
    [...]
\end{center}

\end{prompt}
\vspace{-13pt}
\captionof{figure}{Example survey question presented to human evaluators, showing the source audio/article context, comprehension check, and per-summary quality rating (Summaries 1 and 2 are shown).}
\label{fig:survey_example}

\null
\newpage
\begin{prompt}[colback=black!0!white, colframe=black!98!black]
{Human Evaluation G-Eval survey question}

\textbf{Listen to this audio article, and rate the proposed summary based on the following criteria.}

\vspace{4pt}
\textbf{Audio Article} \\
\textit{[00:00 -- 06:08, waveform playback of source audio]}

\vspace{6pt}
\textbf{Text Article} \\
\begin{quote}
\small
\begin{CJK*}{UTF8}{mj}
\CJKspace
켄드릭 라마는 미국 앨라배마주에서 공연 중이었다 라마는 앨라배마주의 행아웃 축제(Hangout Festival)에서 그의 곡 'M.A.A.D City'를 부르면서 백인 여성을 무대 위로 불러 그의 노래를 따라부르게 했다. 하지만 여성이 라마의 가사에 포함된 단어 '니그로'를 반복적으로 사용하자 라마는 여성을 제지했다. 이에 관중들은 화를 내며 반응했고, 라마는 "한마디만 더 해 봐"라고 말했다. '니그로'는 흑인을 비하하는 단어로, 노예를 부르\ldots
\end{CJK*}
\end{quote}

\vspace{6pt}
\textbf{Proposed Summary} \\
\small A debate has sparked over whether it is acceptable for non-Black fans to sing the n-word after Kendrick Lamar stopped a white woman from using it while performing his song on stage.

\vspace{8pt}
\textbf{Coherence:} \textit{The summary should be well-structured and well-organized. It should build from sentence to sentence into a coherent body of information about a topic. Rate 1--5 points.}
\textit{Overall quality: accurate, engaging, and informative}
\begin{itemize}
    \setlength\itemsep{1pt}
    \item[$\circ$] 1 -- Not Coherent
    \item[$\circ$] 2 
    \item[$\circ$] 3 
    \item[$\circ$] 4 
    \item[$\bullet$] 5 -- Fully Coherent \hfill \textit{(selected)}
\end{itemize}

\vspace{6pt}
\textbf{Consistency:} \textit{A factually consistent summary contains only statements that are entailed by the source document. Penalize summaries that contain hallucinated facts. Rate 1--5 points.}
\begin{itemize}
    \setlength\itemsep{1pt}
    \item[$\circ$] 1 -- Not Consistent
    \item[$\circ$] 2 
    \item[$\circ$] 3 
    \item[$\circ$] 4 
    \item[$\bullet$] 5 -- Fully Consistent \hfill \textit{(selected)}
\end{itemize}

\vspace{6pt}
\textbf{Fluency:} \textit{The quality of the summary in terms of grammar, spelling, punctuation, word choice, and sentence structure. Rate 1--3 points.}
\begin{itemize}
    \setlength\itemsep{1pt}
    \item[$\circ$] 1 -- Not Fluent
    \item[$\circ$] 3 
    \item[$\bullet$] 5 -- Fully Fluent \hfill \textit{(selected)}
\end{itemize}

\vspace{6pt}
\textbf{Relevance:} \textit{The summary should include only important information from the source document. Please penalize summaries which contained redundancies and excess information. Rate 1--5 points.}
\begin{itemize}
    \setlength\itemsep{1pt}
    \item[$\circ$] 1 -- Not Relevant
    \item[$\circ$] 2 
    \item[$\circ$] 3 
    \item[$\circ$] 4 
    \item[$\bullet$] 5 -- Fully Relevant \hfill \textit{(selected)}
\end{itemize}

\end{prompt}
\vspace{-13pt}
\captionof{figure}{Example survey question presented to human evaluators for single-summary rating on four criteria: coherence, consistency, fluency, and relevance, in parallel with our G-Eval evaluation.}
\label{fig:survey_example_criteria}

\null
\newpage
\begin{prompt}[colback=black!0!white, colframe=black!98!black]
{G-Eval: Coherence}
\begin{verbatim}
You will be given one {tgt_lang} summary written for a {src_lang} source audio.

Your task is to rate the summary on one metric.

Please make sure you read and understand these instructions carefully.
Please keep this audio and corresponding transcription open while reviewing,
and refer to it as needed.

Evaluation Criteria:

Coherence (1-5) - the collective quality of all sentences. We align this
dimension with the DUC quality question of structure and coherence whereby
"the summary should be well-structured and well-organized. The summary
should not just be a heap of related information, but should build from sentence
to a coherent body of information about a topic."

Evaluation Steps:

1. Listen to the source audio and read the source text carefully and
    identify the main topic and key points.
2. Read the summary and compare it to the source audio.
    Check if the summary covers the main topic and key points of
    the source audio, and if it presents them in a clear and logical order.
3. Assign a score for coherence on a scale of 1 to 5, where 1 is the
    lowest and 5 is the highest based on the Evaluation Criteria.


Example:
[AUDIO]

Source Text:

{text}

Summary:

{summary}


Evaluation Form (scores ONLY):

- Coherence (1-5):
\end{verbatim}         

\end{prompt}
\vspace{-13pt}
\captionof{figure}{G-Eval coherence evaluation prompt used with Gemini 3.5-Flash.}
\label{fig:coherence_prompt}

\null
\newpage
\begin{prompt}[colback=black!0!white, colframe=black!98!black]
{G-Eval: Consistency}
\begin{verbatim}
You will be given a source audio and corresponding text. You will then be given
one summary written for this source audio.
Your task is to rate the summary on one metric.
Please make sure you read and understand these instructions carefully. Please keep
this audio and source text open while reviewing, and refer to it as needed.

Evaluation Criteria:
Consistency (1-5) - the factual alignment between the summary and the summarized
source. A factually consistent summary contains only statements that are entailed
by the source document. Annotators were also asked to penalize summaries that
contained hallucinated facts. 

Evaluation Steps:

1. Listen to the audio and read the source text carefully and identify the main
    facts and details it presents.
2. Read the summary and compare it to the source document. Check if the summary
    contains any factual errors that are not supported by the source document.
3. Assign a score for consistency based on the Evaluation Criteria.

Example:
[AUDIO]

Source Text:
{text}

Summary:
{summary}

Evaluation Form (scores ONLY):
- Consistency (1-5):
\end{verbatim}         

\end{prompt}
\vspace{-13pt}
\captionof{figure}{G-Eval consistency evaluation prompt used with Gemini 3.5-Flash.}
\label{fig:consistency_prompt}

\begin{prompt}[colback=black!0!white, colframe=black!98!black]
{G-Eval: Fluency}
\begin{verbatim}
You will be given a source audio and corresponding text. You will then
be given one summary written for this source audio.
Your task is to rate the summary on one metric.

Please make sure you read and understand these instructions carefully.
Please keep this audio and source text open while reviewing, and refer to it as needed.

Evaluation Criteria:
Fluency (1-3): the quality of the summary in terms of grammar, spelling, punctuation,
word choice, and sentence structure.

- 1: Poor. The summary has many errors that make it hard to understand or sound unnatural.
- 2: Fair. The summary has some errors that affect the clarity or smoothness of the text,
    but the main points are still comprehensible.
- 3: Good. The summary has few or no errors and is easy to read and follow.

Example:
[AUDIO]
Source Text:
{text}

Summary:
{summary}

Evaluation Form (scores ONLY):
- Fluency (1-3):
\end{verbatim}         
\end{prompt}
\vspace{-13pt}
\captionof{figure}{G-Eval fluency evaluation prompt used with Gemini 3.5-Flash.}
\label{fig:fluency_prompt}

\begin{prompt}[colback=black!0!white, colframe=black!98!black]
{G-Eval: Relevance}
\begin{verbatim}
You will be given a source audio and corresponding text. You will then be
given one summary written for this source audio.

Your task is to rate the summary on one metric.
Please make sure you read and understand these instructions carefully.
Please keep this audio and source text open while reviewing, and refer to it as needed.

Evaluation Criteria:
Relevance (1-5) - selection of important content from the source. The summary
should include only important information from the source document. Annotators
were instructed to penalize summaries which contained redundancies and excess information.

Evaluation Steps:

1. Listen to the audio and read the source text carefully.
2. Compare the summary to the source document and identify the main points
    of the source document.
3. Assess how well the summary covers the main points of the source document,
    and how much irrelevant or redundant information it contains.
4. Assign a relevance score from 1 to 5.

Example:
[AUDIO]

Source Text:
{text}

Summary:
{summary}


Evaluation Form (scores ONLY):
- Relevance (1-5):
\end{verbatim}         

\end{prompt}
\vspace{-13pt}
\captionof{figure}{G-Eval relevance evaluation prompt used with Gemini 3.5-Flash.}
\label{fig:relevance_prompt}

\end{document}